\documentclass[reprint,amsmath,amssymb,aps,prx,superscriptaddress,longbibliography,floatfix]{revtex4-2}

\usepackage{graphicx}
\usepackage{dcolumn}
\usepackage{bm}
\usepackage{bbm}
\usepackage{soul}
\usepackage{enumitem}
\usepackage{makecell}
\usepackage{physics}

\usepackage{braket}
\usepackage{graphicx}
\usepackage{amsmath,verbatim,latexsym,amssymb,indentfirst,mathrsfs,mathtools,amsthm,bbm,bm,hyperref,url}

\newcommand{\Max}{\mathrm{max}}
\newcommand{\Min}{\mathrm{min}}
\newcommand{\vst}{\mathcal{V}}
\newcommand{\q}{{\mathrm{q}}}
\newcommand{\antiq}{{\bar{{\mathrm{q}}}}}
\newcommand{\coup}{{\mathrm{c}}}

\newcommand{\LParen}{ \bm{(} }
\newcommand{\RParen}{ \bm{)} }
\renewcommand{\Tr}{{\rm Tr}}  
\def\id{\mathbf{1}}   

\renewcommand*{\ketbra}[2]{\lvert #1 \rangle\!\langle #2 \rvert}

\renewcommand*{\expval}[1]{\left\langle  #1  \right\rangle}
\usepackage{cancel}
\usepackage{bbold}

\usepackage[dvipsnames]{xcolor}
\usepackage[normalem]{ulem}
\usepackage{physics}

\newcommand{\kater}[1]{{\color{black}#1}}

\newcommand{\xingrui}[1]{{\color{black}#1}}
\newcommand{\response}[1]{{\color{black}#1}}

\newcommand{\reviewercomment}[3]{}
\newcommand{\deletedtext}[1]{}
\newcommand{\deletedtextcontinued}[1]{}
\newcommand{\deletedinline}[1]{}

\begin{document}


\title{\response{Superconducting-antiqubit simulation enables optimal phase estimation via unitary inversion }}

\author{Xingrui Song*}
\affiliation{
School of Applied and Engineering Physics, Cornell University, Ithaca, NY 14853, USA
}

\author{Surihan Sean Borjigin*}
\affiliation{Department of Electrical Engineering and Computer Science, University of California Berkeley, Berkeley, CA, USA, 94720.}

\author{Flavio Salvati*}
\affiliation{
Cavendish Laboratory, Department of Physics, University of Cambridge, Cambridge, CB3 0HE, United Kingdom
}

\author{Yu-Xin Wang}
\affiliation{
Joint Center for Quantum Information and Computer Science, NIST and University of Maryland, College Park, Maryland 20742, USA
}

\author{Nicole Yunger Halpern}
\affiliation{
Joint Center for Quantum Information and Computer Science, NIST and University of Maryland, College Park, Maryland 20742, USA
}
\affiliation{
Institute for Physical Science and Technology, University of Maryland, College Park, Maryland 20742, USA
}

\author{David R. M. Arvidsson-Shukur}
\affiliation{
Hitachi Cambridge Laboratory, J. J. Thomson Avenue, Cambridge CB3 0HE, United Kingdom
}

\author{Kater Murch}
\email{katermurch@berkeley.edu}
\affiliation{Department of Electrical Engineering and Computer Science, University of California Berkeley, Berkeley, CA, USA, 94720.}
\affiliation{Department of Physics, University of California Berkeley, Berkeley, CA, USA, 94720.}

\date{\today}

\begin{abstract}
    A positron is equivalent to an electron traveling backward through time. Casting transmon superconducting qubits as akin to electrons, we simulate a positron with a transmon subject to particular resonant and off-resonant drives. We call positron-like transmons ``antiqubits.'' An antiqubit's effective gyromagnetic ratio equals the negative of a qubit's. This fact enables us to time-invert a unitary implemented on a transmon by its environment. We apply this platform-specific unitary inversion, with qubit--antiqubit entanglement, to achieve a quantum advantage in phase estimation: consider measuring the strength of a field that points in an unknown direction. An entangled qubit--antiqubit sensor offers the greatest possible sensitivity (amount of Fisher information), per qubit, per application of the field. We prove this result theoretically and observe it experimentally. This work shows how antimatter, whether real or simulated, can enable platform-specific unitary inversion and benefit quantum information processing.
\end{abstract}

\maketitle

\section{Introduction}
\label{sec_Introduction}

One can regard a positron as an electron moving backward in time.
This insight has impacted fundamental physics, as championed by Wheeler, Feynman, and St\"{u}ckelberg~\cite{42_Stuckelberg,48_Feynman_Relativistic,49_Feynman_Theory,65_Feynman_Nobel}. Can one apply the insight for practical benefit? One practical challenge, common across quantum computation and metrology, is \emph{unitary inversion}. Consider a quantum system subject to an external field that effects the unitary operation $U_\alpha \coloneqq e^{-i \alpha H} \, .$ (We set $\hbar = 1$.) Often, one wishes to implement $U_\alpha^\dag$. Such unitary inversion can help one detect quantum-information scrambling~\cite{16_Swingle_Measuring,16_Zhu_Measurement,18_NYH_Quasiprobability}, superpose time evolutions \cite{Stromberg2024}, implement higher-order quantum transformations (e.g., evolutions of quantum channels) \cite{Yoshida2023, Bisio_2019}, measure out-of-time-order correlators \cite{corr1, corr2}, and support quantum singular-value transformations~\cite{Gily_n_2019, Martyn2021}.

Conventional unitary inversion can operate imprecisely and requires substantial resources. In a na\"{i}ve approach, one infers the form of $U_\alpha$ from process tomography~\cite{Chuang97, DAriano01, Altepeter03, Song2021}. Then, one constructs a setup intended to apply $U_\alpha^\dag$. Alternatively, one can implement $U_\alpha^\dag$ with a unitary-reversal algorithm~\cite{Yoshida2023, chen2024}. Such an algorithm costs many $U_\alpha$ applications, which serve as resources in quantum computing and metrology: let $d$ denote the system-of-interest Hilbert space's dimensionality. The algorithms require $\mathcal{O}(d^2)$ applications of $U_\alpha$ per $U_\alpha^\dag$ implementation, like tomography.

We propose to invert unitaries by applying the positron insight above. The electron's gyromagnetic ratio, $\gamma$, helps determine the electron's magnetic moment and hence the electron's coupling to an external field and hence $\alpha$. The positron has a gyromagnetic ratio $-\gamma$. If exposed to the same field as an electron, a positron undergoes $e^{-i (-\alpha) H} = U_\alpha^\dag$. Hence one can effectively invert a unitary, and so effectively reverse time, by exchanging matter with antimatter.

\reviewercomment{3}{other comment 1}{The term antiqubit was mentioned in Phys. Rev. Lett. 79, 5194 (1997) as a quantum of negative information. The manuscript should distinguish this original information-theoretic definition from the present use of negative gyromagnetic ratio terminology for a transmon qubit implementing an engineered inverse unitary.}

We experimentally simulate an exchange of a transmon with antimatter. The strategy relies on resonant and off-resonant drives, which negate the effective gyromagnetic ratio of the transmon's pseudospin. This negation inverts the unitary to which a magnetic field subjects the transmon, regardless of the field's direction. We call this process \emph{platform-specific unitary inversion}. Also, we call positron-like transmons \emph{antiqubits}\footnote{The term \emph{antiqubit} was previously used in an information-theoretic sense to denote a quantum of negative information~\cite{Cerf1997}. Here, we use the term in a distinct sense to denote a transmon whose effective gyromagnetic ratio is the negative of an ordinary transmon qubit's, such that an applied field implements $U_\alpha^\dagger$ on the antiqubit and $U_\alpha$ on the qubit.} 

We apply platform-specific unitary inversion and entanglement to achieve a quantum advantage in sensing. Consider a field pointing in an arbitrary direction, which could be unknown. One measures the field strength $\alpha$ in \emph{quantum phase estimation}, a task prevalent in quantum algorithms and metrology~\cite{Colombo_2022, Apellaniz_2018, Ruster_2017, Tang18, Wei21, Stankevic23,smith2023adaptive,Smith24, rovny2025multiqubitnanoscalesensingentanglement, smith2025riskminimizingstatesquantumphaseestimationalgorithm}. To accomplish this task, we entangle a qubit transmon with an antiqubit. We call the entangled pair \emph{synthetic positronium,} after the positronium atom formed from a bound-together electron and positron. The field applies $U_\alpha$ to the qubit and, via platform-specific unitary inversion, $U_\alpha^\dag$ to the antiqubit. This inversion boosts the Fisher information (FI) obtainable, or sensitivity achievable, about the field. We quantify the resources required using the \emph{space--time volume} $\vst$, defined as (number of transmons used)$\times$(number of sequential applications of the \response{field})\response{~\cite{feynman1982,Nielsen11}.} Our strategy achieves the greatest possible amount of FI per two units of space--time volume, 4. [We fix the space--time volume to be $\vst = 2$ throughout this proof-of-principle work for simplicity. However, an extension of our strategy achieves the maximum possible Fisher information, regardless of the space--time volume (Sec.~\ref{sec_Outlook})]. Experimentally, we infer a readout corrected FI of \response{2.98} per two units of space--time volume. In contrast, two entangled qubits can achieve $1$; and a separable qubit--antiqubit state, $4/3$, on average over trials. Hence entanglement and platform-specific unitary inversion, realized with antimatter simulation, enable a quantum advantage in metrology. Furthermore, we prove, that entanglement and effective unitary inversion enable the unique strategy that achieves the optimal FI per two units of space--time volume.

The rest of this paper is organized as follows. Section~\ref{sec_Backgrnd_phase_est} provides background about phase estimation. Section~\ref{sec_Theory_pos_met} theoretically introduces positronium sensing facilitated by platform-specific unitary inversion. 
In Sec.~\ref{sec_Exp_setup}, we describe the experimental setup. Section~\ref{sec_U_inv} explains how we perform platform-specific unitary inversion. The experimental results follow in Sec.~\ref{sec_Pos_exprmt}. Section~\ref{sec_Outlook} outlines opportunities established by this work. \response{Although we implement platform-specific unitary inversion on superconducting qubits, the technique can extend to other platforms: conventional quantum hardware and true antimatter. This versatility can enable applications to quantum algorithms, as well as enhance sensors for diverse stimuli.
}

\section{Background: phase estimation}
\label{sec_Backgrnd_phase_est}

This section introduces a basic task in quantum metrology, phase estimation. Below, we specify the task, then review common metrics for evaluating success: the FI and quantum Fisher information (QFI). We illustrate with a qubit example and three approaches to it. We compare the approaches using a metric that reflects not only the information gained metrologically, but also the resources spent: the space--time volume.

\begin{figure}
    \centering
    \includegraphics[width=0.4 \textwidth]{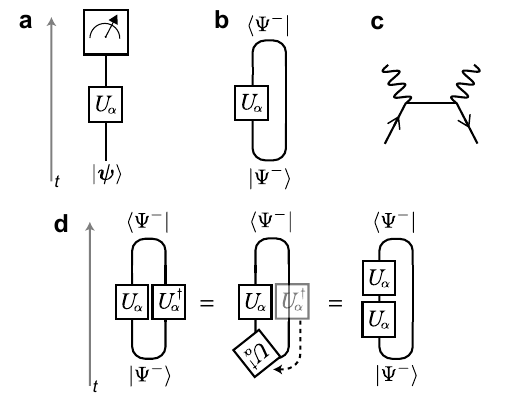}
\caption{{\bf Quantum phase estimation, including platform-specific unitary inversion implemented with synthetic positronium.} 
(a) Circuit diagram for basic phase estimation. Time runs vertically. One prepares a probe in a state $\ket{\psi}$, evolves it under a unitary $U_\alpha$, and measures the probe.  
(b) Agnostic phase estimation employs a two-qubit entangled state to simulate a closed timelike curve. The $\cup$ represents singlet preparation; and the $\cap$, a measurement of whether the state remains a singlet.
(c) Read from bot- tom to top, this diagram represents electron–positron scattering that produces photons. One can view the positron as an electron moving backward through time, as suggested by the arrows' orientations. We leverage this insight to perform platform-specific unitary inversion and positronium sensing.
(d) Positronium sensing involves a qubit probe that undergoes $U_\alpha$ and an antiqubit ancilla that, due to platform-specific unitary inversion, undergoes $U_\alpha^\dagger$ (leftmost diagram). Imagine sliding the $U_\alpha^\dagger$ box leftward along the wire (central diagram). When passing through the $\cup$, the $U_\alpha^\dag$ is inverted. Hence positronium sensing is equivalent to implementing $U_\alpha^2$ on the probe (rightmost diagram). Doubling the number of $U_\alpha$s applied, relative to agnostic sensing, doubles the amount of FI achievable per two units of space-time volume.
}
    \label{fig1}
\end{figure}

The following task exemplifies phase estimation. Let $U_\alpha$ denote an arbitrary unitary parameterized by $\alpha \in \mathbb{R}$. To estimate $\alpha$, we proceed as in Fig.~\ref{fig1}(a): we prepare a probe in $\ket{\psi}$. The unitary maps $\ket{\psi}$ to $U_\alpha \ket{\psi} \eqqcolon \ket{\psi_\alpha}$. We measure some basis $\{ \ket{j} \}$ of the probe. Outcome $j$ occurs with a probability $P_j = |\langle j | \psi_{\alpha} \rangle|^2$.  

The FI $I_\alpha$ quantifies the probability distribution's sensitivity to changes in $\alpha$:
$I_\alpha \coloneqq \sum_{j}{(\partial_{\alpha} P_j)^2/P_j}$. Maximizing the FI over measurements yields the QFI, $\mathcal{I}_\alpha$~\cite{Braunstein94,helstrom76}\response{~\footnote{
\response{To evaluate positronium sensing, we evaluate $\mathcal{I}_\alpha$ on the QFI-maximizing probe state. This QFI equals the channel QFI of~\cite{25_Gardner_Stochastic}.}}}:
\begin{equation}
    I_\alpha \leq
     \mathcal{I}_\alpha = 4 \, 
    \left( \langle \partial_\alpha \psi_{\alpha} | \partial_\alpha \psi_{\alpha} \rangle 
    - \abs{ \langle \psi_{\alpha} | \partial_\alpha \psi_{\alpha} \rangle }^2 \right)   . 
    \label{eq_QFI_bound_form}
\end{equation}
The QFI, too, obeys an upper bound: suppose that $U_{\alpha} = e^{-i \alpha H} \, .$ Denote by $\Delta H$ the difference between the generator's greatest eigenvalue, $E_\Max$, and the least, $E_\Min$: $\Delta H \coloneqq E_\Max - E_\Min$. The QFI obeys 
\begin{equation}
\label{eq_QFI_var}
\mathcal{I}_{\alpha} 
\leq \max_{ \ket{\psi} } \left\{ \bra{\psi} H^2 \ket{\psi} -  \bra{\psi} H \ket{\psi}^2 \right\} 
= (\Delta H)^2 \, .
\end{equation}
The QFI saturates this bound if $\ket{\psi}$ is an equal-weight superposition of an eigenvalue-$E_\Max$ eigenstate and an eigenvalue-$E_\Min$ eigenstate.

We illustrate with a qubit subject to an external field. Denote  by $\hat{\bm{n}} = n_x \hat{\bm{x}} + n_y \hat{\bm{y}} + n_z \hat{\bm{z}}$ the field's direction; and, by $\bm{\sigma} = X \hat{\bm{x}} + Y \hat{\bm{y}} + Z \hat{\bm{z}} \equiv \sigma_x \hat{\bm{x}} + \sigma_y \hat{\bm{y}} + \sigma_z \hat{\bm{z}}$, a vector of the Pauli operators. For future reference, we denote by $\ket{a \pm}$ the $\sigma_a$ eigenstate associated with the eigenvalue $\pm 1$. Also, we define 
$\ket{0} \coloneqq \ket{z+}$ and $\ket{1} \coloneqq \ket{z-}$.
The unitary $U_{\alpha} = e^{-i\alpha \hat{\bm{n}}\cdot\bm{\sigma} / 2 }$ effects a rotation whose angle $\alpha$ we wish to estimate. Since the generator $H = \hat{\bm{n}}\cdot\bm{\sigma}/2$, the QFI is $(\Delta H)^2 = 1$. 

We now present three strategies for inferring about $\alpha$. First, suppose we know the rotation axis $\hat{\bm{n}}$. We can achieve the QFI by preparing even-weight superpositions of the $H$ eigenstates.

Second, suppose we do not know $\hat{\bm{n}}$ and can use only one qubit per trial. We lack information about three parameters: $\alpha$ and the two angles that specify $\hat{\bm{n}}$. In contrast, only two parameters specify a pure qubit state~\footnote{Optimal probe states are pure, as mixed states result from the qubit's leaking information into an environment.}. Due to this discrepancy, we cannot necessarily estimate $\alpha$ from copies of just one qubit state~\cite{Song2024}: if we are unlucky, the Bloch vector points along $\pm \hat{\bm{n}}$. Consequently, the field cannot rotate the qubit, which acquires no information about the field. The optimal strategy is to prepare probes in $X$ eigenstates in one batch of trials, $Y$ eigenstates in another batch, and $Z$ eigenstates in a third batch. On average over trials, this strategy achieves the QFI, $2/3 < 1$~\cite{Song2024}.

Third, suppose that we can use two qubits per trial. \emph{Agnostic phase estimation} leverages entanglement to boost the QFI, which is again achievable~\cite{Song2024}. Figure~\ref{fig1}(b) illustrates the protocol: we prepare a probe qubit and an ancilla qubit in the singlet 
$\ket{\Psi^-} \coloneqq \frac{1}{\sqrt{2}}(\ket{01}-\ket{10})$.
Then, $U_\alpha$ acts on the probe. We measure the positive-operator-valued measure (POVM)~\cite{Nielsen11} 
$\{\ketbra{\Psi^-}{\Psi^-},\id -\ket{\Psi^-}\bra{\Psi^-}\}$. 
The \response{farther $\alpha$ lies from zero (or the nearest multiple of $2\pi$)}, the more $U_\alpha$ perturbs the joint state away from $\ket{\Psi^-}$, and the \response{lesser} our probability of observing the $\ketbra{\Psi^-}{\Psi^-}$ outcome. This strategy yields an FI equal to the QFI, 1. Agnostic sensing requires no knowledge of $\hat{\bm{n}}$ and requires only one $U_\alpha$ application per trial---but requires two qubits per trial.

To compare metrological schemes comprehensively, we need a metric that incorporates not only FI, but also resource costs. Different schemes may require different numbers of unitary applications, and different numbers of particles, per trial. \response{Hence we invoke the \emph{space--time volume} $\vst$~\cite{feynman1982}. 
This measure reflects real-world quantum technologies' practical costs, as quantified with complexity theory. In complexity theory, the total resource tracked equals the product of memory and runtime~\cite{Nielsen11}. In quantum computing, an algorithm's space--time volume equals the number of qubits times the number of circuit layers: qubits and gates serve as resources~\cite{Microsoft_roadmap,25_Barends_Performance,IBM_roadmap,Google_roadmap}. Furthermore, quantum-metrological protocols are algorithms run on quantum hardware. Hence the space--time volume measures the protocol's cost. In metrology, the most challenging gates are the unknown unitaries ($U_\alpha$ and $U_\alpha^\dag$) applied by the field over which the metrologist lacks much control. Hence
}
the space--time volume here equals the number of qubits used, times the number of sequential $U_\alpha$ applications, per trial~\footnote{
Consider applying $U_\alpha$ to each of two qubits in parallel. The unitaries, together, cost no more than one $U_\alpha$. Figure~\ref{fig2}(a) illustrates why: one uniform field implements both unitaries simultaneously.}.
We compare protocols' FIs per $\vst = 2$ units of space--time volume, as some protocols require two qubits each.

Let us apply this metric to phase estimation in the absence of knowledge about $\hat{\bm{n}}$.
One qubit can yield an average FI, per two units of space–time volume, of $I_\alpha = 4/3$. Agnostic sensing yields only 1. \response{This strategy is less efficient than the single-qubit strategy, we see when incorporating both resources---not only time, but also memory---into the cost.
}

We show next how to outperform both these metrological strategies with positronium sensing. It achieves an $I_\alpha$ of 4 per two units of space-time volume, saturating the theoretical limit on this metric (Appendix~\ref{sec_Proof_positronium_QFI}). \response{Positronium sensing therefore quadruples the FI rate achievable in agnostic sensing~\cite{Song2024}. } 
Only effective unitary inversion, combined with a singlet, can achi eve the FI rate of 4 (Appendix~\ref{App_positronium_unique_optimal}). 
Our approach relies on entanglement and on platform-specific unitary inversion applied to a qubit--antiqubit pair.

\section{Theory of positronium sensing}
\label{sec_Theory_pos_met}

Using positronium sensing, one can achieve the optimal FI about $\alpha$ per two units of space–time volume, without knowing $\hat{\bm{n}}$. The scheme leverages two distinct processes that are mathematically equivalent to time reversals (two \emph{effective} time reversals). We introduce the two sequentially, then assess the scheme's efficacy. 

The first effective time reversal results from entanglement manipulation and enables agnostic phase estimation. The previous section reviewed agnostic phase estimation~\cite{Song2024}, depicted in Fig.~\ref{fig1}(b). We can read that figure in two ways. Traversed from bottom to top, the figure shows (i) a probe qubit and ancilla qubit prepared in a singlet, (ii) the probe undergoing a unitary, and (iii) a measurement of whether the qubits remain in a singlet. Alternatively, we can interpret the figure as one qubit's worldline. The worldline forms a loop, or closed timelike curve~\cite{CTC_Deutsch, CTC_Lloyd_1, CTC_Lloyd_2, ArvidssonShukur23}: the qubit reverses temporal direction, relative to the laboratory rest frame, at the $\cup$ and $\cap$. (The qubit also undergoes a transformation, discussed below, at each turning point.) These symbols represent entanglement manipulations in the first reading of the figure. The entanglement manipulations---and the effective time reversals---enable agnostic phase estimation as follows. When performing agnostic phase estimation, we do not know the field direction $\hat{\bm{n}}$. Hence we do not know the optimal state in which to prepare the probe at the beginning of the protocol. We effectively teleport that state backward in time (from the diagram's top to its bottom) \response{\cite{Bennett1993Teleportation, Pirandola2015TeleportationReview}}, via the entanglement manipulations---by effectively reversing time.

During agnostic phase estimation, the field acts only on the probe\response{, not on the ancilla.} Can we boost the FI by acting the field on the ancilla? Doing so na\"ively would \response{prevent the protocol from imprinting the parameter on the two-qubit state}:
$\ket{\Psi^-} 
\mapsto (U_\alpha \otimes U_\alpha) \ket{\Psi^-} 
=\ket{\Psi^-}$.
The second $U_\alpha$ undoes the first one's effect, due to the singlet's rotational invariance under tensor products of identical unitaries. 

Instead, we propose, we should subject the ancilla to $U_\alpha^\dag$. We do so via the second effective time reversal: we convert the ancilla into an antiqubit by negating its effective gyromagnetic ratio. This negation negates $\alpha$, subjecting the (antiqubit) ancilla to 
$e^{-i (-\alpha) \hat{\bm{n}} \cdot \bm{\sigma} / 2} 
= U_\alpha^\dag$. 
This strategy, we call \emph{positronium sensing}. \response{It outperforms agnostic sensing by imprinting information directly on the ancilla, as well as on the probe.} Figure~\ref{fig1}(c) illustrates the fundamental insight that we leverage: interchanging matter with antimatter effectively reverses time.

Figure~\ref{fig1}(d) illustrates how positronium sensing boosts the FI. The leftmost diagram shows the qubit probe undergoing $U_\alpha$ and the antiqubit ancilla undergoing $U_\alpha^\dag$. Consider sliding the $U_\alpha^\dag$ box downward along the wire, as in the central diagram. When the box passes through the $\cup$, it undergoes a universal NOT (for the central diagram to represent the same physics as the leftmost)~\cite{Song2024}: $U_\alpha^\dag 
= e^{i \alpha \hat{\bm{n}} \cdot \bm{\sigma} / 2}
\mapsto e^{-i \alpha \hat{\bm{n}} \cdot \bm{\sigma} / 2}
= U_\alpha \, .$ 
Implementing positronium sensing, therefore, is equivalent to implementing $U_\alpha^2$ on the probe [rightmost diagram in Fig.~\ref{fig1}(d)]. However, truly implementing $U_\alpha^2$ costs two units of space--time volume, whereas our scheme costs one unit.
The effective $U_\alpha^2$ doubles the FI gleaned about $\alpha$: positronium sensing achieves $I_\alpha = 4$ at the cost of $\vst = 2$. This FI equals the corresponding QFI, which is the greatest possible QFI achievable with any strategy that consumes $\leq 2$ units of space--time volume, even if one knows $\hat{\bm{n}}$ (Appendix~\ref{sec_Proof_positronium_QFI}).

\section{Experimental setup}
\label{sec_Exp_setup}

Our experiment features three transmons: a qubit q that acts as ordinary matter, an antiqubit $\antiq$, and a coupler c. (We avoid the phrases \emph{transmon qubit} and \emph{superconducting qubit}, reserving the word \emph{qubit} for q.) The following frequencies specify the gaps between transmons' ground and first-excited energy levels at zero flux (in the absence of drives): 
$\omega_\q / (2\pi) = 4.167$ GHz, 
$\omega_\antiq/(2\pi) = 4.274$ GHz, and 
$\omega_\coup / (2\pi) = 5.250$ GHz.  We modulate $\omega_\mathrm{c}$ to activate  parametric interactions between q and $\antiq$.

We prepare the singlet via the following steps. Each transmon begins in $\ket{z+}$. We subject q to a $\pi$-pulse, preparing it in $\ket{z-}$. Next, we effect a $\sqrt{i\mathrm{SWAP}}$ gate, 
by implementing the resonant parametric interaction for 104 ns. Finally, we correct the antiqubit's phase (Appendix~\ref{sec:parametric}). This process prepares the singlet with a fidelity of 97 \%. 

Similarly, we measure $\{ \ketbra{\Psi^-}{\Psi^-}, \id - 
\ketbra{\Psi^-}{\Psi^-} \}$ in two steps. First, we perform a $\sqrt{i\mathrm{SWAP}} \, .$ Then, we measure each qubit's computational basis. We achieve single-qubit measurement fidelities of $\approx 95 \ \%$, 
for which we correct throughout this work. 

In the absence of extraneous fields (fields other than the quantization field), we set each transmon's Hamiltonian $\propto Z$, in accordance with convention.
Resonant drives effect controlled rotations about the $x$- and $y$-axes. Therefore, transmon $j \in \{\q, \antiq \}$ has a Hamiltonian expressible in terms of effective-magnetic-field components $\Omega_x$, $\Omega_y$ and $\delta_j$:
\begin{equation}
H_j = \Omega_x X + \Omega_y Y + \delta_j Z. \label{eq:magneticH}
\end{equation}
\response{We realize $\Omega_x$ and $\Omega_y$ by applying resonant microwave drives at the transmon's transition frequency. In the rotating frame defined by these drives, the Hamiltonian assumes the form in Eq.~(\ref{eq:magneticH}).}

\reviewercomment{3}{major comment 4}{It is important to note when both transmons are exposed to the external, uncontrolled magnetic field. The manuscript does not explain the flux-to-phase transfer function and the corresponding noise-equivalent sensitivity. It is important to include the flux dependence for the flux-tunable transmon, specifically the first-order sensitivity to a small magnetic-flux perturbation. The probability may have been optimized in response to the engineered phase without optimizing the device as a magnetic-field sensor.}

The signal in Eq.~\eqref{eq:magneticH} is a physical radio-frequency field in the microwave band. This field 
rotates the transmons. The $x$- and $y$-rotation drives reach the 
transmons through their individual drive lines. 
To rotate the transmons about the $z$-axis, we inject a 4.176998~GHz AC Stark tone 
into the shared readout-input line, so both transmons 
rotate simultaneously. The estimated parameter is the dimensionless rotation angle $\alpha$. Magnetic flux does not serve as the sensing signal; the Fisher information reported here quantifies sensitivity to the dimensionless rotation angle $\alpha$ induced by the applied microwave field.
We operate the flux-tunable antiqubit and coupler 
at net-zero-flux sweet spots. There, their transition frequencies 
do not respond to small flux fluctuations at first order in $\delta\Phi$, where $\delta\Phi$ denotes the deviation from zero flux.

\begin{figure*}
    \centering
    \includegraphics[width=.75\textwidth]{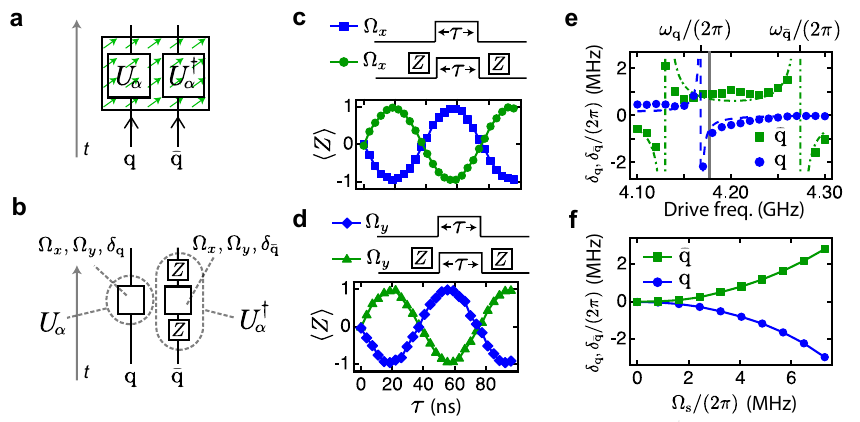}
 \caption{{\bf Platform-specific unitary inversion.} (a) Qubit and antiqubit transmons. An external field (green arrows) effects the unitary $U_\alpha$ on $\q$ and $U^\dagger_\alpha$ on $\antiq$. 
(b) We simulate antimatter by applying two techniques. q undergoes $U_\alpha$ due to a pseudomagnetic field with components $\Omega_x$, $\Omega_y$, and $\delta_\q$. The antiqubit undergoes $U_\alpha^\dag$. The platform-specific unitary inversion results from $Z$ gates and from the field components $\Omega_x$, $\Omega_y \, ,$ and $\delta_\antiq \, .$
(c) The $Z$ gates effectively invert the external-field component $\Omega_x$. To check, we prepare $\ket{y+}$, then apply the field for a time $\tau$. In half the trials, we apply a $Z$ gate before the field application and another $Z$ afterward.
(d) Same as (c), except with $\Omega_y$ and $\ket{x+}$.
(e) An off-resonant drive generates the external field's $z$-component. The drive induces an AC Stark shift of $\delta_\q$ on the qubit and $\delta_\antiq$ on the antiqubit. Dashed lines represent theoretical predictions. Points represent frequency shifts measured with Ramsey measurements.  At 4.177 GHz (solid gray line), $\delta_\q = - \delta_\antiq$. The qubit has a transition frequency $\omega_\q / (2\pi)$; and the antiqubit, one of $\omega_\antiq / (2\pi)$.
(f) AC Stark shifts versus drive amplitude $\Omega_\mathrm{s}$ (Appendix~\ref{sec:antiq}).
} 
    \label{fig2}
\end{figure*}

\section{Platform-specific unitary inversion}
\label{sec_U_inv}

Figure~\ref{fig2}(a) sketches transmon-specific unitary inversion. The qubit and antiqubit pass through the same magnetic field (green arrows).
The qubit undergoes $U_\alpha$. The antiqubit undergoes $U_\alpha^\dag$, having an equal-magnitude, opposite-sign effective gyromagnetic ratio. We effect this gyromagnetic ratio via two control techniques, depicted in Fig.~\ref{fig2}(b): (i) We apply a $Z$ gate before subjecting $\antiq$ to the Hamiltonian (\ref{eq:magneticH}), then apply a $Z$ gate afterward. The gates effectively invert the magnetic field's $x$- and $y$-components. (ii) We apply a single-frequency drive to both transmons. It induces an alternating-current (AC) Stark shift of $\delta_\q$ to the qubit and a shift $\delta_\antiq = - \delta_\q$ to the antiqubit. This strategy effectively inverts the $z$-component of the magnetic field experienced by $\antiq$. \response{The previous experiment~\cite{Song2024} did not involve this strategy, the crux of this paper's experimental innovation, which enables platform-specific unitary inversion.} We now elaborate on these techniques sequentially.

$Z$ gates effectively invert the magnetic field's $x$- and $y$-components. To show how, we leverage the Taylor expansion of 
$U_\alpha = e^{-i \alpha \hat{\bm{n}} \cdot \bm{\sigma} / 2}$ and the Pauli operators' commutation relations:
\begin{align}
   Z e^{- i \alpha \hat{\bm{n}} \cdot \bm{\sigma} / 2} \, Z
   & = e^{- i \alpha \hat{\bm{n}} \cdot (Z \bm{\sigma} Z) / 2} \\
   & = e^{-i \alpha ( -n_x X - n_y Y + n_z Z) / 2} \, .
\end{align}
\response{We implement a $Z$ gate by rotating the qubit through $\pi$ about the $z$-axis, using two on-resonance microwave pulses (Appendix~\ref{subsec:single_qubit_rot}). The $Z$ gates contribute to the platform-specific unitary inversion. However, they incur no space–time overhead under our resource definition: the $Z$ gates do not depend on $\alpha$ and so can be executed outside the sensing period, as follows. The $\sqrt{i\mathrm{SWAP}}$ gates (used to prepare and detect entanglement) induce known $z$-axis rotations.
One can absorb these rotations into the control sequence through parametric-pump phase adjustments or virtual frame updates. The $Z$ gates become virtual, requiring no quantum-hardware time.}

Figure~\ref{fig2}(c) demonstrates this strategy's efficacy. We constructed the figure from the following two protocols, the first applied to q and the second applied to $\antiq$:  
in each of many trials, we prepared q in $\ket{+y}$. We set $\Omega_x/(2\pi)$ to 13.2 MHz for a time $\tau$. In each of many other trials, to simulate antimatter, we performed $Z$ gates before and after the $\Omega_x$ pulse. 
Figure~\ref{fig2}(c) shows the transmons' $\expval{Z}$s plotted against time.
The square blue markers represent qubit data; and the circular green markers, antiqubit data. $\antiq$ rotates oppositely q, as expected. The blue diamonds and green triangles follow from analogous experiments that begin with $\ket{+x}$ and involve rotations about the $y$-axis.

We have shown how to effectively invert a field's $x$- and $y$- components, using $Z$ gates implemented with resonant drives. Using an off-resonant microwave drive, 
we can effectively invert a field's $z$-component. We drive both transmons simultaneously. They begin with different energy gaps and so experience different AC Stark shifts: the qubit's gap changes by an amount $\delta_\q$; and the antiqubit's, by $\delta_\antiq$. Figure~\ref{fig2}(e) shows $\delta_\q$ and $\delta_\antiq$ versus drive frequency. At the magic frequency
4.177~GHz, $\delta_\antiq = - \delta_\q$; the qubit and antiqubit undergo opposite $z$-rotations. Figure~\ref{fig2}(f) shows $\delta_\q$ and $\delta_\antiq$ versus drive amplitude. This figure confirms that one drive frequency induces equal-magnitude, opposite-sign AC Stark shifts on the transmons.


\begin{figure*}
    \centering
    \includegraphics[width=\textwidth]{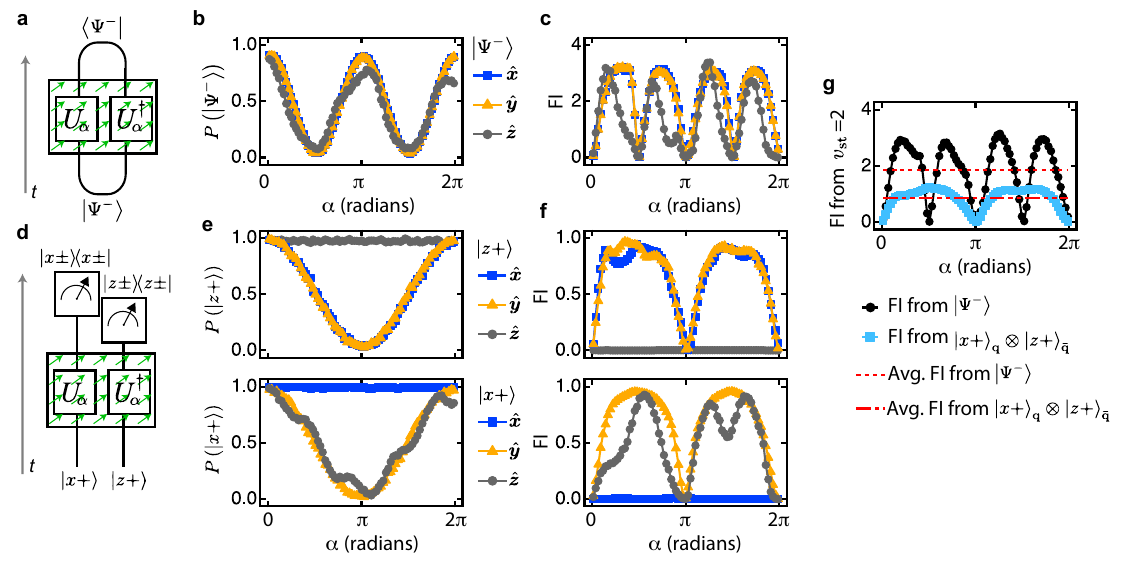}
\caption{\response{{\bf Phase estimation with qubit--antiqubit pairs.} 
(a) Circuit diagram for agnostic phase estimation with synthetic positronium. We prepare $\q$ and $\antiq$ in the singlet $\ket{\Psi^-}$. Then, platform-specific unitary inversion effects $U^\dagger$ on the antiqubit. Finally, we measure whether the pair is in a singlet or not. 
(b) Probability $P(\ket{\Psi^-})$ that the measurement yields the singlet outcome, plotted against $\alpha$. Different colors, marker shapes, and line textures correspond to different rotation axes.
(c) FI inferable via positronium sensing per $\vst {=} 2$ units of space–time volume, versus rotation angle $\alpha$.
(d) Circuit diagram for optimal entanglement-free phase estimation performed with a qubit and an antiqubit.
(e) Probabilities associated with possible outcomes of the entanglement-free experiment's measurements.
(f) FI inferable per $\vst {=} 2$ from the entanglement-free experiment, versus rotation angle $\alpha$.
(g) Average FI inferable per $\vst {=} 2$ versus $\alpha$. Black circular markers come from positronium sensing; and blue squares, from the entanglement-free experiment. The horizontal red lines report the FIs averaged over the $\alpha$ values: 1.85 in positronium sensing and 0.86 in the entanglement-free experiment.
}}
    \label{fig3}
\end{figure*}

\section{Positronium-sensing experiment}
\label{sec_Pos_exprmt}

Figure~\ref{fig3} details the phase-estimation experiment enhanced by synthetic positronium and platform-specific unitary inversion. Below, we detail the protocol. Then, we analyze the possible measurement outcomes' probabilities. From the probabilities, we calculate the FI per two units of space--time volume. We then compare positronium sensing with a protocol that features q and $\antiq$ but no entanglement.

The circuit diagram~\ref{fig3}(a) shows our positronium-sensing protocol's three steps:
first, we prepare q and $\antiq$ in a singlet. Second, we implement $U_\alpha$ on q and $U_\alpha^\dag$ on $\antiq$, using platform-specific unitary inversion. Third, we measure the POVM $\{ \ketbra{\Psi^-}{\Psi^-}, \id - \ketbra{\Psi^-}{\Psi^-} \}$. Let $P(\ket{\Psi^-})$ denote the probability of obtaining the outcome associated with the first projector.

Figure~\ref{fig3}(b) shows $P(\ket{\Psi^-})$ versus $\alpha$. Different colors and marker shapes correspond to different rotation axes $\hat{\bm{n}} = \hat{\bm{x}}, \hat{\bm{y}}, \hat{\bm{z}}$. Regardless of $\hat{\bm{n}}$, $P(\ket{\Psi^-})\approx \cos(2\alpha)$. The 2 comes from the equivalence between (i) positronium sensing and (ii) evolving the qubit under $U_\alpha^2$ (Sec.~\ref{sec_Theory_pos_met}). Two practicalities limit the $P(\ket{\Psi^-})$ curves' contrasts and so the experimentally inferable FI. One practicality consists of the state-preparation and measurement fidelities. The other practicality affects just the $\hat{\bm{n}} = \hat{\bm{z}}$ data: if $\hat{\bm{n}} = \hat{\bm{z}}$, we apply an AC Stark tone, which is not far off resonance. It therefore rotates the transmons slightly about the $x$- and $y$-axes (Appendix~\ref{sec:antiq}).  

. The uncertainty equals the first four maxima's standard deviation.

We calculate the FI from the readout-fidelity-corrected probabilities in Fig.~\ref{fig3}(b) via curve fitting, as described in Appendix~\ref{app_Data_analysis}. Figure~\ref{fig3}(c) shows the FI as a function of $\alpha$. Different colors correspond to different rotational axes (the $x$-, $y$-, and $z$-axes), as do different marker shapes. Figure~\ref{fig3}(g) displays the FI, averaged over the axes, versus $\alpha$. The average FI maximizes near $\alpha \in \{\pi/4\ \mathrm{mod}\ \pi, \, 3\pi/4\ \mathrm{mod}\ \pi\}$. Averaging the FI values obtained at the first four maxima yields an inferred FI of $2.98 \pm 0.13$ per 2 units of phase--space volume. The uncertainty equals the first four maxima's standard deviation. This value is inferred after correcting for readout infidelity and therefore does not represent the FI obtained directly from the raw measurement outcomes. We provide a direct comparison between the FI obtained from the raw and readout-fidelity-corrected data in Appendix~\ref{sec:fidelitycorrection}.



To highlight positronium sensing's entanglement advantage, we compare our strategy with phase estimation that leverages an unentangled qubit--antiqubit sensor. The competitor strategy provides the greatest possible FI achievable if, as in positronium sensing, all trials must begin with identical state preparations~\footnote{If different trials can begin with different state preparations, one can achieve more FI, on average over trials~\cite{Song2024}.}. Figure~\ref{fig3}(d) shows the circuit diagram:
we prepare $\q$ in $\ket{x+}$ and $\antiq$ in $\ket{z+}$. Then, $U_\alpha$ evolves $\q$, while $U_\alpha^\dag$ evolves $\antiq$. Finally, we measure the qubit's $\{ \ketbra{x+}{x+}, \ketbra{x-}{x-} \}$, obtaining the first outcome with a probability $P(\ket{x+})$. We simultaneously measure the antiqubit's $\{ \ketbra{z+}{z+}, \ketbra{z-}{z-} \}$, associated with an analogous $P(\ket{z+})$.

We analyze the competitor protocol as follows. For technical reasons, we assume that the metrologist learns $\hat{\bm{n}}$ after the experiment~\footnote{
Our assumption renders $I_\alpha$ an appropriate measure of how well one can infer $\alpha$ via the competitor protocol. If $\hat{\bm{n}}$ is unknown, the problem falls under the heading of multiparameter estimation; the \emph{FI matrix} should replace the FI \cite{Liu2019} (Appendix~\ref{sec:thy}).}. 
This assumption gives the competitor an advantage over positronium sensing. Nevertheless, we will see, positronium sensing achieves a greater $I_\alpha/\vst$. 
Figure~\ref{fig3}(e) shows the $P(\ket{x+})$ and $P(\ket{z+})$ versus $\alpha$ for each of three rotation axes. If $\q$ rotates about the $y$- or $z$-axis, the qubit's $P(\ket{x+}) \approx \cos(\alpha)$. If $\q$ rotates about the $x$-axis, the qubit's $P(\ket{x+}) = 1$. This probability's $\alpha$-independence implies that we can learn nothing about the field. Analogous statements concern the antiqubit. \response{Figure~\ref{fig3}(f) reports the FI inferable from the entanglement-free experiment, as a function of $\alpha$. We compare that experiment with positronium sensing in Fig.~\ref{fig3}(g), which shows the FI averaged over the rotation axes. In the entanglement-free results [Fig.~\ref{fig3}(f)], the FI peaks near $\alpha \in \{\pi/2, \, 3\pi/2\}$. On average over the peaks, the FI is $I_\alpha = 1.20 \pm 0.03$ per $\vst {=} 2$. This value lies slightly below the theoretically predicted $4/3$~\cite{Song2024} 
and below the positronium-sensing value, $2.98 \pm 0.13$.} \response{The empirically observed FI lies below the theoretical prediction due to input-state infidelity and measurement infidelity (Appendices~\ref{sec:fidelitycorrection} and \ref{sec:noise}).
}
The positronium-sensing and entanglement-free strategies outperform agnostic sensing performed without antimatter~\cite{Song2024}. In the latter experiment, a qubit sensor, entangled with a qubit ancilla, achieved an FI of $I_\alpha = 0.72\pm0.03$, while expending $\vst =2$.


%
%
%
\section{Outlook}
\label{sec_Outlook}

We have applied a fundamental-physics insight to unitary inversion and quantum sensing. According to decades-old particle physics, one can view positrons as electrons traveling backward in time. We exhibited a positron-like transmon whose effective gyromagnetic ratio equals the negative of an ordinary transmon qubit's. This negativity enabled us to invert a unitary implemented by a magnetic field. We leveraged this platform-specific unitary inversion, with entanglement, in phase estimation: using an entangled qubit--antiqubit pair, we measured the strength of a field pointing in an arbitrary direction. Our strategy works as well even if the direction is unknown. Furthermore, the strategy enabled a greater FI, per two units of space--time volume, than two competitor strategies. 

This work establishes \response{five} opportunities for future research. \response{First, we sketch an extension of our scheme that enables the FI to maximizes at small $\alpha$ values of practical interest.} Second, we outline an extension that achieves more FI per unit space--time volume. Third, we sketch a use of actual antimatter in quantum sensing. \response{Fourth, we identify other ordinary-matter platforms amenable to positronium sensing.} Fifth, we identify further applications of platform-specific unitary inversion and synthetic positronium.

\kater{In our experiment, the FI maximizes at rotation angles $\alpha$ near $\pi/4\ \mathrm{mod}\ \pi/2$. However, small $\alpha$ values are often relevant to metrological applications. Some input states and measurements can shift the setup's maximal sensitivity toward $\alpha \approx 0$, while preserving our experiment's space–time efficiency \cite{Xingrui_thesis}.}

By extending our protocol, one can achieve the greatest possible amount of FI per unit space--time volume, not only the greatest possible amount per two units. Consider subjecting one synthetic-positronium atom to the external field $n$ times sequentially. This protocol achieves an FI of $4n^2$---exhibiting Heisenberg scaling---which equals the QFI. The space--time volume used is $2n$, so the FI per unit space--time volume is $2n$, the optimal value.

Our protocol extends not only to $n$ sequential field applications, but also to true antimatter. True positronium offers an advantage over its synthetic counterpart:
the positron's gyromagnetic ratio has precisely the same magnitude as the electron's. Hence imperfect control does not threaten the unitary inversion. Moreover, the unitary inversion costs no control resources. One can perform our protocol's measurement [top of Fig.~\ref{fig3}(a)] on true positronium using current technology, \emph{positronium annihilation-lifetime spectroscopy}~\cite{Cassidy2006}. One measures the time required for a positronium atom to decay into photons. From the time, one can infer the former atom's spin state: the positronium triplets' lifetimes exceed the singlet's lifetime by three orders of magnitude. Yet our protocol's state preparation [bottom of Fig.~\ref{fig3}(a)] poses a challenge for positronium, as positronium singlets decay quickly. One can prepare longer-lived entangled triplets, however~\cite{AEgiS2024}. Using a positronium triplet, one can achieve an entanglement advantage in a metrological task, albeit not the task in the present paper. \response{In~\cite{Wang25}, we will detail the task, detail an antimatter protocol that can achieve it, and argue for the protocol's feasibility.}

\response{One can apply not only true antimatter, but also other ordinary-matter platforms, in positronium sensing. Superconducting-circuit quantum electrodynamics (QED) offers a natural platform for positronium sensing due to its analogy with QED.  In relativistic QED, charge conjugation interchanges electrons and positrons. In superconducting circuits, conjugation by a 
$Z$ gate flips the sign of the transverse Pauli operators, effectively reversing the qubit’s coupling to control fields and enabling the unitary inversion used here. Despite this platform’s suitability to positronium sensing, the protocol can be implemented on other platforms as well. Positronium sensing requires only (i) control typical of quantum-information-processing platforms and (ii) qubits with different gyromagnetic ratios.
(One gyromagnetic ratio ideally equals the negative of the other: $\gamma_1 = - \gamma_2$. However, the protocol requires only distinct gyromagnetic ratios.) Candidate platforms include neutral atoms~\cite{10_Saffman_Quantum}, trapped ions~\cite{19_Bruzewicz_Trapped}, and nitrogen-vacancy centers~\cite{14_Schirhagl_Nitrogen}. The latter have already featured in a CTC-related experiment~\cite{25_Rovny_Multi}. Different platforms have different strengths, such as sensitivities to different stimuli. Quantum metrologists can take advantage of these different strengths, using positronium sensing, due to the protocol's versatility.
}

\response{Beyond these platform considerations, synthetic positronium and platform-specific unitary inversion may find applications beyond metrology.} Unitary-inversion methods have enjoyed considerable interest for years, including recently~\cite{Trev0,Trev1,Trev2,Trev3,Loschmidt0,Loschmidt1,Loschmidt2,Loschmidt3,PlatUniInv1,PlatUniInv0,DDReview,DDNV}. Diverse applications motivate such advances. For instance, efficient unitary inversion underpins  quantum algorithms that require black-box access to $U_\alpha$ and $U_\alpha^\dag$. Example algorithms include amplitude amplification~\cite{AmplitudeAmplification}, phase estimation~\cite{PhaseEstimation}, and quantum singular-value transformations~\cite{Gily_n_2019}. Additionally, unitary inversion enables multiparameter sensing that achieves the best possible precision (the Cram\'{e}r-Rao limit), using a fixed measurement setup~\cite{Miyazaki2022, Wang_2024}. Finally, by efficiently inverting unknown unitaries, an adversary can break certain cryptographic schemes~\cite{crypto1}.  Hence we expect our platform-specific unitary inversion to enable applications across and beyond metrology.

%
%
Appendix~\ref{sec_Proof_positronium_QFI} concerns the FI achievable via positronium sensing. This FI equals the greatest QFI achievable with any strategy, regardless of whether the field direction is known. Appendix~\ref{sec:thy} identifies the greatest possible amount of QFI achievable with synthetic antimatter but without entanglement. The entanglement-free strategy provides a foil for positronium sensing. Appendix~\ref{sec:antiq} details how we realize antiqubits. In Appendix~\ref{sec:exp}, we describe our experimental setup and quantum gates. \kater{Appendix~\ref{sec:fidelitycorrection} describes how readout and gate fidelity affect the FI.} \response{Appendix~\ref{sec:noise} develops a corresponding theoretical model that quantitatively accounts for the experimentally observed singlet-detection statistics, and predicts the corresponding degradation of the FI.}

\begin{acknowledgments}
This work received support from the National Science Foundation: 
QLCI grant OMA-2120757, PHY-2408932,  and by the Gordon and Betty Moore Foundation, grant DOI 10.37807/gbmf11557. 
Also, this research was supported in part by grant NSF PHY-2309135 to the Kavli Institute for Theoretical Physics (KITP).  
\response{Devices were fabricated and provided by the Superconducting Qubits at Lincoln Laboratory (SQUILL) Foundry at MIT Lincoln Laboratory \footnote{
Certain equipment, instruments, software, or materials are identified in this paper to specify the experimental procedure adequately. Such identification is not intended to imply recommendation or endorsement of any product or service by NIST; nor is it intended to imply that the materials or equipment identified are necessarily the best available for the purpose.},
with funding from the Laboratory for Physical Sciences (LPS) Qubit Collaboratory.
}
Y.-X.W. and N.Y.H. thank Ruggero~Caravita and Michael~Doser for input about antimatter.  D.R.M.A.S. thanks W. Salmon for posing the question that led to this result.  F.S. was supported by the Harding Foundation.  Y.-X.W.~acknowledges support from a QuICS Hartree Postdoctoral Fellowship.  
D.R.M.A.S., K.M., and N.Y.H. thank the KITP for its hospitality and thank the organizers of the KITP program ``New directions in quantum metrology,'' out of which this project grew.
\end{acknowledgments}

\section*{Data availability}
The data that support the findings of this study are available from the corresponding authors upon reasonable request.







\clearpage
\appendix

\section{FI achievable via positronium sensing}
\label{sec_Proof_positronium_QFI}

The main text contains three claims about the FI achievable with positronium sensing: (i) Positronium sensing achieves an FI of $I_\alpha = 4$ while consuming $\vst = 2$ units of space--time volume. (ii) This FI equals the corresponding QFI, which equals the greatest QFI achievable when the field direction is unknown. (iii) In every strategy that satisfies property (ii), one must effectively invert the field experienced by $\antiq$. We prove all three claims here.

\reviewercomment{3}{other comment 2}{The use of ``TLS'' for an ideal qubit-like two-level probe is potentially confusing because, in superconducting-circuit literature, TLS commonly denotes a microscopic material defect.}
\deletedtext{We begin by introducing terminology. In the main text, \emph{qubit} means \emph{quantum two-level system that acts like ordinary matter} and contrasts with \emph{antiqubit}. When referring to an object that could be a qubit or an antiqubit, we wrote \emph{transmon}. Here, we often refer to a system that (a) could be a qubit or an antiqubit and (b) could be realized with any suitable physical platform, not only with a transmon. We call such an object a \emph{two-level system} (TLS) throughout this appendix.}
We begin by clarifying terminology. Here, \emph{qubit} and \emph{antiqubit} denote quantum two-level probes that undergo opposite field-induced rotations. Both roles are realized with transmons in our experiment. The analysis below, however, is independent of the physical platform. We therefore use \emph{quantum two-level probe} for an abstract two-dimensional sensing degree of freedom that may play either role, and \emph{probe} thereafter for brevity. This terminology does not refer to a microscopic material two-level defect.

A protocol's space--time volume equals the product (number of probes used)$\times$(number of sequential applications of $U_\alpha \coloneqq e^{-i \alpha H}$). 
Consequently, each of two protocol structures incurs a space--time volume of two. First, one probe may undergo two consecutive applications of $U_{\alpha}$. Second, two probes may simultaneously undergo $U_\alpha$ or $U_\alpha^\dag$. We sequentially calculate the QFIs achievable with these protocol structures.

First, consider subjecting one probe to two sequential unitary applications. Suppose that the rotation axis, $\hat{\bm{n}}$, is known. The optimal probe state is an equal superposition of the eigenstates of 
$H = \hat{\bm{n}}\cdot\bm{\sigma}/2$, which
has eigenvalues $E_{\pm} = \pm 1/2$. The probe, subjected to two sequential unitaries, undergoes
$U_{\alpha}^2=e^{-2i\alpha H}$. 
The effective generator is $2 H$, whose spectral gap is 2. By Eq.~\eqref{eq_QFI_var}, the QFI is $\mathcal{I}_{\alpha} = 4$.

Second, consider two probes evolving in parallel, each undergoing $U_\alpha$ or $U_\alpha^\dag$. In Appendix~\ref{Sec_App_QFI_2TLSs_1U}, we calculate the QFI achievable in this scenario with two probes prepared in a pure state. We apply the formula to separable states and to maximally entangled states in Appendix~\ref{Sec_App_Discussion}. Denote the probes' initial joint state by $\ket{\psi}$. Denote its concurrence---a measure of the state's entanglement \cite{concurrence1998}---by $C(\ket{\psi})$.
In Appendix~\ref{Sec_App_Proof_Concurr}, we prove that the QFI obeys the upper bound
\begin{equation} \label{Eq_App_Concurrence}
\mathcal{I}_{\alpha} 
   \leq 2 \left[ 1 + C(\ket{\psi}) \right] \leq 4.
\end{equation}
The concurrence ranges from $0$, for separable states, to $1$, for maximally entangled states (Bell pairs). Furthermore, pure initial states enable the greatest possible QFI: if the initial state is mixed, quantum information has leaked from the probe to the environment, preventing the metrologist from gathering all possible information from the probe.
Hence 4 is the greatest QFI achievable with any initial state of two probes that experience a field simultaneously.

To achieve this QFI without knowing the rotation axis, one must effectively invert the field, we prove in Appendix~\ref{App_positronium_unique_optimal}. Positronium sensing effectively inverts the field by simulating antimatter.
In Appendix~\ref{App_positronium_achives_QFI}, we prove that positronium sensing can achieve an FI equal to this QFI. Therefore, positronium sensing 
can achieve an FI equal to the greatest possible QFI achievable with two units of space--time volume.

\subsection{QFI achievable with two probes and one application of the external field} \label{Sec_App_QFI_2TLSs_1U}

Here, we calculate the QFI achievable with two probes, each of which undergoes $U_{\alpha}$ or $U^{\dagger}_{\alpha}$. This formula lays the foundation for later appendix subsections: in Appendix~\ref{Sec_App_Discussion}, we apply the formula to separable states and to maximally entangled states. In Appendix \ref{Sec_App_Proof_Concurr}, we use the formula to prove the concurrence bound~\eqref{Eq_App_Concurrence}. We use the formula in Appendix \ref{App_positronium_unique_optimal} to demonstrate that 
unitary inversion and entanglement enable the unique strategy whose FI can achieve the greatest QFI attainable if one does not know the field's direction.

Consider rotating a probe through an unknown angle $\alpha$ about the axis 
$\hat{\bm{n}} = \LParen \sin{(\theta)} \cos{(\phi)}, \sin{(\theta)} \sin{(\phi)}, \cos{(\theta)} \RParen$. 
The unitary $U_\alpha \coloneqq e^{- i\alpha \hat{\bm{n}} \cdot \boldsymbol \sigma/2}$ represents this operation.
Consider subjecting one probe to $U_\alpha$ and another probe to $U_\alpha$ or $U_\alpha^\dag$. The probe pair's joint state evolves under
\begin{widetext}
\begin{align}
  & \mathcal U^{(+)}_\alpha 
  \coloneqq U_\alpha \otimes U_\alpha  
  = e^{- i\alpha \hat{\bm{n}} \cdot \bm{\sigma}/2} \otimes e^{- i\alpha \hat{\bm{n}} \cdot \bm{\sigma}/2}  
  = e^{- i\alpha \left(\hat{\bm{n}} \cdot \bm{\sigma} \otimes \id + \id \otimes \hat{\bm{n}} \cdot \bm{\sigma} \right)/2 }
  \quad \text{or} \\
  & \mathcal U^{(-)}_\alpha 
  \coloneqq U_\alpha \otimes U^{\dagger}_\alpha  
  = e^{-i\alpha \hat{\bm{n}} \cdot \bm{\sigma}/2} \otimes e^{+i\alpha \hat{\bm{n}} \cdot \bm{\sigma}/2} 
  = e^{- i\alpha \left(\hat{\bm{n}} \cdot \bm{\sigma} \otimes \id  - \id \otimes \hat{\bm{n}} \cdot \bm{\sigma} \right)/2  } \, .
\end{align}
\end{widetext}
We synopsize both equations in
\begin{equation}
\mathcal U^{(\pm)}_\alpha 
= e^{- i\alpha \left( 
 \hat{\bm{n}} \cdot \bm{\sigma} \otimes \id \pm \id \otimes \hat{\bm{n}} \cdot \bm{\sigma}  \right)/2 } 
 \equiv e^{- i\alpha \mathcal{H}^{(\pm)}/2 }  . 
\end{equation}
The effective Hamiltonian has the form
$\mathcal{H}^{(\pm)} \coloneqq \hat{\bm{n}} \cdot \bm{\sigma} \otimes \id \pm \id \otimes \hat{\bm{n}} \cdot \bm{\sigma}$.

Let us calculate the QFI one can achieve
by evolving a two-probe state $\ket{\psi}$ under $\mathcal U^{(\pm)}_\alpha$. The calculation does not depend on whether an agent prepared $\ket{\psi}$ using knowledge of $\hat{\bm{n}}$. 

First, we introduce notation. We label the first probe as $A$ and the second as $B$. We define a tensor $T$ in terms of its elements,
\begin{equation}
   \label{eq_T_def}
   T_{ij} \coloneqq \braket{\psi|\sigma_i \otimes \sigma_j|\psi} .
\end{equation}
The probes initially have the Bloch vectors 
\begin{equation}
   \label{eq_Bloch_vs}
   \bm{r}^{(A)} 
   \coloneqq \langle \psi|\boldsymbol\sigma \otimes \id|\psi\rangle  
   \quad \text{and} \quad 
   \bm{r}^{(B)} \coloneqq \langle \psi|\id\otimes \boldsymbol\sigma|\psi\rangle.    
\end{equation}
Projecting the Bloch vectors onto $\hat{\bm{n}}$ yields
\begin{equation}
   {r_{\hat{\bm{n}}}}^{(i)} \coloneqq \hat{\bm{n}} \cdot {\bm{r}}^{(i)}, 
   \quad \text{wherein} \quad 
   i = A,B.
\end{equation}
Finally, let $s \coloneqq \pm 1$. We can express, in terms of the quantities above, the QFI achievable with the probe prepared in $\ket{\psi}$. The QFI equals the variance of $\mathcal{H}^{(\pm)}$, calculated in the rest of this subsection:
\begin{equation} \label{Eq_App_2q}
 \mathcal{I}^{(s)}_{\alpha}
 = \left(\Delta \mathcal{H}^{(s)}\right)^2 
 = 2 \left( 1 + s \, \hat{\bm{n}}^{\textrm{T}} \, T \, \hat{\bm{n}}\right)  
 - \left( r^{(A)}_{\hat{\bm{n}}} + s \, r^{(B)}_{\hat{\bm{n}}} \right)^2  \, .
\end{equation}

Let us prove Eq.~\eqref{Eq_App_2q}. The effective Hamiltonian has a variance
\begin{equation} \label{Eq_App_Spectral_Gap}
   \left( \Delta \mathcal{H}^{(\pm)} \right)^2 
   = \braket{ \psi | \left( \mathcal{H}^{(\pm)} \right)^2 | \psi } - \braket{ \psi | \mathcal{H}^{(\pm)} | \psi }^2.
\end{equation}
We calculate the two terms sequentially. The first has the form
\begin{widetext}
\begin{align}
   \langle \psi | \left(\mathcal{H}^{(\pm)}\right)^2 | \psi \rangle 
   & = \braket{ \psi | \left(\hat{\bm{n}} \cdot \bm{\sigma} \otimes \id \pm \id \otimes \hat{\bm{n}} \cdot \bm{\sigma} \right) ^2 | \psi} \\
   & = \braket{ \psi | \left(\hat{\bm{n}} \cdot \bm{\sigma}\right) ^2  \otimes \id + \id \otimes \left( \hat{\bm{n}} \cdot \bm{\sigma} \right) ^2 \pm 2\, \left( \hat{\bm{n}} \cdot \bm{\sigma}  \otimes \hat{\bm{n}} \cdot \bm{\sigma} \right) | \psi} \\
   & = \braket{ \psi | \left( \id  \otimes \id \right) +  \left( \id  \otimes \id \right) \pm 2\, \left( \hat{\bm{n}} \cdot \bm{\sigma}  \otimes \hat{\bm{n}} \cdot \bm{\sigma} \right) | \psi} \\
   & = 2 \left[1 \pm \braket{ \psi | \left( \hat{\bm{n}} \cdot \bm{\sigma}  \otimes \hat{\bm{n}} \cdot \bm{\sigma} \right) | \psi} \right] \\
   & =2 \left[1 \pm \hat{\bm{n}} \cdot \braket{ \psi | \left(  \bm{\sigma}  \otimes \bm{\sigma} \right) | \psi} \cdot \hat{\bm{n}} \right] \\
   & = 2 \left( 1 \pm \hat{\bm{n}}^{\mathrm{T}} T  \hat{\bm{n}} \right) .
   \label{eq_var_help1}
\end{align}
\end{widetext}
The second term in Eq.~\eqref{Eq_App_Spectral_Gap} has the form 
\begin{align}
\braket{ \psi | \mathcal{H}^{(\pm)} | \psi }^2 & = \braket{\psi | \hat{\bm{n}} \cdot \bm{\sigma} \otimes \id \pm \id \otimes \hat{\bm{n}} \cdot \bm{\sigma} | \psi }^2 \\
& = \left( \braket{\psi | \hat{\bm{n}} \cdot \bm{\sigma} \otimes \id  | \psi } \pm \braket{\psi |  \id \otimes \hat{\bm{n}} \cdot \bm{\sigma} | \psi } \right)^2  \\
& = \left( \hat{\bm{n}} \cdot \braket{\psi |  \bm{\sigma} \otimes \id  | \psi } \pm \hat{\bm{n}} \cdot \braket{\psi |  \id \otimes  \bm{\sigma} | \psi } \right)^2  \\
& = \left( \hat{\bm{n}} \cdot \bm{r}^{(A)} \pm \hat{\bm{n}} \cdot \bm{r}^ {(B)} \right)^2  \\
& = \left( \bm{r}_{\hat{\bm{n}}}^ {(A)} \pm  \bm{r}_{\hat{\bm{n}}}^ {(B)} \right)^2 \, .
\label{eq_var_help2}
\end{align}
Inserting Eqs.~\eqref{eq_var_help1} and~\eqref{eq_var_help2} into Eq.~\eqref{Eq_App_Spectral_Gap} yields Eq.~\eqref{Eq_App_2q}, the formula we aimed to prove.

\subsection{QFI achievable with two probes prepared in a maximally entangled state or in a separable state} \label{Sec_App_Discussion}
In this subsection, we apply Eq.~\eqref{Eq_App_2q} to two cases. First, we calculate the QFI achievable with two probes prepared in a maximally entangled state, such as the singlet used in positronium sensing. Then, we calculate the QFI achievable with any separable state. This analysis illuminates how the QFI depends on the initial state's entanglement.

\textbf{Maximally entangled states}: Suppose the probes are prepared in a maximally entangled $\ket{\psi}$. 
Each probe's reduced state is maximally mixed:
$\Tr_B ({\ketbra{\psi}{\psi}} ) 
= \frac{1}{2}\id$, and $\Tr_A({\ketbra{\psi}{\psi}}) 
=\frac{1}{2} \id$.
Since the Pauli operators are traceless, therefore,
$\bm r^{(A)} 
= \frac{1}{2} \Tr{\left(\boldsymbol\sigma \otimes \id \right)} 
= \frac{1}{2} \Tr{\left(\boldsymbol\sigma \right)} = \bm{0}$. The Bloch vector $\bm r^{(B)} $ vanishes analogously.
Hence the Bloch vectors' projections vanish,
$\bm{r}^{(A)}_{\hat{\bm{n}}} = \bm{r}^{(B)}_{\hat{\bm{n}}}= 0$,
regardless of $\hat{\bm{n}}$.
Equation~\eqref{Eq_App_2q} reduces to 
\begin{equation} \label{Eq_App_2q_bell}
   \mathcal{I}^{(s)} _{\alpha}
   = 2 \left( 1 + s \, \hat{\bm{n}}^{\textrm{T}} \, T \, \hat{\bm{n}}\right)  .  
\end{equation}

In positronium sensing, the two-probe state $\ket{\psi} = \ket{\Psi^-}$ is the singlet. The qubit 
probe and antiqubit ancilla rotate oppositely, so $s=-1$. This state's $T = - \id$, so  Eq.~\eqref{Eq_App_2q_bell} simplifies to 
\begin{equation}
   \mathcal{I}^{(-)} _{\alpha}
   = 2 \left[1 - \hat{\bm{n}}^{\textrm{T}} \, (- \id) \, \hat{\bm{n}} \right] 
   = 2 \left(1 + |\hat{\bm{n}}|^2 \right)= 4.
\end{equation}
Hence the QFI does not depend on the rotation axis. 

Suppose that $\ket{\psi}$ is any other maximally entangled state. $T$ is a diagonal matrix with a determinant of $1$. However, $T \not\propto \id$, unlike if $\ket{\psi}$ is a singlet. Consequently, the QFI depends on the rotation axis. 
$\mathcal{I}^{(s)} _{\alpha}$ achieves its maximum value (which equals 4, we show shortly) when the rotation axis
is an eigenvalue-$s$ eigenvector of $T$:
$T \hat{\bm{n}} = s \hat{\bm{n}}$. Under this condition, the QFI evaluates to
\begin{widetext}
\begin{equation}
\mathcal{I}^{(s)} _{\alpha}= 2 \left( 1 + s \, \hat{\bm{n}}^{\textrm{T}} \, T \, \hat{\bm{n}}\right) = 2 \left( 1 + s \, \hat{\bm{n}}^{\textrm{T}} \, s \, \hat{\bm{n}}\right) = 2 \left(1+ s^2 \hat{\bm{n}}^2\right) = 4  . 
\end{equation}
\end{widetext}
In  Appendix~\ref{Sec_App_Proof_Concurr}, we identify the states $\ket{\psi}$ that satisfy the maximum-QFI condition $T\hat{\bm{n}} = s\hat{\bm{n}}$.

\textbf{Separable states:} Now, suppose that the two probes are prepared in any separable $\ket{\psi}$.  By Eqs.~\eqref{eq_T_def} and~\eqref{eq_Bloch_vs}, $T = \bm{r}^{(A)} {\bm{r}^{(B)}}^\mathrm{T} \, .$
We substitute the right-hand side into the QFI formula~\eqref{Eq_App_2q}:
\begin{widetext}
\begin{align}
   \mathcal{I}^{(s)} _{\alpha} 
   & = 2 \left( 1 + s \, \hat{\bm{n}}^{\textrm{T}} \, \bm{r}^{(A)}{\bm{r}^{(B)}}^\mathrm{T} \, \hat{\bm{n}}\right)  - \left( r^{(A)}_{\hat{\bm{n}}}  + s r^{(B)}_{\hat{\bm{n}}}\right) ^2      
   = 2 \left( 1 + s \, \, r_{\hat{\bm{n}}}^{(A)} r_{\hat{\bm{n}}}^{(B)} \right)  - \left( r^{(A)}_{\hat{\bm{n}}}  + s r^{(B)}_{\hat{\bm{n}}}\right) ^2     \\
   & = 2 \left[ 1 - \, \, \left({r_{\hat{\bm{n}}}^{(A)}}^2 +{r_{\hat{\bm{n}}}^{(B)}}^2 \right)\right] \leq 2 . 
\end{align} 
\end{widetext}
This QFI maximizes when $r_{\hat{\bm{n}}}^{(A)} = r_{\hat{\bm{n}}}^{(B)} = 0$: both probes' initial Bloch vectors point orthogonally to $\hat{\bm{n}}$.

\subsection{Proof of the bound~\eqref{Eq_App_Concurrence} on the QFI achievable with two probes} \label{Sec_App_Proof_Concurr}

In this subsection, we bound the QFI $\mathcal{I}^{(s)}_{\alpha}$ in terms of the two probes' initial concurrence, $C (\ket{\psi})$. We reproduce the main result, Eq.~\eqref{Eq_App_Concurrence}, here for convenience:
\begin{equation} \label{Eq_Concurr_Bound}
\mathcal{I}^{(s)}_{\alpha} \leq 2[1+C (\psi)].
\end{equation}
The concurrence $C$ quantifies entanglement~\cite{concurrence1998}, ranging from $0$ (achieved by separable states) to $1$ (achieved by maximally entangled states). We also identify the two-probe states that saturate Eq.~\eqref{Eq_Concurr_Bound}. 

Our strategy is as follows. First, we consider a general two-probe state $\ket{\psi}$ that has a fixed concurrence $C(\ket{\psi}) = C_0$. We parameterize $\ket{\psi}$ using local unitaries acting on a simpler reference state $\ket{\chi}$ that has the concurrence $C_0$.
This parameterization leverages the concurrence's invariance under local unitary operations. We compute $\ket{\chi}$'s QFI, using Eq.~\eqref{Eq_App_2q}. To simplify the analysis, we move to a rotated frame, in which the QFI depends only on the relative rotation between the probes' states, $U_{\text{rel}} \coloneqq U_A ^{-1} U_B$. Finally, we reformulate the QFI as a quadratic form that reveals the optimal relative rotation's structure. This reformulation yields the bound~\eqref{Eq_Concurr_Bound}, as well as a complete characterization of the states $\ket{\psi}$ that saturate it.

\subsubsection{General pure two-probe state and concurrence}

A general two-probe pure state has the form
\begin{widetext}
\begin{equation}
\ket{\psi} = a\ket{00} + b\ket{01} + c\ket{10} + d\ket{11}, \quad \text{wherein} \quad 
a, b, c, d \in \mathbb{C}
\quad \text{and} \quad
|a|^2 + |b|^2 + |c|^2 + |d|^2 = 1 .
\end{equation}
\end{widetext}
This state has the concurrence \cite{concurrence2012} 
\begin{equation}
    C_0  \coloneqq  C( \ket{\psi} ) = |ad - bc|
    \in [0, 1] .
\end{equation}
Denote by $\ket{\chi}$ a reference state that has the same concurrence, $C(\ket{\chi}) = C_0$:
\begin{equation} \label{Eq_App_chi}
\ket{\chi} \coloneqq \sqrt{\lambda_1} \, \ket{00} + \sqrt{\lambda_2} \, \ket{11},
\end{equation}
wherein $\lambda_{1,2} = \frac{1\pm\sqrt{1-C_0^2}}{2}$. 
From $\ket{\chi}$, one can generate every concurrence-$C_0$ two-probe state $\ket{\chi}$ via single-probe unitaries $U_A$ and $U_B$ \cite{Fan_2003}:
\begin{equation} \label{Eq_General_State}
\ket{\psi} = (U_A \otimes U_B)\ket{\chi}.
\end{equation}

\subsubsection{QFI achievable with fixed-concurrence states}

Let us compute the QFI achievable with a general pure two-probe state $\ket{\psi}$ of fixed concurrence $C_0$. Directly evaluating the QFI is algebraically cumbersome. Yet we can simplify the calculation because every such state decomposes as $\ket{\psi} = (U_A \otimes U_B)\ket{\chi}$ in terms of the reference state $\ket{\chi}$ introduced in Eq.~\eqref{Eq_App_chi}. Evaluating the QFI achievable with $\ket{\chi}$ is simple: the state's correlation tensor $T$ is diagonal, and its Bloch vectors $\bm{r}^{(A)}, \bm{r}^{(B)} = \hat{\bm{z}}$. We incorporate the local unitaries' effects into the QFI using the unitaries, $R_A$ and $R_B$, that represent the corresponding rotations on the Bloch sphere. A simple QFI expression results; we optimize it over the field directions $\hat{\bm{n}}$ and over the concurrence-$C_0$ states.

To calculate the QFI of $\ket{\psi}$, we use \eqref{Eq_App_2q}:
\begin{equation}
 \label{eq_I_help_app_1}
 \mathcal{I}^{(s)}_{\alpha}
 = \left(\Delta \mathcal{H}^{(s)}\right)^2 
 = 2 \left( 1 + s \, \hat{\bm{n}}^{\textrm{T}} \, T \, \hat{\bm{n}}\right)  
 - \left( r^{(A)}_{\hat{\bm{n}}} + s \, r^{(B)}_{\hat{\bm{n}}} \right)^2  \, .
\end{equation}
First, we compute $T$, $r^{(A)}_{\hat{\bm{n}}}$, and $r^{(B)}_{\hat{\bm{n}}}$:
\begin{widetext}
\begin{align}
T_{ij} & \coloneqq \braket{\psi|\bm{\sigma}_i\otimes\bm{\sigma}_j|\psi} 
= \braket{\chi| \left( U_A^{\dagger} \otimes U_B^{\dagger} \right) \bm{\sigma}_i\otimes\bm{\sigma}_j (U_A \otimes U_B)|\chi} 
= \braket{\chi| U_A^{\dagger} \bm{\sigma}_i U_A \otimes U_B ^{\dagger} \bm{\sigma}_j U_B|\chi} \\
& = (R_A)_{ik}(R_B)_{j\ell} \braket{\chi|\sigma_k\otimes\sigma_\ell|\chi} .
\label{eq_T_phi_help1}
\end{align}
\end{widetext}
The matrices $R_A, R_B \in \mathrm{SO}(3)$ describe how the local unitaries $U_A$ and $U_B$ rotate the Pauli vectors $\bm{\sigma}_{i,j}$ across the Bloch sphere:
\begin{equation}
 U_X^{\dagger} \bm{\sigma}_i U_X 
 = (R_X)_{ij}  \, \bm{\sigma}_j 
 \, , \quad \text{wherein} \quad X \in \{A, B\} .   
\end{equation}
Directly calculating the final factor in Eq.~\eqref{eq_T_phi_help1}, $\braket{\chi|\sigma_k\otimes\sigma_\ell|\chi}$, yields 
\begin{equation} \label{Eq_T}
T =  R_A
\begin{pmatrix}
C_0 & 0 & 0 \\
0 & -C_0 & 0 \\
0 & 0 & 1
\end{pmatrix}
R_B^\text{\textrm{T}}
\equiv R_A \, D \, R_B^\text{\textrm{T}} \, .
\end{equation}
We have defined the diagonal matrix $D \coloneqq \text{diag}(C_0, -C_0, 1)$. 
Having calculated the $T$ in Eq.~\eqref{eq_I_help_app_1}, we compute the Bloch-vector projections $r^{(A)}_{\hat{\bm{n}}} \coloneqq \hat{\bm{n}} \cdot \bm{r}^{(A)}$ and $r^{(B)}_{\hat{\bm{n}}} \coloneqq \hat{\bm{n}} \cdot \bm{r}^{(B)}$. We can express the $i^{\rm th}$ component of ${\bm{r}^{(A)}}$ in terms of the rotation matrix $R_A$ and the concurrence $C_0$:
\begin{widetext}
\begin{align}
{\bm{r}^{(A)}} _i
& \coloneqq 
\braket{\psi | \bm{\sigma}_i \otimes \id | \psi} 
= 
\braket{\chi | U_A^{\dagger} \bm{\sigma}_i U_A \otimes \id | \chi} 
= (R_A)_{ik}\braket{\chi | \sigma_k \otimes \id | \chi} 
= \sqrt{1-C_0^2}\,(R_A\,\hat{\bm{z}})_i \, .
\end{align}
\end{widetext}
$\bm{r}^{(B)}$ obeys an analogous equation. In summary,
\begin{align} 
   \label{Eq_ra}
   \bm r^{(A)} = \sqrt{1-C_0^2}R_A\,\hat{\bm{z}}, 
   \quad \text{and} \quad
   \bm r^{(B)}
   = \sqrt{1-C_0^2}R_B\, \hat{\bm{z}}. 
\end{align}
Let us substitute the calculated quantities from Eqs.~\eqref{Eq_T} and~\eqref{Eq_ra} into the QFI formula~\eqref{Eq_App_2q}:
\begin{widetext}
\begin{align}
\mathcal{I}^{(s)}_{\alpha} 
&= 2 \left( 1 + s \, \hat{\bm{n}}^{\textrm{T}} \, T \, \hat{\bm{n}}\right)  
 - \left( r^{(A)}_{\hat{\bm{n}}} + s
 r^{(B)}_{\hat{\bm{n}}} \right)^2  
 = 2 + 2 s \, \hat{\bm{n}}^{\textrm{T}} \, T \, \hat{\bm{n}}
 -  \left[ \hat{\bm{n}} \cdot \left(\bm{r}^{(A)} + s
 \bm{r}^{(B)} \right) \right]^2 \\
 &= 2 + 2 s \, \hat{\bm{n}}^\text{\textrm{T}} \left( R_A \, D \, R_B^\text{\textrm{T}} \right) \hat{\bm{n}}  -\left( 1- C_0^2 \right)  \left[\hat{\bm{n}} ^\text{\textrm{T}}   \left( R_A \hat{\bm{z}} + s R_B \hat{\bm{z}} \right)  \right]^2 \label{Eq_QFI_rot}.
\end{align}
\end{widetext}

\subsubsection{QFI achievable with any pure two-probe state, calculated in a rotated reference frame}

Equation~\eqref{Eq_QFI_rot} shows a QFI dependent on the Bloch-sphere rotations $R_A,R_B\in$ SO(3). They represent the single-probe unitaries $U_A$ and $U_B$. 
We now analyze how these rotations affect the QFI. According to Eq.~\eqref{Eq_General_State}, every pure, concurrence-$C_0$ two-probe state $\ket{\psi}$ results from applying single-probe unitaries to a concurrence-$C_0$ reference state $\ket{\chi}$. Those unitaries decompose into (i) two identical single-probe unitaries $U_{\rm id}$ and (ii) a relative rotation $U_{\text{rel}}$:
\begin{equation} \label{Eq_Rot_Decomp}
\ket{\psi} = (U_{\text{id}} \otimes U_{\text{id}})(\id \otimes U_{\text{rel}})\ket{\chi} .
\end{equation}
We now show that the greatest possible QFI (the maximum over field directions $\hat{\bm{n}}$) is invariant under identical rotations of both probes. Let $S \in$ SO(3)  denote an arbitrary rotation matrix: $S S^{\textrm{T}} = S^{\textrm{T}} S = \id$. We define a rotated reference frame by applying $S$ to the original rotation matrices ($R_A$ and $R_B$) and to the field direction ($\hat{\bm{n}}$):
\begin{equation} \label{Eq_Mapping_Rotated_Frame}
   R_A \mapsto S R_A \, , \qquad 
   R_B \mapsto S R_B \, , 
   \qquad \text{and} \quad 
   \hat{\bm{n}} \mapsto S \hat{\bm{n}}.
\end{equation}
The QFI is invariant under this rotation. We prove this claim by transforming each component of Eq.~\eqref{Eq_QFI_rot} that is not overtly a scalar:
\begin{widetext}
\begin{align}
& \hat{\bm{n}}^\text{\textrm{T}} \left( R_A \, D \,  R_B^\text{\textrm{T}} \right) \hat{\bm{n}} 
\mapsto (S\hat{\bm{n}})^\text{\textrm{T}}  
\left[ S R_A \, D \,  (S R_B)^\text{\textrm{T}} \right] \left(S \hat{\bm{n}}\right) 
= \hat{\bm{n}}^\text{\textrm{T}}  S^\text{\textrm{T}} S \left( R_A \, D \,  R_B^\text{\textrm{T}} \right) S^\text{\textrm{T}} S  \hat{\bm{n}}
=\hat{\bm{n}}^\text{\textrm{T}} \left( R_A \, D \,  R_B^\text{\textrm{T}} \right) \hat{\bm{n}} , \; \mathrm{and}\\
& \hat{\bm{n}} ^\text{\textrm{T}}  \left( R_A \hat{\bm{z}} + s R_B \hat{\bm{z}} \right)  
\mapsto (S\hat{\bm{n}})^\text{\textrm{T}}   \left( S R_A \hat{\bm{z}} + s S R_B \hat{\bm{z}} \right)=\hat{\bm{n}}^\text{\textrm{T}} S^\text{\textrm{T}}  S \left( R_A \hat{\bm{z}} + s R_B \hat{\bm{z}} \right)
= {\hat{ \bm{n}}}^\text{\textrm{T}}  \left( R_A \hat{\bm{z}} + s R_B \hat{\bm{z}} \right) .
\end{align}
\end{widetext}

Without loss of generality, we select $S = R_A^{-1}$. By the mapping in Eq. \eqref{Eq_Mapping_Rotated_Frame}, $R_A \mapsto S R_A = \id$, and $R_B \mapsto S R_B = R_A^{-1} R_B \eqqcolon R_{\text{rel}}$. 
To calculate the effect of $S$ on the QFI, we apply these mappings to Eq. \eqref{Eq_QFI_rot}:
\begin{widetext}
\begin{align} 
\mathcal{I}^{(s)}_{\alpha} & =
  2 + 2 s \, \left( S\hat{\bm{n}}\right)^\text{\textrm{T}} \left[ \left(S R_A \right) \, D \, \left( S R_B \right) ^\text{\textrm{T}} \right] \left( S \hat{\bm{n}} \right) -\left( 1- C_0^2 \right)  \left\{\left( S \hat{\bm{n}} \right)^\text{\textrm{T}}   \left[ \left( S R_A \right)  \hat{\bm{z}} + s \left( S R_B \right) \hat{\bm{z}} \right]  \right\}^2 \\
  & = 2 +2  \, \left\{ s\, \left( S\hat{\bm{n}} \right)^\text{\textrm{T}} \left(
D
R_{\text{rel}}^\text{\textrm{T}} \right) \left( S \hat{\bm{n}} \right) -\left( 1- C_0^2 \right)  \left[ \left( S \hat{\bm{n}}  \right)^\text{\textrm{T}}   \left( \hat{\bm{z}} + s R_{\text{rel}} \hat{\bm{z}} \right)  \right]^2 \right\} \label{Eq_Redefine_n}\\
& \leq 2 +2  \, \max_{\hat{\bm{n}}}\left\{ s\, \hat{\bm{n}}^\text{\textrm{T}} \left(
D
R_{\text{rel}}^\text{\textrm{T}} \right) \hat{\bm{n}} -\left( 1- C_0^2 \right)  \left[\hat{\bm{n}} ^\text{\textrm{T}}  \left( \hat{\bm{z}} + s R_{\text{rel}} \hat{\bm{z}} \right)  \right]^2 \right\} \label{Eq_Rb} .
\end{align}
\end{widetext}
Equation~\eqref{Eq_Redefine_n} reveals that rotating the reference frame, using $S$, amounts to redefining $\hat{\bm{n}}$. By maximizing over $\hat{\bm{n}}$ in Eq.~\eqref{Eq_Rb},  we removed the $S$-dependence. The bound depends only on the relative rotation, $R_{\text{rel}}$.

\subsubsection{Greatest QFI achievable with any fixed-concurrence two-probe state}

Here, we identify the external-field directions $\hat{\bm{n}}$ and relative rotation $R_{\text{rel}}$ that maximize the QFI achievable with a concurrence-$C_0$ state.
One might hope to maximize the first term inside the curly braces in Eq.~\eqref{Eq_Rb} while keeping the second (nonpositive) term zero. This strategy is 
impossible, except when $C_0=1$. We thus rewrite Eq.~\eqref{Eq_Rb} in a form that points to the optimal relative rotation $R_{\text{rel}}$. 

To rewrite the bound, we expand the term proportional to $(1-C_0^2)$. Then, we express the diagonal matrix as $D$ as $\text{diag}(C_0, -C_0, 1) =C_0 \hat{\bm x}\hat{\bm x}^{\textrm{T}} - C_0 \hat{\bm y}\hat{\bm y}^{\textrm{T}}   + \hat{\bm z}\hat{\bm z}^{\textrm{T}}$.
The bound assumes the form
\begin{widetext}
\begin{align} 
\mathcal{I}^{(s)}_{\alpha}   
& \leq 2 +2  \, \max_{\hat{\bm{n}}}\left\{ s\, \hat{\bm{n}}^\text{\textrm{T}} \left(
D R_{\text{rel}}^\text{\textrm{T}} \right) \hat{\bm{n}} -\left( 1- C_0^2 \right)  \left[\hat{\bm{n}}^\text{\textrm{T}}   \left( \hat{\bm{z}} + s R_{\text{rel}} \hat{\bm{z}} \right)  \right]^2 \right\} \\
& = 2 +2  \, \max_{\hat{\bm{n}}}\left\{ s\, \hat{\bm{n}}^\text{\textrm{T}} \left( D R_{\text{rel}}^\text{\textrm{T}} \right) \hat{\bm{n}} -\left( 1- C_0^2 \right)  \hat{\bm{n}}^\text{\textrm{T}}    \left( \hat{\bm{z}} + s R_{\text{rel}} \hat{\bm{z}} \right) \left( \hat{\bm{z}} + s R_{\text{rel}} \hat{\bm{z}} \right)^\text{\textrm{T}}  \hat{\bm{n}}  \right\} \\
& = 2+ 2  \, \max_{\hat{\bm{n}}} \hat{\bm{n}}^\text{\textrm{T}}\Big\{ s \left( \hat{\bm z}\hat{\bm z}^{\textrm{T}}
R_{\text{rel}}^\text{\textrm{T}}  \right) + s C_0 \left( \hat{\bm x}\hat{\bm x}^{\textrm{T}} - \hat{\bm y}\hat{\bm y}^{\textrm{T}}\right)
R_{\text{rel}}^\text{\textrm{T}}  \notag \\
& \quad \quad - (1-C_0^{2}) \left[ \hat{\bm z}\hat{\bm z}^{\textrm{T}} +s\hat{\bm z}(R_{\text{rel}}\hat{\bm z})^{\textrm{T}} +s(R_{\text{rel}}\hat{\bm z})\hat{\bm z}^{\textrm{T}} +(R_{\text{rel}}\hat{\bm z})(R_{\text{rel}}\hat{\bm z})^{\textrm{T}} \right] \Big\} \hat{\bm{n}}   \\
& = 2+ 2  \, \max_{\hat{\bm{n}}} \hat{\bm{n}}^\text{\textrm{T}}\Big\{ s C_0^2 \left( \hat{\bm z}\hat{\bm z}^{\textrm{T}}
R_{\text{rel}}^\text{\textrm{T}}  \right) + s C_0 \left( \hat{\bm x}\hat{\bm x}^{\textrm{T}} - \hat{\bm y}\hat{\bm y}^{\textrm{T}}\right)
R_{\text{rel}}^\text{\textrm{T}}  \notag \\
& \quad \quad - (1-C_0^{2})
    \left[
      \hat{\bm z}\hat{\bm z}^{\textrm{T}}
      +s(R_{\text{rel}}\hat{\bm z})\hat{\bm z}^{\textrm{T}}
      +(R_{\text{rel}}\hat{\bm z})(R_{\text{rel}}\hat{\bm z})^{\textrm{T}}
    \right] \Big\} \hat{\bm{n}} \\
& = 2+ 2  \, \max_{\hat{\bm{n}}} \hat{\bm{n}}^\text{\textrm{T}}\left\{  s C_0 \left( \hat{\bm x}\hat{\bm x}^{\textrm{T}} - \hat{\bm y}\hat{\bm y}^{\textrm{T}} + C_0 \hat{\bm z}\hat{\bm z}^{\textrm{T}}\right)
R_{\text{rel}}^\text{\textrm{T}}   - (1-C_0^{2})
    \left[
      \hat{\bm z}\hat{\bm z}^{\textrm{T}}
      +s(R_{\text{rel}}\hat{\bm z})\hat{\bm z}^{\textrm{T}}
      +(R_{\text{rel}}\hat{\bm z})(R_{\text{rel}}\hat{\bm z})^{\textrm{T}}
    \right] \right\} \hat{\bm{n}} . \label{Eq_Rrel_apparent}
\end{align}
\end{widetext}
The term proportional to $(1 - C_0^2)$
yields a nonpositive contribution to the QFI, due to two facts. First, $-(1 - C_0^2)$ is nonpositive for all $C_0 \in [0, 1]$. Second, the quadratic form is non-negative for all $\hat{\bm{n}}$: 
\begin{widetext}
\begin{align} \hat{\bm{n}}^{\textrm{T}}
    \left[
      \hat{\bm z}\hat{\bm z}^{\textrm{T}}
      +s(R_{\text{rel}}\hat{\bm z})\hat{\bm z}^{\textrm{T}}
      +(R_{\text{rel}}\hat{\bm z})(R_{\text{rel}}\hat{\bm z})^{\textrm{T}}
    \right] \hat{\bm{n}} & = |\hat{\bm{n}}^{\textrm{T}} \hat{\bm z}|^2 + |\hat{\bm{n}}^{\textrm{T}} R_{\text{rel}}\hat{\bm z}|^2 + s (\hat{\bm{n}}^{\textrm{T}} R_{\text{rel}}\hat{\bm z}) (\hat{\bm z}^{\textrm{T}} \hat{\bm n} )\\
    & = \left(\hat{\bm{n}}^{\textrm{T}} \hat{\bm z} + \frac{s}{2}\hat{\bm{n}}^{\textrm{T}} R_{\text{rel}}\hat{\bm z}\right)^2 + \left(1- \frac{s^2}{4}\right) |\hat{\bm{n}}^{\textrm{T}} R_{\text{rel}}\hat{\bm z}|^2 
     \geq 0 . \label{Eq_ineq_negative_term}
\end{align}
\end{widetext}
This inequality saturates only if both terms in Eq.~\eqref{Eq_ineq_negative_term} vanish. The second term vanishes when $\hat{\bm{n}}^{\textrm{T}} R_{\text{rel}}\hat{\bm z} = 0$, such that $\hat{\bm{n}} \perp R_{\text{rel}}\hat{\bm z}$. 
Under this condition, the first term in Eq.~\eqref{Eq_ineq_negative_term} vanishes when $\hat{\bm{n}}^{\textrm{T}} \hat{\bm z}$, such that $\hat{\bm{n}} \perp  \hat{\bm z}$. 
In summary, the inequality~\eqref{Eq_ineq_negative_term} saturates if and only if $\hat{\bm{n}} \perp \hat{\bm{z}}, R_{\text{rel}}\hat{\bm z}$. 
This condition is satisfied, for example, when
$\hat{\bm{n}} \perp \hat{\bm{z}}$ and $R_{\text{rel}} \hat{\bm{n}} = \pm \hat{\bm{n}}$. 

Having identified under which condition the nonpositive contribution to the QFI vanishes, we demonstrate that this condition maximizes the term proportional to $sC_0$ in Eq.~\eqref{Eq_Rrel_apparent}. The analysis decomposes into two cases:
\begin{itemize}
    \item $\bm{C_0\neq 1}$: The contribution proportional to $sC_0$ in Eq.~\eqref{Eq_Rrel_apparent}, $sC_0\,  \hat{\bm{n}}^\text{\textrm{T}} \left( \hat{\bm x}\hat{\bm x}^{\textrm{T}} - \hat{\bm y}\hat{\bm y}^{\textrm{T}} + C_0 \hat{\bm z}\hat{\bm z}^{\textrm{T}}\right) 
R_{\text{rel}}^\text{\textrm{T}} \hat{\bm{n}}$, 
assumes its maximum value, $+1$, if the field direction $\hat{\bm{n}}$ is an eigenvector, associated with the eigenvalue $s' \coloneqq \pm 1$, of $R_{\text{rel}}^\text{\textrm{T}}$: 
$R_{\text{rel}}^\text{\textrm{T}} \hat{\bm{n}}
= s' \hat{\bm{n}}$.
If $s s' = -1$, then 
$\hat{\bm{n}} = \hat{\bm y}$.
If $s s' = 1$, then 
$\hat{\bm{n}}= \hat{\bm x}$.
Under each of these two conditions, the term proportional to $(1-C_0^2)$ in Eq.~\eqref{Eq_Rrel_apparent} vanishes.  The QFI simplifies to
\begin{equation}
\mathcal{I}^{(s)}_{\alpha} = 2(1+C_0) .
\end{equation}
Each other $R_{\text{rel}}$ or $\hat{\bm{n}}$ leads to a smaller QFI: the term proportional to $sC_0$ in Eq.~\eqref{Eq_Rrel_apparent} shrinks, and the term proportional to $(1-C_0^2)$ contributes a negative value.

\item $\bm{C_0 =1}$: The term proportional to $(1 - C_0^2)$ in Eq.~\eqref{Eq_Rrel_apparent} vanishes. The maximal QFI is therefore determined entirely by the contribution proportional to $sC_0$, $sC_0 \, \hat{\bm{n}}^\text{\textrm{T}} \left( \hat{\bm x}\hat{\bm x}^{\textrm{T}} - \hat{\bm y}\hat{\bm y}^{\textrm{T}} + \hat{\bm z}\hat{\bm z}^{\textrm{T}}\right) 
R_{\text{rel}}^\text{\textrm{T}} \hat{\bm{n}}$. 
This expression assumes its maximum value, $+ 1$, when $R_{\text{rel}}^{\textrm{T}} \hat{\bm{n}} = s'\hat{\bm{n}}$. 
If $s s'=-1$, then $\hat{\bm{n}} = \hat{\bm{y}}$. If 
$s s'=+1$, then $\hat{\bm{n}}$ lies in the $xz$-plane.
Under each of these two conditions, the QFI achieves its maximal value, $\mathcal{I}_{\alpha}^{(s)} = 4$. 
The QFI maximizes, for example, if $s=-1$ and $R_{\text{rel}}^\text{\textrm{T}}  = \text{diag} (-1, 1, -1)$. This $R_{\text{rel}}$ represents a $\pi$ rotation about the $y$-axis.
The achievable QFI is optimal in this case, regardless of $\hat{\bm{n}}$.
\end{itemize}
In summary, we have shown that the greatest QFI achievable with a concurrence-$C_0$ input state $\ket{\psi}$ is $\mathcal{I}^{(s)}_{\alpha} = 2(1 + C_0)$, if $C_0 < 1$, or $\mathcal{I}^{(s)}_{\alpha} = 4$, if $C_0 = 1$. We have therefore proved Eq.~\eqref{Eq_Concurr_Bound}.

\subsubsection{Optimal input states}

We now characterize the optimal two-probe input states consistent with a fixed concurrence $C_0$.
We define the optimal input states $\ket{\psi}$ as those that, while having a concurrence $C_0$, saturate the upper bound~\eqref{Eq_Concurr_Bound}. They achieve the greatest possible QFI, $\mathcal{I}^{(s)}_{\alpha} = 2(1+C_0)$. 

We pursue the following strategy.
The previous subsection specifies the relative rotation matrix $R_{\text{rel}}$ that maximizes the QFI.
We construct the single-probe unitary $U_{\text{rel}}$ that implements this $R_{\text{rel}}$. By substituting the unitary's form into Eq.~\eqref{Eq_Rot_Decomp}, we construct the optimal input states.
We enumerate all the possible cases, distinguished by (i) the concurrence's value ($C_0 \neq 1$ or $C_0=1$) and (ii) whether the probes rotate in the same direction ($s = \pm 1$).
Then, we show that positronium sensing exemplifies one of these cases and so achieves the greatest possible QFI.
Throughout this subsection, $U_{\text{id}}$ denotes an arbitrary single-probe unitary, introduced in Eq.~\eqref{Eq_Rot_Decomp}.

Recall that, when $C_0\neq 1$, the QFI is maximized when $R_{\text{rel}}^\text{\textrm{T}} \hat{\bm{n}} = s' \hat{\bm{n}}$, wherein 
 $\hat{\bm{n}}= \hat{\bm y}$ (if $s s' =-1$) or $\hat{\bm{n}} = \hat{\bm x}$ (if $s s' =1$).
Hence, if $C_0\neq 1$ and $s = -1$, the QFI is maximized when $R_{\text{rel}} \hat{\bm{y}}= \hat{\bm{y}} $ or $R_{\text{rel}} \hat{\bm{x}}= -\hat{\bm{x}} $. 
First, suppose that $R_{\text{rel}}  \hat{\bm{y}} = \hat{\bm{y}}  $: the relative rotation is through an arbitrary angle $\phi$ about the $y$-axis.
The unitary $U_{\text{rel}} = \exp\left(-i \frac{\phi}{2} \, \sigma_y\right)$ implements this rotation.  
Now, suppose that $R_{\text{rel}}  \hat{\bm{x}} = -\hat{\bm{x}}  $: the relative rotation is through an angle $\pi$ about an axis perpendicular to $\hat{\bm{x}}$. The unitary 
$U_{\text{rel}} 
= \exp\LParen -i \frac{\pi}{2} \left[ \cos{(\phi)} \, \sigma_y + \sin{(\phi)} \, \sigma_z \right] \RParen$ 
implements this rotation. The rotation axis, perpendicular to $\hat{\bm{x}}$, lies in the $yz$-plane, at an arbitrary angle $\phi$ to the $y$-axis. 
By Eq.~\eqref{Eq_Rot_Decomp}, the QFI is maximized if
\begin{widetext}
\begin{align} \label{Eq_Optimal}
\ket{\psi} =
\begin{cases} 
\left[U_{\text{id}} \otimes U_{\text{id}} \exp\left(-i \frac{\phi}{2} \, \sigma_y\right) \right] \ket{\chi}  \\[0.5cm]
\left\{ U_{\text{id}} \otimes U_{\text{id}} 
\exp\LParen -i \frac{\pi}{2} \left[ \cos{(\phi)} \, \sigma_y + \sin{(\phi)} \, \sigma_z \right] \RParen \right\} \ket{\chi}  ,
\end{cases}
\quad \text{if} \quad s = - 1 .
\end{align}
\end{widetext}

Now, suppose that $C_0\neq 1$ and $s = 1$. 
The QFI is maximized when  $R_{\text{rel}}  \hat{\bm{x}} = \hat{\bm{x}}$ or $R_{\text{rel}}  \hat{\bm{y}} = - \hat{\bm{y}}$. 
We now identify the unitary that implements each of these possible relative rotations. 
If $R_{\text{rel}}  \hat{\bm{x}} =  \hat{\bm{x}}$, then 
$U_{\text{rel}} =  \exp\left(-i \frac{\phi}{2} \, \sigma_x \right)$. 
If $R_{\text{rel}}  \hat{\bm{y}} = - \hat{\bm{y}}  $, then 
$U_{\text{rel}} 
= \exp\LParen -i \frac{\pi}{2} 
\left[ \cos{(\phi)} \, \sigma_x + \sin{(\phi)} \, \sigma_z \right] \RParen$. By Eq.~\eqref{Eq_Rot_Decomp}, the optimal input state has the form
\begin{widetext}
\begin{align} \label{Eq_Optimal2}
\ket{\psi} =
\begin{cases} 
\left[U_{\text{id}} \otimes U_{\text{id}} \exp\left(-i \frac{\phi}{2} \, \sigma_x\right) \right] \ket{\chi} \\[0.5cm]
\left\{U_{\text{id}} \otimes U_{\text{id}} 
\exp\LParen -i \frac{\pi}{2} \left[ \cos{(\phi)} \, \sigma_x + \sin{(\phi)} \, \sigma_z \right] \RParen \right\} \ket{\chi}  ,
\end{cases}
\quad \text{if} \quad s =1 . 
\end{align}
\end{widetext}

Finally, suppose that $C_0=1$. We identify similarly the states that saturate the QFI bound Eq.~\eqref{Eq_Concurr_Bound}:
\begin{equation} \label{Eq_Generate_Positronium}
\ket{\psi} = 
\begin{cases}
\left[ U_{\text{id}} \otimes U_{\text{id}} \exp\left(-\frac{i\phi}{2} \, \sigma_y\right) \right] \ket{\chi}\\[10pt]
\left[ U_{\text{id}} \otimes 
\left( -i U_{\text{id}} \,\hat{\bm n}\cdot\bm\sigma \right) \right] \ket{\chi},
\end{cases}    
\quad \text{if} \quad s = - 1 ,
\end{equation}
and
\begin{widetext}
\begin{equation}
\ket{\psi} = 
\left\{ U_{\text{id}} \otimes U_{\text{id}} 
\exp \left( -i\frac{\theta}{2} \left[ \cos(\phi) \,\sigma_x + \sin(\phi) \, \sigma_z \right] \right) \right\} \ket{\chi},  
\quad \text{if} \quad s = 1 . 
\end{equation}
\end{widetext}

We now show that positronium sensing exemplifies one of the optimal-QFI conditions above. If $C_0 = 1$ and $s=-1$ [Eq.~\eqref{Eq_Generate_Positronium}], 
one can achieve the greatest possible QFI (under the concurrence constraint), regardless of $\hat{\bm{n}}$, using the relative rotation $R_{\text{rel}}^\text{\textrm{T}}  = \text{diag} (-1, 1, -1)$, a $\pi$ rotation about the $y$-axis. 
The unitary $U_{\text{rel}} = \exp\left(-i \frac{\pi}{2} \sigma_y\right) = - i \sigma_y$ implements this rotation. 
Since $C_0=1$,  $\ket{\chi} = \ket{\Phi^{+}}$. According to Eq. \eqref{Eq_Rot_Decomp}, the relative rotation transforms $\ket{\Phi^+}$ into the singlet: $(\id \otimes -i \sigma_y) \ket{\Phi^+} = \ket{\Psi^{-}}$. The singlet is invariant under the tensor products of identical rotations: $(U_{\text{id}} \otimes U_{\text{id}}) \ket{\Psi^{-}} = e^{i \gamma} \ket{\Psi^{-}}$, wherein $\gamma \in [0, 2\pi)$. Thus, the optimal input state is the singlet:
\begin{widetext}
\begin{equation}
   \ket{\psi} 
   = (U_{\text{id}} \otimes U_{\text{id}} )  
   \left[ \id \otimes (-i \sigma_y) \right] 
   \ket{\Phi^+} 
   = (U_{\text{id}} \otimes U_{\text{id}} ) \ket{\Psi^{-}} = e^{i \gamma} \ket{\Psi^{-}} .
\end{equation}
\end{widetext}
This protocol (initializing the singlet $\ket{\Psi^-}$ and rotating the probes oppositely) defines our positronium-sensing protocol. Therefore, positronium sensing achieves the maximal QFI,
$\mathcal{I}_\alpha^{(-)}=4$, regardless of the direction $\hat{\bm n}$. 
Below, we prove that, if $\hat{\bm{n}}$ is unknown and one can consume only two units of space--time volume, to achieve the greatest possible QFI, one must (i) entangle two probes and (ii) effectively invert the field experienced by one probe.

\subsection{Effective unitary inversion, combined with entanglement, is the only strategy whose QFI achieves the greatest possible QFI attainable if (i) $\hat{\bm{n}}$ is unknown and (ii) 2 units of space–time volume are available}
\label{App_positronium_unique_optimal}

Let us prove that effective unitary inversion and entanglement enable the unique optimal sensing strategy available when (i) $\hat{\bm{n}}$ is unknown and (ii) the sensing can consume only $\vst = 2$ units of space--time volume. By \emph{optimal}, we mean that the strategy's QFI equals the greatest possible QFI achievable,
$\mathcal{I}_{\alpha}=4$. This equality does not depend on the rotation axis.

Two protocol structures incur  space--time volumes $\vst=2$, as explained at the beginning of Appendix \ref{sec_Proof_positronium_QFI}. First, one probe may undergo two sequential applications of $U_{\alpha}=e^{-i\alpha H}$. In this scenario, the QFI attains its maximum value, 4, if the input state is any equal-weight superposition of the eigenstates of $H = \hat{\bm{n}}\cdot\bm{\sigma}/2$.  These eigenstates depend on $\hat{\bm{n}}$, so this strategy is not deterministically achievable if $\hat{\bm{n}}$ is unknown.
Second, two probes may experience the external field simultaneously, such that each undergoes $U_\alpha$ or $U_\alpha^\dag$. 
We continue to analyze this scenario below. We show that the only optimal strategy meets two conditions: (i) The input state $\ket{\psi}$ is a singlet. (ii) One probe undergoes $U_\alpha$, while the other undergoes $U_\alpha^\dag$.

To identify the optimal strategy, we consider a general pure two-probe state:
\begin{widetext}
\begin{equation} \label{Eq_App_gen_2qubit_state}
|\psi\rangle = a|00\rangle + b|01\rangle + c|10\rangle + d|11\rangle, 
\quad \text{wherein} \quad 
a, b, c, d \in \mathbb{C}
\quad \text{and} \quad
|a|^2 + |b|^2 + |c|^2 + |d|^2 = 1.
\end{equation}
\end{widetext}
This state's $T$ matrix depends on the coefficients' real ($\mathfrak{R}$) and imaginary ($\mathfrak{I}$) components as
\begin{widetext}
\begin{equation}
\label{eq_T_optimal}
T = \left(
\begin{array}{ccc}
2\mathfrak{R}(ad^* + bc^*) & 2\mathfrak{I}(bc^* - ad^*) & 2\mathfrak{R}(ac^* - bd^*) \\
-2\mathfrak{I}(ad^* + bc^*) & 2\mathfrak{R}(bc^* - ad^*) & 2\mathfrak{I}(bd^* - ac^*) \\
2\mathfrak{R}(ab^* - cd^*) & 2\mathfrak{I}(cd^* - ab^*) & |a|^2 - |b|^2 - |c|^2 + |d|^2
\end{array}
\right)   .
\end{equation} 
\end{widetext}
The optimal $\hat{\bm{n}}$-independent strategy implies, by Eq.~\eqref{Eq_App_2q_bell}, that
$\mathcal{I}^{(s)}_{\alpha} 
= 2 \left( 1 + s \, \hat{\bm{n}}^{\textrm{T}} \, T \, \hat{\bm{n}}\right)   = 4$, $\forall \hat{\bm{n}}$.
This equation simplifies algebraically to $\hat{\bm{n}}^{\textrm{T}} \, T \, \hat{\bm{n}} = s  $. Since this equation holds for all $\hat{\bm{n}}$ (by the strategy's independence from $\hat{\bm{n}}$),
$T  = s \id  $. 
This condition is equivalent, by Eq.~\eqref{eq_T_optimal}, to the system of equations
\begin{equation} \label{Eq_App_System}
 \begin{cases}
&|a|^2 - |b|^2 - |c|^2 + |d|^2 = s\\[5pt]
&2\mathfrak{R}(b c^* + a d^*) = s\\[5pt]
&2\mathfrak{R}(b c^* - a d^*) = s .
\end{cases}   
\end{equation}
$T$ has a determinant of one. Therefore, if $T$ satisfies the equations above, its off-diagonal elements vanish. By the second and third equations, 
$\mathfrak{R}( a d^*) = 0$, and $\mathfrak{R}(b c^*) = s/2$. Furthermore, 
by the Cauchy–Schwarz inequality, $\left|\mathfrak{R}(b c^*) \right| \leq |b| \cdot |c|$. By the normalization condition in Eq.~\eqref{Eq_App_gen_2qubit_state}, $ |b| \cdot |c|  \leq \frac{1}{2}$. These conditions, with $s = \pm 1$, implies a chain of inequalities:
\begin{equation}
   \label{eq_ineq_chain}
   \frac{1}{2} 
   = \left| \frac{s}{2} \right| 
   = \left|\mathfrak{R}(b c^*) \right| 
   \leq |b| \cdot |c|  
   \leq \frac{1}{2} \, .
\end{equation}
Since the leftmost and rightmost expressions equal each other, the two inequalities are saturated.
To saturate the first inequality, $b$ and $c$ must have the same phase: 
$b = |b| e^{i \gamma}$, and $c = |c| e^{i \gamma}$. To saturate the second inequality, $b$ and $c$ must satisfy $|b| = |c| = 1/\sqrt{2}$. The normalization of $\ket{\psi}$ [Eq.~\eqref{Eq_App_gen_2qubit_state}] implies $a = d = 0$, which satisfies the earlier requirement $\mathfrak{R}( a d^*) = 0$ [Eq.~\eqref{Eq_App_System}].

Consider substituting $|b| = |c| = 1/\sqrt{2}$ and $a = d = 0$ in the first equality in~\eqref{Eq_App_System}.
The condition $s = -1$ results ($s = +1$ leads to a contradiction). 
Only the $s{=} -1$ state $\ket{\psi}$---the singlet---satisfies the $\hat{\bm{n}}$-independent-optimality condition $T = s \id$:
$|\psi\rangle = \frac{e^{i \gamma}}{\sqrt{2}} \left( \ket{10} - \ket{01}\right) .$

\subsection{Proof that positronium sensing achieves an FI of $I_{\alpha} = 4$ while consuming a space--time volume $\vst = 2$} \label{App_positronium_achives_QFI}

Here, we prove that positronium sensing attains an FI of $I_{\alpha} = 4$ while consuming exactly two units of space--time volume. This FI equals the corresponding QFI, as proved in Appendix~\ref{Sec_App_Discussion}.

Positronium sensing involves three steps, as explained in Sec.~\ref{sec_Pos_exprmt} of the main text.
First, we prepare a synthetic-positronium atom (a qubit $q$ and an antiqubit $\bar{q}$) in a singlet,
$\ket{\Psi^-} = \frac{1}{\sqrt{2}} \left( \ket{01} - \ket{10} \right).$ 
Second, we apply the unitary $U_{\alpha}$ to $q$ and $U_{\alpha}^{\dagger}$ to $\bar{q}$. The joint state evolves to $\ket{\psi_{\alpha}} = (U_{\alpha} \otimes U_{\alpha}^{\dagger}) \ket{\Psi^-} 
    = \frac{1}{\sqrt{2}} \left(e^{-i \alpha}\ket{01} - e^{i \alpha}\ket{10}\right)$.
Finally, we projectively measure the POVM 
$\{\ketbra{\Psi^-}{\Psi^-},\, \id - \ketbra{\Psi^-}{\Psi^-}\}.$

Having reviewed the protocol, we calculate the FI achievable with it. The measurement yields the $\ketbra{\Psi^-}{\Psi^-}$ outcome with a probability
\begin{equation}
    P(\ket{\Psi^-}) = \left| \braket{\Psi^-|\psi_{\alpha}} \right|^2 = \cos^2{(\alpha)}.
\end{equation}
This probability serves as the $P_{j=1}$ in the FI formula $I_{\alpha} 
= \sum_j \left( \partial_{\alpha} P_j \right)^2 / P_j$
(reviewed in Sec.~\ref{sec_Backgrnd_phase_est} of the main text).
The other probability is $P_2 = 1 - P_1$. We substitute for $P_2$ into the FI formula:
\begin{widetext}
\begin{align}
    I_{\alpha} 
    &= \sum_j{\frac{\left( \partial_{\alpha} P_j\right)^2}{P_\response{j}}} 
    = \frac{\left( \partial_{\alpha} P_1\right)^2}{P_1} + \frac{\left( \partial_{\alpha} P_2\right)^2}{P_2} 
    = \frac{\left( \partial_{\alpha} P_1\right)^2}{P_1} + \frac{\left[ \partial_{\alpha}(1-P_1)\right]^2}{1-P_1} 
    = \frac{\left( \partial_{\alpha} P_1\right)^2}{P_1} + \frac{\left( \partial_{\alpha}P_1\right)^2}{1-P_1} 
    = \frac{\left( \partial_{\alpha} P_1\right)^2}{P_1 (1-P_1)} \, .
\end{align}
\end{widetext}
The remaining probability, $P_1 = P(\ket{\Psi^-}) = \cos^2{(\alpha)}$, has a derivative
$\partial_{\alpha} P(\ket{\Psi^-}) = -2 \cos{(\alpha)}\sin{(\alpha)}.$
Substituting into the FI formula yields
\begin{align}
    I_{\alpha} 
    = \frac{\left[\partial_{\alpha} P(\ket{\Psi^-})\right]^2}{P(\ket{\Psi^-})\left[1 - P(\ket{\Psi^-})\right]} 
    = \frac{4 \cos^2{(\alpha)}\sin^2{(\alpha)}}{\cos^2{(\alpha)}\sin^2{(\alpha)}} 
    = 4.
\end{align}
We proved in Appendix~\ref{Sec_App_Discussion} that the QFI, too, equals 4. Applying positronium sensing, one can achieve an FI equal to the corresponding QFI (the greatest possible FI).

\section{QFI achievable with synthetic antimatter and no entanglement} \label{sec:thy}

Positronium sensing exploits antimatter simulation and entanglement (Sec.~\ref{sec_Theory_pos_met} of the main text). We now isolate the entanglement's contribution to the metrological advantage. To do so, we evaluate the QFI attainable with (i) synthetic antimatter and (ii) the input state used in our separable-state experiment, reported on in the main text.
We calculate the QFI available to a metrologist who does not know the field orientation $\hat{\bm{n}}$ during the experiment.

First, we briefly review the separable-state protocol implemented experimentally. Two transmons are prepared in a product state: the qubit $\q$ is initialized in $\ket{x+}$; and the antiqubit $\antiq$, in $\ket{z+}$. (More generally, the two probes may begin in any two orthogonal pure states.) The external field applies $U_\alpha = e^{i \alpha H}$ to $\q$ and $U_\alpha^\dag = e^{-i \alpha H}$ to $\antiq$. The generator has the form $H = \left(\bm{\sigma} \cdot \hat{\bm{n}}\right)/2$, and $\hat{\bm{n}}$ denotes the unknown field direction. We measure the qubit's $\{\ketbra{x+}{x+}, \ketbra{x-}{x-}\}$, which has a probability $P(\ket{x+})$ of yielding the $x+$ outcome. Simultaneously, we measure the antiqubit's 
$\{\ketbra{z+}{z+}, \ketbra{z-}{z-}\}$,
which yields the $z+$ outcome with a probability $P(\ket{z+})$. From these probabilities, one can estimate $\alpha$.

In the main text, we assumed that the metrologist learns $\hat{\bm{n}}$ after the experiment concludes. This assumption simplified the analysis without affecting our main conclusion: entangled antimatter outperforms unentangled antimatter. Here, we operate under a more stringent asumption: the direction $\hat{\bm{n}}$ remains unknown at all times. The motivation is the need to compare the entanglement-free and positronium-sensing strategies fairly. One can use positronium sensing without knowing the external field's direction. Therefore, we assume that one does not know the direction when using unentangled synthetic antimatter.

\subsection{Background: multiparameter quantum metrology}
\label{app_backgrnd_multiparam}

To assess the entanglement-free strategy available to a metrologist ignorant of $\hat{\bm{n}}$, we must
review another metrological tool, the \emph{quantum Fisher-information matrix} (QFIM). The metrologist aims to estimate the rotation angle $\alpha$. However, their ignorance about $\hat{\bm{n}}$ naturally elevates the task to multiparameter estimation of not only $\alpha$, but also the polar angle $\theta$ and azimuthal angle $\phi$ that define $\hat{\bm{n}}$. In classical statistics, $\theta$ and $\phi$ act as nuisance parameters \textemdash variables that influence the estimation of $\alpha$ but are not themselves of interest  \cite{Hayashi2020}. To accommodate them in one's analysis, one extends the FI framework to the multiparameter framework, defining an effective FI that marginalizes the nuisance contributions. Similarly, we  calculate an effective single-parameter QFI, $\mathcal{I}_{\alpha, \mathrm{eff}} \, ,$ that captures the average precision with which one can estimate $\alpha$ without knowing the field direction.

In multiparameter quantum metrology, the QFIM quantifies a strategy's effectiveness~\cite{paris2009quantum, Liu2019}. To introduce the QFIM, we define $\boldsymbol{\vartheta}=(\vartheta_1, \vartheta_2, \dots ,\vartheta_k)$ as the vector of $k$ parameters to be estimated. Let $\rho_{\boldsymbol{\vartheta}}$ denote a state dependent on the vector. 
Associated with each parameter $\vartheta_i$ is an operator $L_i \, ,$ the \emph{symmetric logarithmic derivative} (SLD). It is defined implicitly through
\begin{equation} \label{Eq_App_SLD}
\partial_{\vartheta_i}\rho_{\boldsymbol{\vartheta}}
= \frac{1}{2}\bigl(\rho_{\boldsymbol{\vartheta}}L_i + L_i\rho_{\boldsymbol{\vartheta}}\bigr).
\end{equation}
Using the SLDs, we can express the QFIM achievable with this state. The QFIM is a
$k\times k$ positive-semidefinite matrix with the elements
\begin{widetext}
\begin{equation}
   \mathcal{M}_{i j}(\boldsymbol{\vartheta}) =
   \frac12\,
   \mathrm{Tr} \Bigl(
     \rho_{\boldsymbol{\vartheta}}
     \bigl\{L_i, L_j\bigr\}
   \Bigl),
   \quad \text{wherein} \quad
   i,j = 1, 2, \ldots ,k .
\label{eq:QFIM_def}
\end{equation}
\end{widetext}
The $\{A,B\} \coloneqq AB+BA$ denotes the anticommutator of the operators $A$ and $B$.

Having defined the QFIM, we review its relevant properties.
The QFIM generalizes the single-parameter QFI. If $k = 1$,
\(\mathcal{M}_{ii}\) reduces to the QFI about \(\vartheta_i\).
However, when one estimates multiple parameters simultaneously, the QFI does not quantify a strategy's effectiveness.
Rather, a matrix-valued bound captures the limitations on simultaneous parameter estimation: the \emph{quantum Cram\'{e}r–Rao bound} (QCRB).
To derive the QCRB, we review estimators and covariance matrices. Consider a measurement that yields outcome $k$ with a probability $p(k | \bm{\vartheta})$.
An estimator is a function $\hat{\bm{\vartheta}}(k)$ that maps each $k$ to a candidate estimate of $\bm{\vartheta}$. We focus on locally unbiased estimators, which satisfy the conditions
\begin{widetext}
\begin{equation} \label{Eq_LocalUnbiased}
\sum_k [\bm{\vartheta}_i - \hat{\bm{\vartheta}}_i(k)]\, p(k|\bm{\vartheta}) = 0
\quad \text{and} \quad 
\sum_k \hat{\bm{\vartheta}}_i(k) \,  \frac{\partial p(k|\bm{\vartheta})}{\partial \bm{\vartheta}_j } = \delta_{ij}.
\end{equation}
\end{widetext}
According to the first condition, the estimator tracks the parameter’s true value faithfully, to first order around the point $\bm{\vartheta}$. The second constraint excludes pathological estimators. We can illustrate pathological estimators with a function that returns a fixed value, irrespectively of the measurement outcome. We quantify an unbiased estimator's accuracy $\hat{\bm{\vartheta}}$ with the covariance matrix
\begin{equation} \label{Eq_Covariance}
\mathrm{Cov}(\hat{\bm{\vartheta}}) 
\coloneqq \sum_k 
\left[ \hat{\bm{\vartheta}}(k) - \bm{\vartheta} \right]
\left[\hat{\bm{\vartheta}}(k) - \bm{\vartheta} \right]^\textrm{T} p(k|\bm{\vartheta}).
\end{equation}
To express the QCRB concisely, we introduce the Loewner (positive-semidefinite) ordering: if $A$ and $B$ denote Hermitian operators, then $A \succeq B$ means that $A - B$ is positive-semidefinite.  In terms of this definition, the multiparameter quantum QCRB has the form
\begin{equation} \label{Eq_QCRB_multip}
\mathrm{Cov}(\hat{\bm{\vartheta}}) \succeq 
\mathcal{M}^{-1}(\bm{\vartheta}) .
\end{equation}
We apply this formalism to our setting in the next subsection.

\subsection{Application of multiparameter-quantum-metrology toolkit}
\label{app_apply_multiparam}

Let us apply the previous subsection's formalism to our setting. The QFIM for the three unknown parameters ($\alpha$, $\theta$, and $\phi$) assumes the form
\begin{equation} \label{Eq_General_QFIM}
    \mathcal{M} = \begin{pmatrix}
        \mathcal{M}_{\alpha \alpha} & \mathcal{M}_{\alpha \theta} & \mathcal{M}_{\alpha \phi}\\
        \mathcal{M}_{\theta \alpha} & \mathcal{M}_{\theta \theta} & \mathcal{M}_{\theta \phi}\\
        \mathcal{M}_{\phi \alpha} & \mathcal{M}_{\phi \theta} & \mathcal{M}_{\phi \phi}
    \end{pmatrix}.
\end{equation}
Define $\mathcal{M}_{\hat{\bm{n}}\hat{\bm{n}}}$ as the lower-right $2 \times 2$ block, associated with the nuisance parameters. 
Also, define  $\mathcal{M}_{\alpha \hat{\bm{n}}}^{\textrm{T}} \coloneqq (\mathcal{M}_{\alpha \theta}, \mathcal{M}_{\alpha \phi})$. This two-element vector quantifies the correlations between $\alpha$ and $\hat{\bm{n}}$. 
In terms of these quantities, we can express the QFIM compactly: 
\begin{equation} \label{Eq_General_QFIM_Block}
    \mathcal{M} = \begin{pmatrix}
        \mathcal{M}_{\alpha \alpha} & \mathcal{M}_{\alpha \hat{\bm{n}}}^{\mathrm{T}} \\
    \mathcal{M}_{\hat{\bm{n}} \alpha } &\mathcal{M}_{\hat{\bm{n}}\hat{\bm{n}}}
    \end{pmatrix} .
\end{equation}

Now, we bound the precision with which one can estimate $\alpha$ without knowing $\hat{\bm{n}}$. We adapt the multiparameter QCRB [Eq.~\eqref{Eq_QCRB_multip}] by invoking two properties of matrices. One is the Loewner ordering. The other is the non-negativity of every positive-semidefinite matrix's diagonal elements. From these ingredients, we derive
$\mathrm{Var}(\hat{\bm{\alpha}}) 
\geq \left(\mathcal{M}^{-1}\right)_{\alpha \alpha}$. 
We refine this bound using the Schur complement formula~\cite{APE-SM, Hayashi2020}:
{\color{black}
\begin{equation} \label{Eq_QCRB_Nuisance}
\mathrm{Var}(\hat{\bm{\alpha}}) \geq \left(\mathcal{M}^{-1}\right)_{\alpha \alpha} =
\left( \mathcal{M}^{-1}_{\alpha \alpha} -  \mathcal{M}_{\alpha \hat{\bm{n}}}^{\textrm{T}}\mathcal{M}_{\hat{\bm{n}} \hat{\bm{n}}}^{-1} \mathcal{M}_{\hat{\bm{n}} \alpha }  \right)^{-1}    .
\end{equation}
}
The QFIM is positive-semidefinite, so the second term in Eq.~\eqref{Eq_QCRB_Nuisance} is non-negative. Therefore, 
we can achieve less precision in the presence of nuisance parameters than when performing single-parameter estimation. In the latter case,
$\mathrm{Var}(\hat{\alpha}) \geq \mathcal{M}_{\alpha \alpha}^{-1}$. The two bounds coincide if and only if  $\mathcal{M}_{\alpha \hat{\bm{n}}} = 0$---if and only if the QFIM is block-diagonal:
\begin{equation} \label{Eq_Simple_QFIM}
    \mathcal{M} = \begin{pmatrix}
        \mathcal{M}_{\alpha \alpha} & 0\\
0&\mathcal{M}_{\hat{\bm{n}}\hat{\bm{n}}}
    \end{pmatrix} \quad \text{for all} \quad  (\alpha, \hat{\bm{n}}) .
\end{equation}
If the QFIM lacks this structure, we can define an effective QFI $\mathcal{I}_{\alpha,\mathrm{eff}}$ for $\alpha$. We average over a prior distribution for $\hat{\bm{n}}$:
{\color{black}
\begin{widetext}
\begin{equation}
   \label{eq_I_eff_def}
  \mathcal{I}_{\alpha, \text{eff}} 
  \coloneqq  \left\langle \left(\mathcal{M}^{-1}\right)_{\alpha \alpha}\right\rangle_{\hat{\bm{n}}} = \left\langle \left( \mathcal{M}^{-1}_{\alpha \alpha} -  \mathcal{M}_{\alpha \hat{\bm{n}}}^{\textrm{T}}\mathcal{M}_{\hat{\bm{n}} \hat{\bm{n}}}^{-1} \mathcal{M}_{\hat{\bm{n}} \alpha }  \right)^{-1} \right\rangle_{\hat{\bm{n}}} .
\end{equation}
\end{widetext}
}
If $\hat{\bm{n}}$ is unknown, a natural prior is the uniform distribution over the sphere.

In summary, $\mathcal{I}_{\alpha, \text{eff}}$ is useful under two simultaneous conditions: 
(i) The QFIM does not exhibit the block-diagonal structure in Eq.~\eqref{Eq_Simple_QFIM}. (ii) The metrologist estimating $\alpha$ does not know the values of $\theta$ and $\phi$.
The entanglement-free experiment depicted in Fig.~\ref{fig3}(c) meets condition (i). Condition (ii) is an assumption introduced in this appendix so that we can fairly compare the entanglement-free strategy with positronium sensing, in which $\hat{\bm{n}}$ is unknown.

\subsection{Calculation of the effective QFI achievable without entanglement and without knowledge of the field direction}

We now apply the  QFIM's properties to assess the entanglement-free antimatter strategy applicable without knowledge of $\hat{\bm{n}}$.
As detailed earlier, the entanglement-free protocol begins with the qubit–antiqubit pair in the state $\ket{x+}_\q\otimes\ket{z+}_\antiq$. The external field evolves the state to
\begin{equation} 
  \ket{\psi_{\alpha}}
   = U_{\alpha} \ket{x+}_\q \otimes  U^{\dagger}_{\alpha} \ket{z+}_\antiq .
\end{equation}  
Every product state's QFIM is additive \cite{Liu2019}:
\begin{equation} \label{QFIM_Exp_Fig3c}
   \mathcal{M} \left(\ket{\psi_{\alpha}} \right)
   = 
   \mathcal{M} \left(U_{\alpha} {|x+\rangle}_\q \right)+ \mathcal{M}\left({U^{\dagger}_{\alpha}|z+\rangle}_\antiq \right) .
\end{equation}
We can calculate $\mathcal{M}(U_{\alpha} {|x+\rangle}_\q)$ and $\mathcal{M}({U^{\dagger}_{\alpha}|z+\rangle}_\antiq)$ straightforwardly.
They reveal that $\mathcal{M}(\ket{\psi_{\alpha}})$ has nonzero off-diagonal elements. 
They encode correlations between $\alpha$ and the nuisance parameters. These correlations confirm the usefulness of the effective QFI, $\mathcal{I}_{\alpha, \mathrm{eff}}$ [Eq.~\eqref{eq_I_eff_def}].

To calculate $\mathcal{I}_{\alpha, \mathrm{eff}} \, ,$ we calculate $(\mathcal{M}^{-1})_{\alpha \alpha} \, .$ Its value follows from Eq.~\eqref{QFIM_Exp_Fig3c}, which we can evaluate straightforwardly:
\begin{equation} 
   \label{eq_thry_help2}
   (\mathcal{M}^{-1})_{\alpha \alpha}= 
   \frac{1}{8} \left[ 7 + \cos(2\theta) + 2 \cos(2\phi) \sin^2(\theta) \right]  .
\end{equation}
Because $\hat{\bm{n}}$ is unknown, we average $(\mathcal{M}^{-1})_{\alpha \alpha}$
uniformly over the possible field directions $\hat{\bm{n}}$. 
Denote this average by $\langle\cdot\rangle_{\hat{\bm n}}$.
To evaluate it, we integrate each side of Eq.~\eqref{eq_thry_help2} over the unit sphere: 
\begin{equation}
   \left\langle \left(\mathcal{M}^{-1} \right)_{\alpha \alpha} \right\rangle_{\hat{\bm{n}}} 
   = \frac{1}{4\pi} \int_0^{2\pi} \int_0^{\pi} 
    \left(\mathcal{M}^{-1} \right)_{\alpha \alpha} \,  
   \sin(\theta) \, d\theta \, d\phi = \dfrac{5}{6} \, .
\end{equation}
By the QCRB [Eq.~\eqref{Eq_QCRB_multip}], 
\begin{equation}
   \left\langle \text{Var}(\alpha) \right\rangle_{\hat{\bm{n}}}
   \geq \left\langle (\mathcal{M}^{-1})_{\alpha \alpha} \right\rangle_{\hat{\bm{n}}} 
   = \frac{5}{6} \, .
\end{equation}
The effective QFI follows from Eq.~\eqref{eq_I_eff_def}:
\begin{equation}
   \label{eq_I_eff_val}
   \mathcal{I}_{\alpha, \text{eff}}
   = \left[ \left\langle (\mathcal{M}^{-1})_{\alpha \alpha}  \right\rangle_{\hat{\bm{n}}} \right]^{-1} 
   = \left( \frac{5}{6} \right)^{-1} 
   = \frac{6}{5} 
   = 1.2 .
\end{equation}

This effective QFI is slightly lower than the optimal QFI achievable without entanglement but with a space--time volume $\vst=2$,
$4/3 \approx 1.33$. One can achieve this optimal QFI using a tensor product of two copies of the optimal input state used when the probe consists of just one qubit. This state is a mixture of three mutually unbiased states
(Supplementary~Note~C.4 of \cite{APE-SM}).
One should expect our experiment's entanglement-free strategy to perform suboptimally, since the QFIM~\eqref{QFIM_Exp_Fig3c} is not diagonal. 

Finally, we compare the entanglement-free synthetic-antimatter strategy with positronium sensing. Positronium sensing achieves a QFIM the form~\eqref{Eq_Simple_QFIM}, and $\mathcal{I}_{\alpha} =4$. By Eq.~\eqref{eq_I_eff_val}, entanglement enhances the achievable precision significantly.

\section{Experimental realization of qubits and antiqubits} 
\label{sec:antiq} 

This appendix reviews how we realize antiqubits.
As discussed in the main text, we apply $Z$ gates to $\antiq$ to effectively reverse the rotation induced by the field's $x$- and $y$-components. Appendix~\ref{subsec:single_qubit_rot} details the $Z$-gate implementation. Also, we effectively invert the $z$-component of the field experienced by the antiqubit. To do so, we apply an AC Stark shift $\delta_\q$ to the qubit and a shift $\delta_\antiq$ to the antiqubit. We choose the drive frequency such that $\delta_\q = -\delta_\antiq$. q and $\antiq$ rotate about the $z$-axis oppositely.

To begin, we calculate the AC Stark shifts conferred upon $\q$ and $\antiq$ by an off-resonant drive. Recall that the qubit has a frequency $\omega_\q$. Denote by $\alpha_\mathrm{q}$ the qubit's anharmonicity, the difference
(gap between second and third energy levels)$-$(gap between first and second energy levels).
Consider subjecting the qubit to an oscillating external field of amplitude $\Omega_\mathrm{s}$ and frequency $\omega_\mathrm{s}$. The field has a detuning $\Delta_{\q\mathrm{s}} \coloneqq \omega_\q - \omega_\mathrm{s}$.
The qubit acquires an AC Stark shift~\cite{Carroll2022}
\begin{equation}
\delta_\mathrm{q} = \frac{\alpha_\q \, \Omega_\mathrm{s}^2}{2 \Delta_{\mathrm{qs}} (\alpha_\q + \Delta_{\mathrm{qs}})} \, .
\label{eq:omega_q}
\end{equation} 
The antiqubit obeys an analogous result. q and $\antiq$ have different frequencies, so a ``magic'' drive frequency induces equal-magnitude, opposite-sign Stark shifts on the transmons.  Every magic frequency depends on the device parameters. Table~\ref{tab:sim} specifies the parameters' values. 
Two drive frequencies enable $\delta_\q =-\delta_\antiq$; 
one is $4.19742$ GHz. During the experimental calibration, we found that $\delta_\q =-\delta_\antiq$ at $4.176998$ GHz. 
The inferred magic frequency differs slightly from the prediction because q and $\antiq$ experience different field amplitudes: $\antiq$ experiences a field strength that is $\approx 1.78$ times larger. In Fig.~\ref{fig2}(f), the tick marks along the $x$-axis show the $\Omega_{\rm s}$ values (the drive amplitudes experienced by q) inferred from Eq.~\eqref{eq:omega_q}. 
The inferred magic frequency is scarcely detuned from the qubit's transition frequency: $\Delta_{\q \mathrm{s}}/(2\pi) = -9.52$ MHz. Therefore, the oscillating field does not only induce $\delta_\q$ and $\delta_\antiq$. Also, the field drives off-resonant rotations about the qubit's $x$- or $y$-axis, depending on the drive's phase. These imperfections affect the gray disks, inferred after rotations about the $z$-axis, in Fig.~\ref{fig3}(b,e). The gray disks approximately form a curve that wobbles more than the figure's other approximate curves.

We infer the AC Stark shift by performing simultaneous Ramsey measurements on q and $\antiq$.  To do so, we rotate each transmon through an angle $\pi/2$ about the $x$-axis.
Then, we apply the Stark tone to the shared readout-input line for an amount of time that varies from batch of trials to batch of trials.
Each transmon then undergoes another $\pi/2$ rotation about a different axis (relative to the first $\pi/2$ rotation) to produce a synthetic detuning. 
Finally, we read out the transmons' states. Figure~\ref{fig2}(e) displays the measured frequency shifts, as well as theoretical predictions. Using the measured data, we identified a magic frequency. 
To obtain the results in Fig.~\ref{fig3}, we used frequency shifts 
$|\delta_j| = \Omega_x = \Omega_y = 2\pi (2.13 \ \mathrm{MHz})$, wherein $j \in \{\q, \antiq \}$. 
To rotate each transmon through an angle $\alpha \in [0, 2\pi]$, we drove them for amounts of time $\in [0, 470]$~ns.

\begin{table*}[t]
\caption{Device parameters. Transition frequencies, anharmonicities, readout-resonator frequencies, and coherence times for the three transmons.}
\label{tab:sim}
\centering
\begin{ruledtabular}\begin{tabular}{lccccc} 
 & \thead{$\omega_\mathrm{q,\antiq,c}/(2\pi)$ (GHz)} & \thead{$\alpha/(2\pi)$ (MHz)} & 
 \thead{Resonator (GHz)}  & 
 \thead{$T_1\ (\mu\mathrm{s})$} & 
 \thead{$T_2^*\ (\mu\mathrm{s})$} \\ [0.5ex]
\hline 
\thead{Qubit} & 4.16748 & $-146.916$ & 6.69 & 28 & 35 \\ [0.5ex] 
\thead{Antiqubit} & 4.27398 & $-144.658$ & 6.88 & 17 & 22 \\ [0.5ex] 
 \thead{Coupler} & 5.24975 & $-152.384$ & 7.08 & 14 & 16 \\ [0.5ex] 

\end{tabular}\end{ruledtabular}

\end{table*}

\section{Experimental setup} \label{sec:exp}

The experiments were performed on a superconducting device with three transmon circuits. Table~\ref{tab:sim} specifies the device parameters. We denote the qubit transmon by q; the antiqubit, by $\antiq$; and the coupler, by c. The qubit's frequency is fixed, whereas c and $\antiq$ are flux-tunable. We operate c and $\antiq$ with net zero flux threading their superconducting-quantum-interference-device (SQUID) loops. Each transmon has a  ground state $\ket{g}$, first excited state $\ket{e}$, and second excited state $\ket{f}$.
Dispersively coupled readout resonators read out all the transmons' states.
The device was fabricated in the SQUILL foundry at MIT Lincoln Laboratory \footnote{
Certain equipment, instruments, software, or materials are identified in this paper to specify the experimental procedure adequately. Such identification is not intended to imply recommendation or endorsement of any product or service by NIST; nor is it intended to imply that the materials or equipment identified are necessarily the best available for the purpose.}.
We mounted the device on a dilution refrigerator's mixing-chamber stage, with attenuated coaxial lines for implementing qubit drives and flux biases. Reference~\cite{Gaikwad_2024} provides details about the overall setup at the hardware level.
\reviewercomment{3}{other comment 3}{The section about the experimental setup is not sufficiently self-contained. It does not provide a full room-temperature and cryogenic wiring schematic for this particular experiment.}

The room-temperature and cryogenic signal chains are similar to those used in Ref.~\cite{Gaikwad_2024}. That reference depicts only the cryogenic wiring, so Fig.~\ref{fig:full_wiring} shows both portions of the setup used here. A Quantum Machines OPX supplies the intermediate-frequency control envelopes and digitizes the heterodyne readout. Independent microwave sources provide the 4.241~GHz shared qubit and antiqubit local oscillator, the 6.982~GHz readout local oscillator, and the traveling-wave parametric-amplifier pump. Cascaded two-way splitters distribute the shared qubit and antiqubit local oscillator to the two drive mixers and to one input of an OPX-controlled microwave switch. A second splitter distributes the readout local oscillator between readout downconversion and the switch's other input. The switch selects the local oscillator for a single upconversion mixer shared by readout and AC Stark generation, and the mixer receives its I/Q envelopes from the same OPX DAC outputs in both modes. In the AC Stark configuration, the mixer generates the 4.176998~GHz tone directly on the shared readout-input line, so the tone is applied simultaneously to the qubit and antiqubit. The antiqubit and coupler fast-flux lines combine OPX DAC outputs with DC current sources through bias tees.

\begin{figure}
    \centering
    \includegraphics[width=\columnwidth]{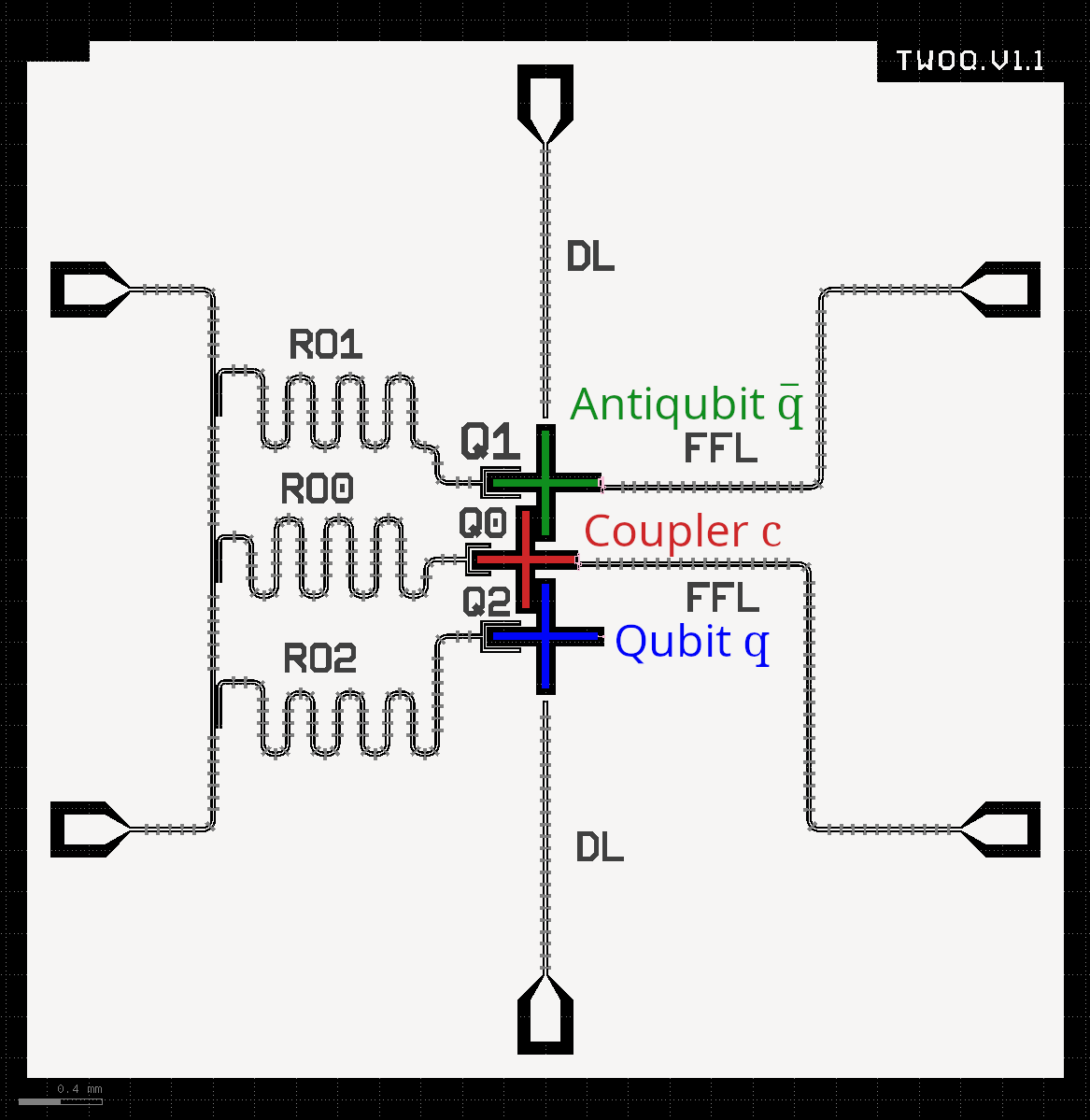}
\caption{\textbf{Device layout.} \kater{Layout of the three-transmon chip. A fixed-frequency qubit ($\q$, blue) couples to a flux-tunable antiqubit ($\antiq$, green) via a flux-tunable coupler ($\coup$, red). Each transmon couples dispersively to a readout resonator used in single-shot measurement.}}
    \label{fig:device_layout}
\end{figure}

\begin{figure*}[t]
    \centering
    \includegraphics[width=\textwidth,height=0.65\textheight,keepaspectratio]{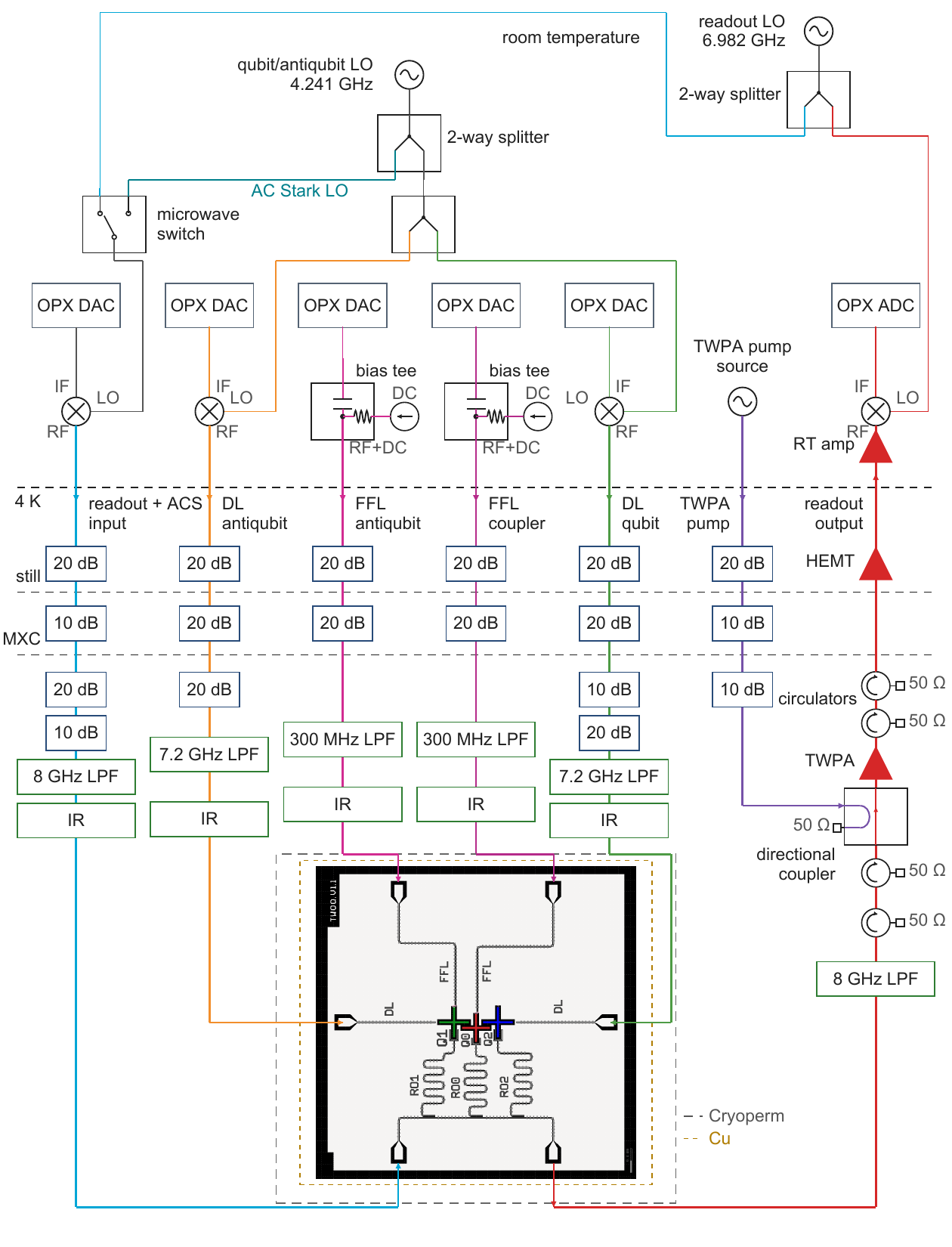}
    \caption{\textbf{Experimental wiring.} Room-temperature and cryogenic signal chain, including the OPX-controlled local-oscillator selection for the mixer shared by readout and AC Stark upconversion.}
    \label{fig:full_wiring}
\end{figure*}

The coaxial input lines are thermalized by cryogenic attenuators at 4 K, the still, and the mixing chamber. The antiqubit and qubit drive lines have 60 dB and 70 dB of total attenuation, respectively, followed by 7.2 GHz low-pass filters and infrared filters. Each fast-flux line has 40 dB of attenuation, a 300 MHz low-pass filter, and an infrared filter. The readout-input line has 60 dB of attenuation, an 8 GHz low-pass filter, and an infrared filter, while the pump line has 40 dB of attenuation. The device is mounted in a copper enclosure within a Cryoperm magnetic shield. The readout-output chain contains an 8 GHz low-pass filter, a directional coupler for pump injection, a traveling-wave parametric amplifier, circulators terminated in 50~$\Omega$, a 4 K HEMT amplifier, and a room-temperature amplifier before I/Q downconversion and OPX digitization. Figure~\ref{fig:device_layout} shows the chip connectivity.

\subsection{Dispersive readout}

The qubit and antiqubit couple to their respective readout resonators~\cite{Boissonneault2009} dispersively. We can thereby perform simultaneous high-fidelity single-shot readouts~\cite{Walter2017}. Our readout scheme is heterodyne: we multiplex the readout signal by simultaneously sending in two pulses that have different frequencies. We separate the two readouts in the frequency domain for processing. To implement low-noise amplification, we use a traveling-wave parametric amplifier based on SNAILs (Superconducting Nonlinear Asymmetric Inductive eLements) \cite{Ranadive2022}. 
To increase the  signal-to-noise ratio, we apply $\pi$-pulses that promote q and $\antiq$ from their $\ket{e}$ to the $\ket{f}$ levels. 
We optimize the readout signal-to-noise ratio over readout amplitude, frequency, and amplifier-bias settings, using a gradient-free optimization algorithm~\cite{nevergrad}. We construct the multicomponent-pulse-integration envelopes via linear-discriminant analysis. Ultimately, we achieve a qubit-readout fidelity of $97.8\ \%$ and an antiqubit-readout fidelity of $95.0\ \%$, using a random forest classifier \cite{scikit-learn}.  After this calibration, we correct all tomography results for the finite readout fidelities,
using the iterative Bayesian-update correction method \cite{Nachman2020}.

{\color{black}
\subsection{Readout postprocessing}
\label{sec_Readout_postproc}

This subsection details the readout processing. For simplicity, we first consider readout from one resonator. Readout occurred throughout the times $t_n \in [0, \, 16~\mu\mathrm{s})$. We bin the times and index the bins with $n \in [0, 15999]$. The raw digitized waveform has two quadratures, $I^{\mathrm{raw}}[n]$ and $Q^{\mathrm{raw}}[n]$, sampled at 1~GHz. 

During measurement, the population on a transmon's (first) excited state can decay. Therefore, the signal-to-noise ratio decreases toward the end of the recording process. We therefore integrate the readout with a time-dependent weight function that emphasizes the early, high-contrast portion of the signal. 

We design the weights as follows. First, we reduce the data's dimensionality: we coarse-grain the record into a bank of 24 rectangular windows $w^{\mathrm{rect}}_j[n]$, wherein $0 \leq j \leq 23$. This process produces 48 ``features'' (Fig.~\ref{fig:integration_weights_windows}). We choose narrower windows at the beginning of the record to capture transient dynamics exhibited before the excited-state population decays. Let $\omega_{\mathrm{IF}}$ and $\phi_\mathrm{IF}$ denote the readout pulse's intermediate frequency and phase. The coarse-grained features are 
\begin{widetext}
\begin{equation}
\begin{aligned}
    & I^{\mathrm{rect}}_j \coloneqq \sum_n w^{\mathrm{rect}}_j[n]\left\{ I^{\mathrm{raw}}[n] \cos(\omega_\mathrm{IF}t_n+\phi_\mathrm{IF}) + Q^{\mathrm{raw}}[n] \sin(\omega_\mathrm{IF}t_n+\phi_\mathrm{IF}) \right\}
    \quad \text{and} \\
    & Q^{\mathrm{rect}}_j \coloneqq \sum_n w^{\mathrm{rect}}_j[n]\left\{ -I^{\mathrm{raw}}[n] \sin(\omega_\mathrm{IF}t_n+\phi_\mathrm{IF}) + Q^{\mathrm{raw}}[n] \cos(\omega_\mathrm{IF}t_n+\phi_\mathrm{IF}) \right\}.
\end{aligned}
\end{equation}
\end{widetext}

In order to capture the targeted signal dynamics, we identify a new basis of orthogonal vectors that represent the $I^{\mathrm{rect}}_j$ and $Q^{\mathrm{rect}}_j$ features. We represent the vectors as discrete integration weights ($w^I$, $w^J$, $w^K$, and $w^L$). We calculate the weights by applying Linear Discriminant Analysis (LDA) to the $(I^{\mathrm{rect}}_j, Q^{\mathrm{rect}}_j)$ data. The first basis vector, $w^I$, equals the primary axis identified by the LDA solution. The second basis vector, $w^J$, follows from rotating $w^I$ counterclockwise through $90^\circ$ in the $I$–$Q$ plane. To remove the signal along the $w^I$ and $w^J$ vectors, we project the input data onto the subspace orthogonal to the $w^I$–$w^J$ plane. We apply a second round of LDA to this residual data to obtain the third basis vector, $w^K$. The fourth basis vector, $w^L$, results from rotating $w^K$ counterclockwise through $90^\circ$. The rotated basis vectors ($w^J$ and $w^L$) are often less informative than the other vectors ($w^I$ and $w^K$). However, including the rotated vectors (i) makes the representation robust with respect to phase changes and (ii) facilitates the subsequent optimization of the readout frequency and amplifier settings.

\begin{figure*}[htbp]
    \centering
    \includegraphics[width=0.9\textwidth]{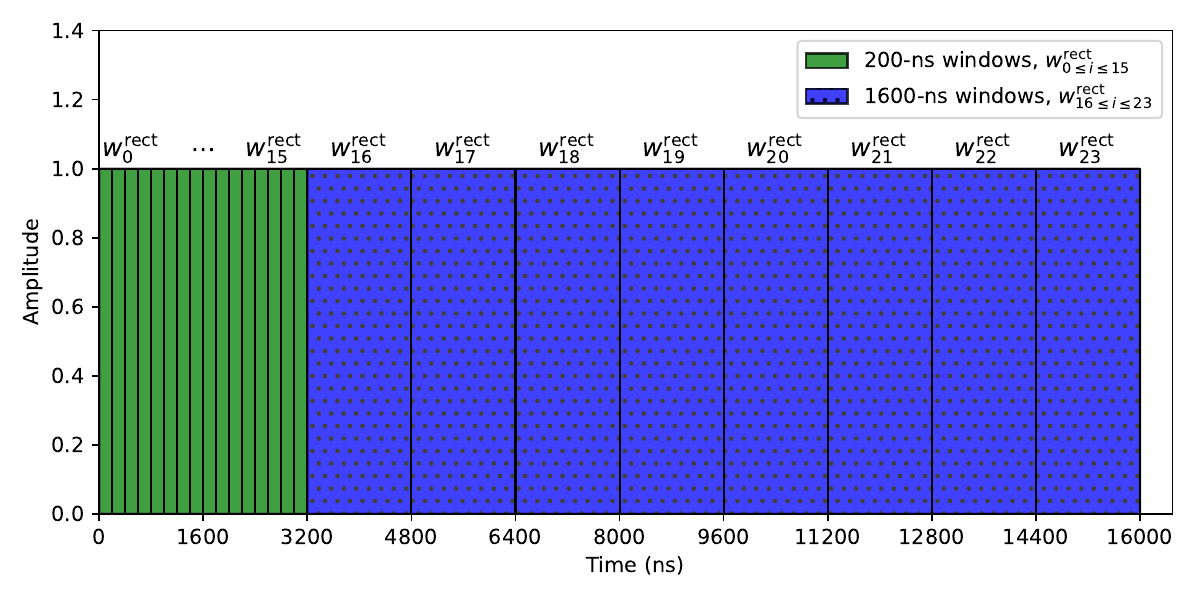}
\caption{\xingrui{\textbf{Rectangular windows for calibrating integration weights.} Bank of 24 rectangular windows, $w^{\mathrm{rect}}_j[n]$, used to coarse-grain the raw heterodyne record into 48 features, $I_j^{\mathrm{rect}}$ and $Q_j^{\mathrm{rect}} \, .$ The window widths vary over time. We narrow the windows at early times to emphasize the transient dynamics realized before the transmon's excited-state population decays.}}
    \label{fig:integration_weights_windows}
\end{figure*}

\begin{figure*}[htbp]
    \centering
    \includegraphics[width=0.8\textwidth]{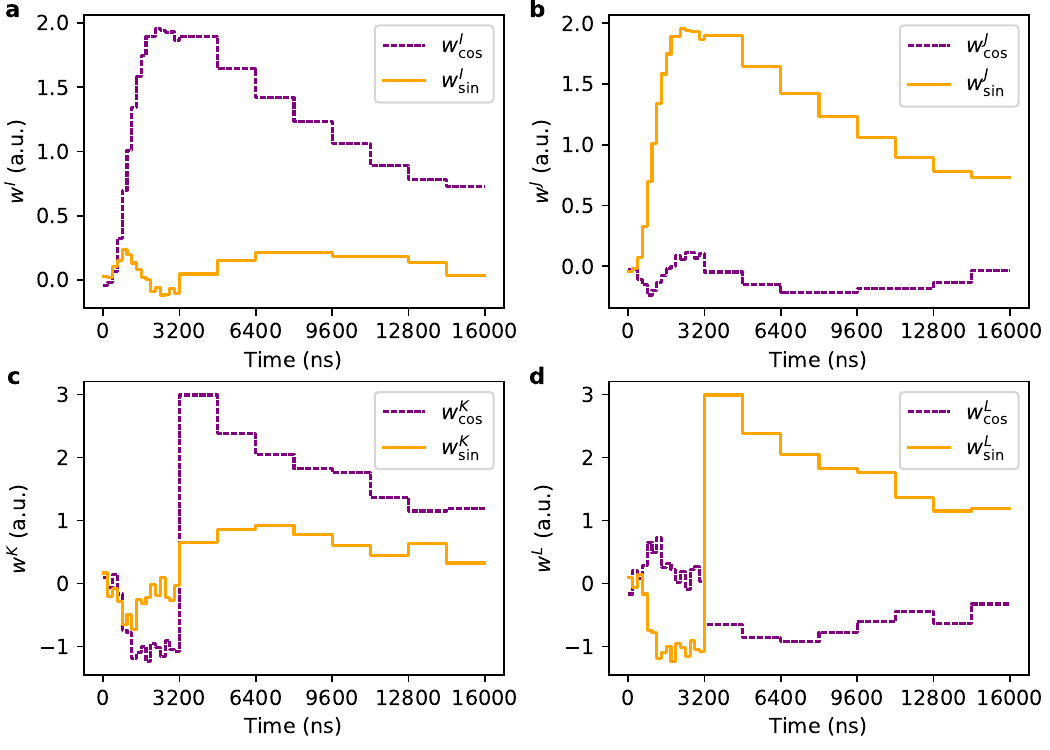}
\caption{\xingrui{\textbf{Readout integration weights.} Panels (a)–(d) show the temporal profiles of the four optimized integration weights ($w^I$, $w^J$, $w^K$, and $w^L$). We used these weights to compress the raw heterodyne record into four scalar features ($I$, $J$, $K$, and $L$) to perform state discrimination. We constructed the weights using Linear Discriminant Analysis and $90^\circ$ rotations.}}
    \label{fig:integration_weights}
\end{figure*}

After the foregoing analysis, we obtain the four scalar readout components ($I$, $J$, $K$, and $L$) by projecting the input data, $I^{\mathrm{raw}}$ and $Q^{\mathrm{raw}}$, onto the corresponding time-domain components of the basis vectors: if $X \in \{I, J, K, L\}$, then
\begin{widetext}
\begin{equation}
    X = \sum_n \left( 
    w^{X}_{\cos}[n]I^{\mathrm{raw}}[n] \cos(\omega_{\mathrm{IF}}t_n+\phi_\mathrm{IF}) 
    + w^{X}_{\sin}[n]Q^{\mathrm{raw}}[n] \sin(\omega_{\mathrm{IF}}t_n+\phi_\mathrm{IF}) \right).
\end{equation}
\end{widetext}
Figure~\ref{fig:integration_weights} displays the four integration weights as functions of time. Panel (a) corresponds to $w^I$; (b) to $w^J$; (c) to $w^K$; and (d) to $w^L$.

We infer the two-transmon state using a Histogram-based Gradient Boosting Classification Tree from scikit-learn. When analyzing any given shot, we concatenate the four integrated features extracted from the two readout resonators. We feed the concatenated features into the classifier, which outputs (i) a state assignment in $\{\ket{00},\ket{01},\ket{10},\ket{11}\}$ and (ii) a joint probability distribution over these possible states. We call the probabilities \emph{raw}. We can characterize the state assignments' fidelities using a \emph{confusion matrix}, 
\begin{widetext}
\begin{equation}
C \;=\;
\begin{pmatrix}
91.1\% & 5.2\% & 3.4\% & 0.2\% \\
6.8\%  & 92.0\% & 0.4\% & 4.2\% \\
1.9\%  & 0.1\% & 89.2\% & 5.5\% \\
0.2\%  & 2.7\% & 7.1\% & 90.0\%
\end{pmatrix},
\qquad
\text{(rows: measured; columns: prepared; basis } \{|00\rangle,|01\rangle,|10\rangle,|11\rangle\}\text{).}
\end{equation}
\end{widetext}
To correct for the effect of imperfect readout fidelity, we apply the confusion matrix's inverse to the raw probabilities. We call the result \emph{readout-fidelity-corrected} data in Appendix~\ref{app_Data_analysis}.
}

\subsection{Single-qubit rotations}
\label{subsec:single_qubit_rot}

We use four types of single-qubit rotations in this project:
\begin{itemize}
    \item Single-qubit $\pi/2$ rotations: We use 
    $\pi/2$ rotations about the $x$- and $y$-axes 
    in quantum state tomography and in the arbitrary-axis rotations detailed below. Each $\pi/2$ rotation lasts for
    44 ns.
    We construct pulse envelopes from cosine waveforms. As a function of the time $t$ (expressed in ns), the pulse waveform has the form $\propto [\cos(2\pi t/(44\ \mathrm{ns})-\pi)+1]$. 
    We estimate these gates' fidelity to be 98.8 \%.
    
    \item Single-qubit $\pi$ rotations: We implement these rotations in the same manner as the $\pi/2$ rotations. However, each pulse lasts for 88 ns. 

    \item Single-qubit $\pi$ rotations in the $\{\ket{e},\ket{f}\}$ manifold: 
    These rotations are implemented in the same manner as the $\{\ket{g},\ket{e}\}$ rotations described in the previous two bullet points.
    The $\pi/2$ gates last for 32 ns each; and the $\pi$ gates, 64 ns.

    \item Single-qubit rotations about the $z$-axis, through arbitrary angles $\alpha$:
    Denote by $R(\beta, \phi)$ a rotation through an angle $\beta$ about the $\LParen \cos(\phi) \, \hat{\bm{x}} + \sin(\phi) \, \hat{\bm{y}} \RParen$-axis. The corresponding rotation operator is represented by the matrix, relative to the $Z$-eigenbasis,
\begin{align}
     R(\beta, \phi) &=
    \begin{pmatrix}
        \cos \left(\frac{\beta}{2} \right) & 
        -i e^{-i \phi} \sin \left( \frac{\beta}{2} \right) \\
        -i e^{i \phi} \sin \left(\frac{\beta}{2} \right) & 
        \cos \left(\frac{\beta}{2} \right)
    \end{pmatrix}.
\end{align}
In terms of this rotation, we define the \emph{physical} rotation through an angle $\alpha$ about the $z$-axis,
\begin{align}
    R_z(\alpha) &= R\left(\pi, \frac{\alpha}{2}\right) R\left(\pi, 0\right). \label{Eq_RZ_as_R_X} \\
\end{align}

\end{itemize}

\subsection{Parametric gates} \label{sec:parametric}
   
We implement parametric gates by modulating the coupler's frequency: define $\Delta_{\antiq\q} \equiv \omega_\antiq - \omega_\q$. We apply a microwave tone to c's fast flux line at a frequency $\Delta_{\antiq\q}/2$, modulating c's energy gaps. Denote the modulation depth by $\alpha$. The coupler's time-dependent frequency is $\tilde{\omega}_\mathrm{c}(t) = \omega_\mathrm{c} - \alpha \cos^2(\Delta_{\antiq\q} t/2)$.  During this modulation, $\q$ and $\antiq$ come into parametric resonance.
When calibrated, the parametric resonance induces qubit--antiqubit coupling at a rate of 54.0 MHz. At this rate, an $\sqrt{i\mathrm{SWAP}}$ gate lasts 104 ns. 

We estimate the gate's fidelity using quantum state tomography. To do so, we measure the expectation values of $9$ two-Pauli products: 
$\langle \sigma_\q \sigma_\antiq\rangle$. 
We use maximum-likelihood estimation to infer the qubit--antiqubit density matrix's components~\cite{James2001}.   

We implement as follows the protocol shown in Fig.~\ref{fig3}(a). To prepare an entangled state, we perform one $\sqrt{i\mathrm{SWAP}}$ gate. Then, we projectively measure whether the system is in a singlet, $\ket{\Psi^-}$. That is, we measure $\{ \Pi_0, \mathbb{1} - \Pi_0 \}$ 
(as in the main text, $\Pi_0 \coloneqq \ketbra{ \Psi^- }{ \Psi^- }$).
To measure this POVM, we apply a second $\sqrt{i\mathrm{SWAP}}$ gate, which maps the singlet to 
$\ket{g}_\mathrm{q}\ket{e}_\mathrm{\antiq}$.
The qubits have different energies, so they accumulate different relative phases between the two $\sqrt{i\mathrm{SWAP}}$ gates. To remove this phase difference, we apply another physical rotation to the antiqubit before the second $\sqrt{i\mathrm{SWAP}}$ gate.  We implement this phase correction with the physical rotation (\ref{Eq_RZ_as_R_X}) described above. 

\subsection{Data analysis}
\label{app_Data_analysis}

\begin{figure*}[t]
  \centering
  \includegraphics[width=1\linewidth]{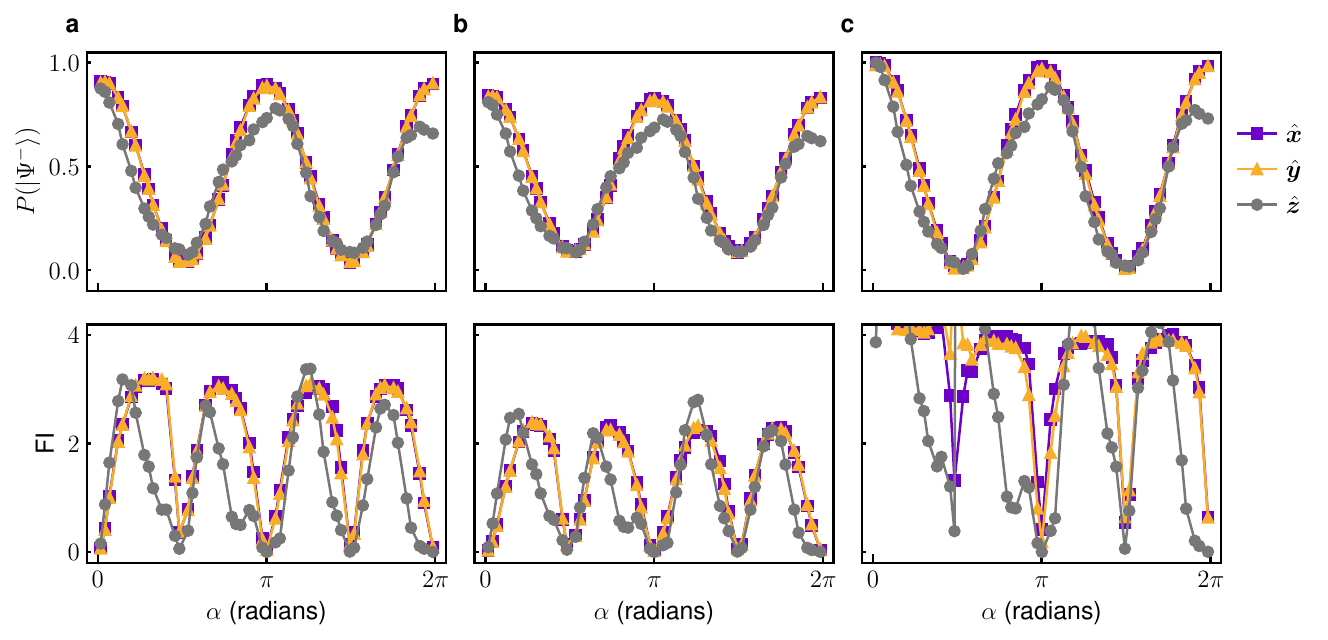}
  \caption{\kater{
  \textbf{Singlet-state probabilities and Fisher information.} The top panels show the measurement's probability of yielding the singlet outcome, $P\!\left(|\Psi^-\rangle\right)$, as a function of $\alpha$. Different colors and marker shapes correspond to different rotation axes. The leftmost column shows readout-fidelity-corrected data; the middle column, raw data; and the rightmost column, gate-fidelity-corrected data. The bottom panels show the inferred FI as a function of $\alpha$. 
  }
}
  \label{fig:probcorrected}
\end{figure*}

\begin{figure*}[t]
  \centering
  \includegraphics[width=1\linewidth]{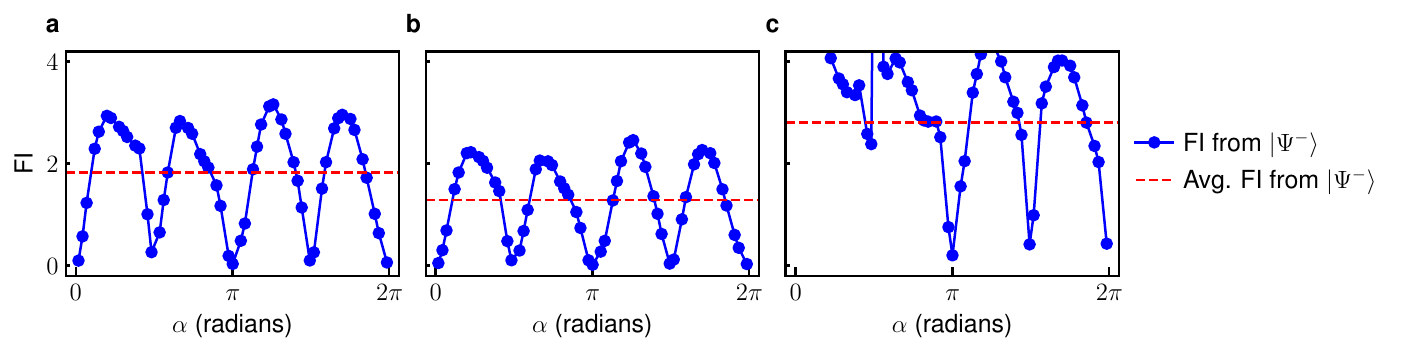}
  \caption{\kater{
  \textbf{Average Fisher information versus rotation angle.}
  To produce each panel, we averaged the FI obtained from rotations about the $x$-, $y$-, and $z$-axes. Panel (a) shows readout-fidelity-corrected data; (b), raw data; and (c), gate-fidelity-corrected data. The horizontal lines show the FI averaged over all $\alpha$ values. 
  The gate-fidelity-corrected data lead the FI to diverge. We discard the singularities when averaging the FI over $\alpha$.
  }
}
  \label{fig:SI_fig2}
\end{figure*}

\kater{The FI quantifies by how rapidly measured probabilities change as $\alpha$ changes. Empirically inferring the FI, we perform three steps:
(i) measuring state probabilities $P(\alpha)$,
(ii) varying $\alpha$,
and (iii) calculating the derivative $dP(\alpha)/d\alpha$ from the measured data. Below, we describe the procedure followed to implement each of these steps. We apply these procedures to the entanglement-free and positronium-sensing experiments:
\begin{enumerate}

   %
   \item \emph{Measured probabilities and readout fidelity-correction:} We infer state probabilities  from the dispersive readout. Appendix~\ref{sec_Readout_postproc} describes the integration weights, classifiers, and fidelity correction applied. To identify the readout-fidelity correction's effect on the FI, we obtain raw probabilities by applying the confusion matrix to the readout-fidelity-corrected data (because we do not save the raw probabilities). We analyze the two data sets independently.




   %
   \item \emph{Time-to-phase conversion:} We vary the parameter $\alpha$ by applying the Hamiltonian \eqref{eq:magneticH} for a duration that varies across trials. When the qubit is prepared in a superposition of the eigenstates of \eqref{eq:magneticH}, the resulting time evolution generically has oscillatory dynamics in one of the Pauli expectation values. By experimentally determining the oscillation frequency through quantum state tomography, we calibrate $\alpha$. 

   \item \emph{Local fitting and probability models:} To calculate the FI, we must calculate the derivatives $dP(\alpha)/d\alpha$. We calculate the derivatives from the measured probabilities using a sliding-window local fitting procedure. The window is of size 0.52~rad. Within each window, we approximate $P(\alpha)$ with a cosine, evaluate the derivative analytically, and average overlapping estimates.

Within each window, we model the local probability as
\begin{equation}
P(\alpha) = A \cos(B \alpha) + C, 
\end{equation}
where $A$ is an amplitude scaling factor, $B$ is a frequency scaling factor, and $C$ is an offset. This model captures local behavior without attributing global periodicity or uniform
visibility to the full data set.
If the prepared state is an eigenstate of the applied rotation, the probability is flat, so the FI vanishes.

\end{enumerate}
We apply this analysis independently to the data associated with the $x$-, $y$-, and $z$-rotation axes.

\section{Correction of Fisher information for gate fidelity and readout fidelity} \label{sec:fidelitycorrection}

Consistently with prior work, we correct our data for the calibrated readout infidelity (described above) before evaluating the FI. This strategy allows us to highlight the metrological scheme's advantage. However, the raw readout fidelity would affect the FI achieved in a practical sensing experiment. In this appendix, we analyze the FI obtained in positronium sensing and in entanglement-free sensing, with and without the correction.

Figure~\ref{fig:probcorrected}'s top row displays the final measurement's probability of yielding the singlet outcome. The bottom row shows the empirically inferred FI. We produced different columns from different datasets: the leftmost column follows from the \emph{readout-fidelity-corrected} data, as does Fig.~\ref{fig3}. The middle column follows from the \emph{raw} data. The rightmost column follows from the \emph{gate-fidelity-corrected} data, which results from correcting for the entangling gate's infidelity. The gate-fidelity-corrected data are obtained by rescaling the fitted probability model to have ideal contrast, such that the probability oscillates between 0 and 1, and then recalculating the FI from the rescaled model. At points where the fitted probability reaches 0 or 1, direct numerical evaluation of the FI can produce singular or nonfinite values because its denominator vanishes. These numerical artifacts are not physically meaningful and are excluded when calculating the average FI.} The raw data leads to a lesser FI, as expected (Fig.~\ref{fig:probcorrected}a, compared to Fig.~\ref{fig:probcorrected}b). Once we correct for the entangling-gate fidelity and readout fidelity, the probability oscillation's contrast approaches unity (Fig.~\ref{fig:probcorrected}c). Furthermore, the FI nearly reaches its maximal value, 4. Figure~\ref{fig:SI_fig2} displays the extracted FI, averaged over the rotation axes, versus $\alpha$ in each case.

{\color{black}
\section{Theoretical model for effect of state-preparation fidelity and single-transmon-readout errors on the Fisher information} \label{sec:noise}

In this appendix, we develop a theoretical noise model to quantify how experimental imperfections reduce the FI achieved in positronium sensing. Two noise sources affect the FI most: (i) The imperfect entangling-gate fidelity distorts the singlet preparation. (ii) The imperfect single-transmon-measurement fidelities distort the measurement of whether the transmons are in a singlet. In Sec.~\ref{sec:prepnoisemodel}, we model state-preparation errors using a Werner mixture and derive the resulting FI. In Sec.~\ref{sec:readout_noise}, we incorporate single-transmon-readout errors into the analysis. These results quantitatively predict the experimentally observed singlet-detection statistics and hence the diminution of the FI. 

This appendix resembles and contrasts with Appendix~\ref{sec:fidelitycorrection}. In both appendices, we incorporate experimental errors into our FI analysis. Here, experimental parameter values inform the analysis; but the goal is to predict the experimentally inferred FI, and the model is mostly abstract. In Appendix~\ref{sec:fidelitycorrection}, the goal is to calculate the FI from experimental data.

\subsection{Model for imperfect state preparation} \label{sec:prepnoisemodel}

As discussed in the main text, we aim to prepare the singlet $\ket{\Psi^-} =\ketbra{\Psi^-}{\Psi^-}$ using $\sqrt{i\mathrm{SWAP}}$ gates. In this section, we quantify the singlet preparation's infidelity. We model the realized state with a Werner mixture.

To characterize the prepared state, we perform quantum state tomography. We denote the reconstructed density operator by $\rho_{\exp}$.
We quantify the singlet-preparation quality with the Uhlmann fidelity between $\rho_{\exp}$ and the ideal singlet \cite{Nielsen2012}:
\begin{equation}
   \mathcal{F}
   \coloneqq \left(\mathrm{Tr} \sqrt{\sqrt{\rho_{\exp}}\,\sigma\,\sqrt{\rho_{\exp}}} \, \right)^2 .
\label{eq:fidelity_def_UI}
\end{equation}
The experimentally observed fidelity is $\mathcal{F} = 0.97$.

To model this fidelity's effect on the measurement, we approximate the realized state with a Werner mixture,
\begin{equation}
   \rho^{\mathcal{(F)}}
   \coloneqq \mathcal{F}\,\ketbra{\Psi^-}{\Psi^-}
   + \frac{\id - \mathcal{F}}{3}\Bigl(\id-\ketbra{\Psi^-}{\Psi^-}\Bigr).
\label{eq:B9_Werner_UI}
\end{equation}
This model captures the mixing of (i) the singlet and (ii) the maximally mixed state defined on the orthogonal (triplet) subspace. 

\subsection{State-preparation fidelity's effect on Fisher information} \label{sec:effect_fidelity_FI}

Let us compute the FI, assuming that the initial singlet is prepared with a fidelity $\mathcal{F}$.
Section \ref{sec_Backgrnd_phase_est} of the main text introduced the unitary
$U_{\alpha} \coloneqq e^{-i\alpha\,\hat{\bm{n}}\cdot\boldsymbol{\sigma}/2} \, ,$
which rotates a qubit about an unknown axis $\hat{\bm{n}}$ through an angle $\alpha$. The unitary evolves the singlet approximation to
\begin{equation}
   \rho^{\mathcal{(F)}}_\alpha
   \coloneqq (U_{\alpha}\otimes U_{\alpha}^\dagger)\,\rho^{\mathcal{(F)}}\,(U_{\alpha}^\dagger\otimes U_{\alpha}). \label{eq:rho_alpha_UI}
\end{equation}
Let $\Pi_{\text{s}} \coloneqq \ketbra{\Psi^-}{\Psi^-}$ denote the projector onto the singlet subspace. Upon preparing $\rho^{\mathcal{(F)}}_\alpha \, ,$ we measure the POVM $\{\Pi_{\text{s}},\id - \Pi_{\text{s}}\}$. We obtain the ``singlet'' outcome with a probability
\begin{equation}
\tilde p^{\mathcal{(F)}}_{\text{s}}(\alpha)=\mathrm{Tr}\left(\Pi_{\text{s}}\rho^{\mathcal{(F)}}_\alpha\right). \label{eq:ps_def_UI}
\end{equation}
Substituting in from Eq.~\eqref{eq:rho_alpha_UI} yields
\begin{widetext}
\begin{align}
\tilde   p^{\mathcal{(F)}}_{\text{s}}(\alpha)
   &=\mathcal{F}\,
   \left|\braket{\Psi^-|U_{\alpha}\otimes U_{\alpha}^\dagger|\Psi^-}\right|^2
   +\frac{1-\mathcal{F}}{3}\,
   \mathrm{Tr}\left( \Pi_{\text{s}} 
   \left[ U_{\alpha}\otimes U_{\alpha}^\dagger \right]
   \left[ \id-\Pi_{\text{s}} \right]
   \left[ U_{\alpha}^\dagger\otimes U_{\alpha} \right] \right).
\label{eq:ps_intermediate}
\end{align}
\end{widetext}
Substituting in the identity 
$\braket{\Psi^-| U_{\alpha} \otimes U_{\alpha}^\dagger|\Psi^-}
   =\frac{1}{2}\,\mathrm{Tr} \left(U_{\alpha}^2 \right) 
   = \cos^2 ({\alpha})$
yields
\begin{equation}
\tilde p^{\mathcal{(F)}}_{\text{s}}(\alpha)
=\frac{1-\mathcal{F}}{3}+\frac{4\mathcal{F}-1}{3}\cos^2 \left( \alpha \right) \label{eq:ps_prob} . 
\end{equation}

We now quantify the information provided by the measurement about the parameter $\alpha$. To do so, we compute the FI, using the formula in the third paragraph of the main text's Sec.~\ref{sec_Backgrnd_phase_est}. The measurement outcome is binary, and the success probability is $p_{\text{s}}(\alpha)$; so 
\begin{equation}
I_\alpha
= \left(\partial_\alpha \tilde   p^{\mathcal{(F)}}_{\text{s}}(\alpha)\right)^2 / [\tilde   p^{\mathcal{(F)}}_{\text{s}}(\alpha)(1-\tilde   p^{\mathcal{(F)}}_{\text{s}}(\alpha))] \, .
\label{eq:FI_binary_UI}
\end{equation}
To apply this formula, we differentiate Eq.~\eqref{eq:ps_prob}:
\begin{equation}
\partial_\alpha \tilde   p^{\mathcal{(F)}}_{\text{s}}(\alpha)=\frac{1-4\mathcal{F}}{3}\,\sin(2\alpha). \label{eq:der_p}
\end{equation}
Let us substitute from Eqs.~\eqref{eq:der_p} and \eqref{eq:ps_prob} into Eq.~\eqref{eq:FI_binary_UI}. The FI obtainable about $\alpha$, despite imperfect singlet preparation, is
\begin{widetext}
\begin{equation}
   I_\alpha
   = -\frac{4(1-4\mathcal{F})^2\,\sin^2(2\alpha)}
{\left[-5+2\mathcal{F}+(-1+4\mathcal{F})\cos(2\alpha)\right]
\Bigl[1+2\mathcal{F}+(-1+4\mathcal{F})\cos(2\alpha)\Bigr]} \, .
\label{eq:FI_UI_factored}
\end{equation}
\end{widetext}

Equation~\eqref{eq:FI_UI_factored} follows from Eq.~(B10) of~\cite{Song2024}, once one replaces $\alpha$ with $2\alpha$ and $I_\alpha$ with $4 I_\alpha$. In the ideal limit, $\mathcal{F}\to 1$. Consequently, the FI saturates at $I_\alpha=4$, independently of $\alpha$, as expected from the main text's Sec.~\ref{sec_Theory_pos_met}. In the next two subsections, we analyze the single-transmon-readout noise's effect on the FI.

\subsection{Model for single-transmon-readout noise}
\label{sec:readout_noise}

In this section, we model imperfections in the readout of each transmon. In the experiment, we measure whether the two transmons are in a singlet or not. To do so, we measure the Bell basis in two steps. First, we map the Bell basis to the computational basis via a unitary---a CNOT followed by a Hadamard. Then, we measure the computational basis projectively. This procedure maps the singlet $\ket{\Psi^-}$ to the measurement outcome $11$. We implement the POVM $\{ \Pi_{\text{s}},\id - \Pi_{\text{s}} \}$ by labeling the outcomes 00, 01, and 10 as ``not singlet.''

We model the qubit-readout error as independent of the antiqubit-readout error. Let $b_i \in\{0,1\}$ denote outcome that would result from measuring the computational basis of transmon $i\in\{ {\rm q},\bar {\rm q} \}$ ideally. Let $b_i'$ denote the bit recorded by the measurement device.
The mechanism for reading out transmon $i$, we posit, exhibits a bit-flip error and a fidelity $F_i$:
\begin{equation}
   p(b_i'{=}b_i)=F_i \, ,
   \quad \text{and} \quad
   p(b_i'{\neq} b_i)=1-F_i \, .
\end{equation}

Define the two-bit strings 
$b \coloneqq b_{\rm q} \, b_{\bar {\rm q}}$ and 
$b' \coloneqq b_{\rm q}' \, b_{\bar {\rm q}}' \, .$
Denote by $p(b' | b)$ the probability of obtaining the $b'$ outcome, given that an ideal measurement would yield $b$. We call $p(b' | b)$ the \emph{assignment model}.
By the single-transmon-readout mechanisms' independence,
\begin{equation}
   p(b'|b)
   =p(b_{\rm q}'|b_{\rm q})\:
   p(b_{\bar {\rm q}}'|b_{\bar {\rm q}}) .
\end{equation}
Let $w_b \coloneqq p(b'{=}11\,|\,b)$ denote the probability of recording the singlet outcome ($b'=11$), given the ideal outcome $b\in\{00,01,10,11\}$. We can express these probabilities in terms of the single-transmon-readout fidelities $F_q$ and $F_{\bar {\rm q}}$:
\begin{widetext}
\begin{equation}
   w_{11} = F_{\rm q} F_{\bar {\rm q}}  ,\quad
   w_{10} = F_{\rm q} (1-F_{\bar {\rm q}}), \quad
   w_{01} = (1-F_{\rm q})F_{\bar {\rm q}},
   \quad \text{and} \quad
   w_{00} = (1-F_{\rm q})(1-F_{\bar {\rm q}}). \label{Eq_w}
\end{equation}
\end{widetext}

\subsection{Fisher information subject to state-preparation and single-qubit-readout errors}

Having specified the assignment model $p(b'|b)$, we use it to predict the FI obtained from the experiment, which features imperfect singlet preparation and imperfect single-transmon readout. First, we imagine preparing the singlet perfectly and measuring it using perfect single-transmon readout. We calculate the ideal probabilities $p_b$, which inform the experimentally observed probabilities,
$\tilde p_{b'}=\sum_b p(b'|b) \: p_b$.
The experimental measurement yields the ``singlet'' outcome with a probability $\tilde p_{11}=\sum_b p(b'{=}11|b)\: p_b$.
This probability depends on the rotation axis $\hat{\bm{n}}$, which we assume obeys an isotropic probability distribution. We calculate the FI expected to be available from the experiment, on average over rotation axes. 

Let us calculate the outcome probabilities obtained from an ideal singlet and ideal single-transmon readout. Denote the evolved singlet by 
$\rho_{\alpha} 
\coloneqq U_{\alpha} \ketbra{\Psi^{-}}{\Psi^{-}}U_{\alpha}^{\dagger} $. Denote by $( \Pi_{00},\Pi_{01},\Pi_{10},\Pi_{11} )$ the list of Bell projectors associated with the ideal measurement outcomes $( 00,01,10,11 )$. As mentioned in the previous subsection, we obtain these outcomes from 
$( \ket{\Phi^+},\ket{\Psi^+},\ket{\Phi^-},\ket{\Psi^-} )$ via a CNOT, then a Hadamard, and then a computational-basis measurement. Consider performing an ideal Bell measurement on $\rho_{\alpha}$. The outcome probabilities share the form
\begin{equation}
   p_b
   \coloneqq \Tr(\Pi_b\,\rho_{\alpha}),
   \quad \text{wherein} \quad
   b\in\{00,01,10,11\} .
\end{equation}
To write out the probabilities' forms explicitly, we recall that $\hat{\bm{n}}=(n_x,n_y,n_z)$ and $n_x^2+n_y^2+n_z^2=1$:
\begin{widetext}
\begin{align}
& p_{11}
=\Tr\left(\Pi_{11}\,\rho_{\alpha}\right)
=\cos^2 \left( \alpha \right), \qquad
p_{10}
=\Tr\left(\Pi_{10}\,\rho_{\alpha}\right)
=\sin^2 \left( \alpha \right)\,n_x^2 \, , \nonumber\\
& p_{00}
=\Tr\left(\Pi_{00}\,\rho_{\alpha}\right)
=\sin^2 \left( \alpha \right)\,n_y^2 \, , 
\quad \text{and} \quad 
p_{01}
=\Tr\left(\Pi_{01}\,\rho_{\alpha}\right)
=\sin^2 \left( \alpha \right)\,n_z^2 \, .
\label{eq:bell_probs_pure}
\end{align}
\end{widetext}

In the experiment, we prepare the singlet with a fidelity $\mathcal{F}$, defined in Sec.~\ref{sec:prepnoisemodel}. Consider evolving the experimentally realized initial state under $U_\alpha \, .$ We model resulting state with
\begin{equation}
   \rho^{(\mathcal{F})}_\alpha
   \coloneqq \mathcal{F}\,\rho_{\alpha}
+\frac{1-\mathcal{F}}{3}\left(\id-\rho_{\alpha}\right).
\end{equation}
Consider performing an ideal Bell measurement of this state. The measurement yields the outcomes $b \in \{ 00, 01, 10, 11 \}$ with probabilities
\begin{equation}
   p^{(\mathcal{F})}_{b}
   \coloneqq \Tr \left( \Pi_b \, \rho^{(\mathcal{F})}_\alpha \right)
   = \frac{1-\mathcal{F}}{3}+\frac{4\mathcal{F}-1}{3}\,p_{b} \, .
\label{eq:bell_probs_werner}
\end{equation}
Now, consider performing on $\rho^{(\mathcal{F})}_\alpha$ a Bell measurement plagued by imperfect single-transmon readout. The measurement yields the outcome $b'{=}11$ with a probability
\begin{widetext}
\begin{equation}
   \tilde p_{\text{s}}(\alpha) 
   \coloneqq \tilde p_{11} 
   = \sum_{b\in\{00,01,10,11\}} p(b'{=}11\,|\,b)\, 
   p^{(\mathcal{F})}_b
   = \sum_{b\in\{00,01,10,11\}} w_b\, p^{(\mathcal{F})}_b \, .
\label{eq:pshat_def}
\end{equation}
\end{widetext}
Let us substitute in for $p_b^{\mathcal{F}}$ from Eq.~\eqref{eq:bell_probs_werner}. Then, we substitute in for the ideal probabilities $p_b$ (on which $p_b^{\mathcal{F}}$ depends) from Eq.~\eqref{eq:bell_probs_pure}. The measurement yields the ``singlet'' outcome with a probability
\begin{widetext}
\begin{equation}
   \tilde p_{\text{s}}(\alpha)
   = \frac{1-\mathcal{F}}{3}
   +\frac{4\mathcal{F}-1}{3}
   \left[ w_{11}\cos^2 \left( \alpha \right) 
          +\sin^2 \left( \alpha \right)\,
          \bigl(w_{10}n_x^2+w_{00}n_y^2+w_{01}n_z^2\bigr)
\right].
\label{eq:pshat_general}
\end{equation}
\end{widetext}
The singlet-outcome probability depends on the rotation axis. 

Suppose that the probability distribution over the possible axes $\hat{\bm{n}}$ is isotropic. Let $\expval{.}_{\hat{\bm{n}}}$ denote an average over the possible axes. The squared axis components $n_{x,y,z}^2$ average to
$\langle n_x^2\rangle_{\hat{\bm{n}}}
=\langle n_y^2\rangle_{\hat{\bm{n}}}
=\langle n_z^2\rangle_{\hat{\bm{n}}}=1/3$. Let us introduce these averages into Eq.~\eqref{eq:pshat_general}. Also, we substitute into Eq.~\eqref{eq:pshat_general} the probabilities $w_b$ from Eq. \eqref{Eq_w}. We denote by $\eta \coloneqq F_q F_{\bar q}$ the joint probability that both transmons are read out correctly. The singlet-outcome probability averages to
\begin{equation}
\langle \tilde p_{\text{s}}(\alpha) \rangle_{\hat{\bm{n}}}
=\frac{1-\eta}{3}
+\frac{4\eta-1}{3}\left[
\frac{1-\mathcal{F}}{3}+\frac{4\mathcal{F}-1}{3}\cos^2 \left( \alpha \right)
\right].
\label{eq:ps_readout_alpha}
\end{equation}

Having calculated the average singlet-outcome probability, we calculate the (axis-averaged) FI from it. Differentiating Eq.~\eqref{eq:ps_readout_alpha} gives
\begin{widetext}
\begin{equation}
\partial_{\alpha}\langle \tilde p_{\text{s}}(\alpha) \rangle_{\hat{\bm{n}}}
=\frac{4\eta-1}{3}\,\partial_\alpha p_{\text{s}}(\alpha)
=\frac{(4\eta-1)(1-4\mathcal{F})}{9}\,\sin(2\alpha).
\label{eq:dps_readout_alpha}
\end{equation}
\end{widetext}
Let us substitute from Eqs.~\eqref{eq:dps_readout_alpha} and \eqref{eq:ps_readout_alpha} into the FI formula Eq.~\eqref{eq:FI_binary_UI}. The FI obtainable about $\alpha$, despite imperfect singlet preparation and singlet-transmon-readout noise, on average over axis angles, is 
\begin{equation}
\tilde I_\alpha
= \frac{(4\eta-1)^2(4\mathcal{F}-1)^2\,\sin^2(2\alpha)}
{81 \langle \tilde p_{\text{s}}(\alpha) \rangle_{\hat{\bm{n}}} \bigl[1-\langle \tilde p_{\text{s}}(\alpha) \rangle_{\hat{\bm{n}}} \bigr]} \, .
\label{eq:FI_readout}
\end{equation}

We check this result in two ways. First, we consider the ideal-readout limit, $\eta\to 1$. In this regime, $\tilde p_{\text{s}}\to p_{\text{s}} \, ,$ and $\tilde I_\alpha\to I_\alpha \, ,$ as expected. Second, we plot Eqs.~\eqref{eq:ps_readout_alpha} and~\eqref{eq:FI_readout} in Fig.~\ref{fig:SI_Fig_3}. We substitute in experimentally obtained values for the accuracy parameters: $\mathcal{F} = 0.970$, $F_{q} = 0.978$, and $F_{\bar q} = 0.95$. 

\reviewercomment{1}{4}{The curves in Fig.~9 do resemble those in Fig.~3(b), but there is still a quantitative mismatch, as the maximal FI is higher than in the experiment. I think that the authors should acknowledge this difference and give a possible explanation as to why it does not fully match. I acknowledge the effort in the modelling, though, and I think it is enough to understand a large part of the experimental limitations.}

\reviewercomment{3}{5}{The reported measured probability oscillations have high contrast even for the entangled state, while the FI is reduced from the ideal value of 4.0 to $2.98\pm0.13$. The model includes the input-state infidelity and measurement infidelity. However, it does not include contributions from relaxation, dephasing, device-modelled gate errors, and unitary inversion.}

Although the model reproduces the qualitative behavior of the experimental data, it predicts a somewhat larger maximal FI. This quantitative discrepancy likely arises because the model does not include the cumulative errors from the full experimental sequence, which contains two $\sqrt{\mathrm{iSWAP}}$ gates and several single-qubit $\pi/2$ and $\pi$ rotations, as well as decoherence and calibration drift. In addition, the experimental FI is extracted from locally fitted probability curves. These fitted curves can underestimate their derivatives and therefore the FI. Including these additional imperfections would further reduce the modeled FI and improve its quantitative agreement with the experiment.

\begin{figure*}[t]
    \centering
    \includegraphics[width=\columnwidth]{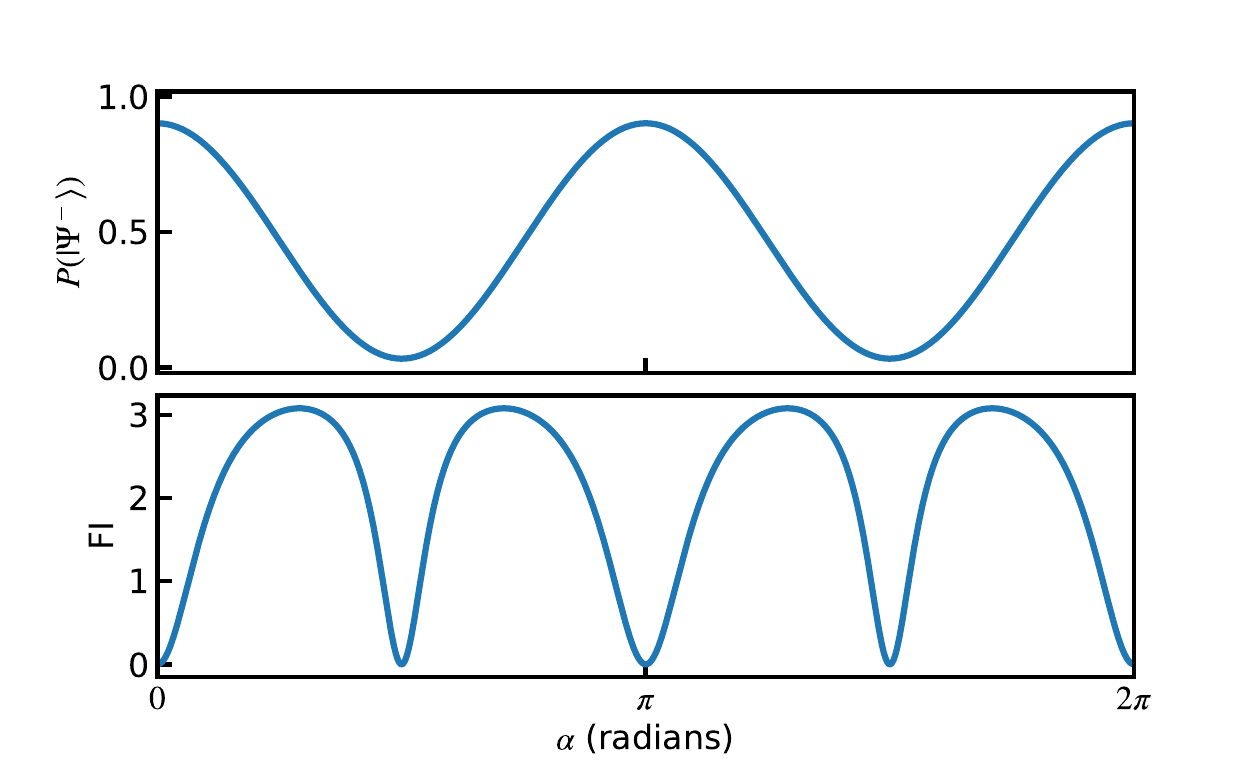}
    \caption{\response{Theoretical predictions for quantities obtained from positronium sensing, on average over field axes. The average singlet-outcome probability (top panel) is calculated from Eq.~\eqref{eq:ps_readout_alpha}; and the average FI (bottom panel), from Eq.~\eqref{eq:FI_readout}. The singlet-preparation fidelity is $\mathcal{F} = 0.970$, and the single-transmon-readout fidelities are $F_{q} = 0.978$ and $F_{\bar q} = 0.95$.  The FI attains its maximum, 3.08, at $\alpha = 52^\circ$. These curves resemble the experimentally measured counterparts in Fig.~\ref{fig:probcorrected}(b). }}
    \label{fig:SI_Fig_3}
\end{figure*}

}


\clearpage

\nocite{Cerf1997,25_Gardner_Stochastic}
\bibliography{agnostic_sensing}

@article{Fan_2003,
   title={Quantify entanglement by concurrence hierarchy},
   volume={36},
   ISSN={1361-6447},
   url={http://dx.doi.org/10.1088/0305-4470/36/14/316},
   DOI={10.1088/0305-4470/36/14/316},
   number={14},
   journal={Journal of Physics A: Mathematical and General},
   publisher={IOP Publishing},
   author={Fan, Heng and Matsumoto, Keiji and Imai, Hiroshi},
   year={2003},
   pages={4151–4158} }

@article{concurrence1998,
  title = {Entanglement of Formation of an Arbitrary State of Two Qubits},
  author = {Wootters, William K.},
  journal = {Phys. Rev. Lett.},
  volume = {80},
  issue = {10},
  pages = {2245--2248},
  numpages = {0},
  year = {1998},
  publisher = {American Physical Society},
  doi = {10.1103/PhysRevLett.80.2245},
  url = {https://link.aps.org/doi/10.1103/PhysRevLett.80.2245}
}

@misc{concurrence2012,
      title={Concurrence for a two-qubits mixed state consisting of three pure states in the framework of SU(2) coherent states}, 
      author={S. Salimi and A. Mohammadzade and K. Berrada},
      year={2012},
      eprint={1005.2083},
      archivePrefix={arXiv},
      primaryClass={quant-ph},
      url={https://arxiv.org/abs/1005.2083}, 
}

@misc{APE-SM,
  title        = {Supplemental Material for "Agnostic Phase Estimation"},
  howpublished = {\url{http://link.aps.org/supplemental/10.1103/PhysRevLett.132.260801}},

  year         = {2024},
}

@article{CTC_Deutsch,
  title = {Quantum mechanics near closed timelike lines},
  author = {Deutsch, David},
  journal = {Phys. Rev. D},
  volume = {44},
  issue = {10},
  pages = {3197--3217},
  numpages = {0},
  year = {1991},
  publisher = {American Physical Society},
  doi = {10.1103/PhysRevD.44.3197},
  url = {https://link.aps.org/doi/10.1103/PhysRevD.44.3197}
}

@article{CTC_Lloyd_1,
  title = {Closed Timelike Curves via Postselection: Theory and Experimental Test of Consistency},
  author = {Lloyd, Seth and Maccone, Lorenzo and Garcia-Patron, Raul and Giovannetti, Vittorio and Shikano, Yutaka and Pirandola, Stefano and Rozema, Lee A. and Darabi, Ardavan and Soudagar, Yasaman and Shalm, Lynden K. and Steinberg, Aephraim M.},
  journal = {Phys. Rev. Lett.},
  volume = {106},
  issue = {4},
  pages = {040403},
  numpages = {4},
  year = {2011},
  publisher = {American Physical Society},
  doi = {10.1103/PhysRevLett.106.040403},
  url = {https://link.aps.org/doi/10.1103/PhysRevLett.106.040403}
}

@article{CTC_Lloyd_2,
  title = {Quantum mechanics of time travel through post-selected teleportation},
  author = {Lloyd, Seth and Maccone, Lorenzo and Garcia-Patron, Raul and Giovannetti, Vittorio and Shikano, Yutaka},
  journal = {Phys. Rev. D},
  volume = {84},
  issue = {2},
  pages = {025007},
  numpages = {11},
  year = {2011},
  publisher = {American Physical Society},
  doi = {10.1103/PhysRevD.84.025007},
  url = {https://link.aps.org/doi/10.1103/PhysRevD.84.025007}
}

@misc{AmplitudeAmplification,
   title={Quantum amplitude amplification and estimation},
   ISSN={0271-4132},
   url={http://dx.doi.org/10.1090/conm/305/05215},
   DOI={10.1090/conm/305/05215},
   journal={Quantum Computation and Information},
   publisher={American Mathematical Society},
   author={Brassard, Gilles and Høyer, Peter and Mosca, Michele and Tapp, Alain},
   year={2002},
   pages={53–74} }

@article{PhaseEstimation,
   title={On low-depth algorithms for quantum phase estimation},
   volume={7},
   ISSN={2521-327X},
   url={http://dx.doi.org/10.22331/q-2023-11-06-1165},
   DOI={10.22331/q-2023-11-06-1165},
   journal={Quantum},
   publisher={Verein zur Forderung des Open Access Publizierens in den Quantenwissenschaften},
   author={Ni, Hongkang and Li, Haoya and Ying, Lexing},
   year={2023},
   pages={1165} }

@article{Miyazaki2022,
  doi = {10.22331/q-2022-03-10-665},
  url = {https://doi.org/10.22331/q-2022-03-10-665},
  title = {Imaginarity-free quantum multiparameter estimation},
  author = {Miyazaki, Jisho and Matsumoto, Keiji},
  journal = {{Quantum}},
  issn = {2521-327X},
  publisher = {{Verein zur F{\"{o}}rderung des Open Access Publizierens in den Quantenwissenschaften}},
  volume = {6},
  pages = {665},
  year = {2022}
}

@article{Wang_2024,
   title={Achieving the Multiparameter Quantum Cram\'{e}r-Rao Bound with Antiunitary Symmetry},
   volume={133},
   ISSN={1079-7114},
   url={http://dx.doi.org/10.1103/PhysRevLett.133.210801},
   DOI={10.1103/physrevlett.133.210801},
   number={21},
   journal={Physical Review Letters},
   publisher={American Physical Society (APS)},
   author={Wang, Ben and Zheng, Kaimin and Xie, Qian and Zhang, Aonan and Xu, Liang and Zhang, Lijian},
   year={2024}}

@article{corr1,
  title = {Measuring Out-of-Time-Order Correlators on a Nuclear Magnetic Resonance Quantum Simulator},
  author = {Li, Jun and Fan, Ruihua and Wang, Hengyan and Ye, Bingtian and Zeng, Bei and Zhai, Hui and Peng, Xinhua and Du, Jiangfeng},
  journal = {Phys. Rev. X},
  volume = {7},
  issue = {3},
  pages = {031011},
  numpages = {12},
  year = {2017},
  publisher = {American Physical Society},
  doi = {10.1103/PhysRevX.7.031011},
  url = {https://link.aps.org/doi/10.1103/PhysRevX.7.031011}
}

@article{corr2,
author = {Gärttner, Martin and Bohnet, Justin and Safavi-Naini, Arghavan and Wall, Michael and Bollinger, John and Rey, Ana},
year = {2016},
pages = {},
title = {Measuring out-of-time-order correlations and multiple quantum spectra in a trapped ion quantum magnet},
volume = {13},
journal = {Nature Physics},
doi = {10.1038/nphys4119}
}

@article{PlatUniInv0,
author = {P. Schiansky and T. Str\"{o}mberg and D. Trillo and V. Saggio and B. Dive and M. Navascu\'{e}s and P. Walther},
journal = {Optica},
number = {2},
pages = {200--205},
publisher = {Optica Publishing Group},
title = {Demonstration of universal time-reversal for qubit processes},
volume = {10},
year = {2023},
url = {https://opg.optica.org/optica/abstract.cfm?URI=optica-10-2-200},
doi = {10.1364/OPTICA.469109},
}

@article{PlatUniInv1,
  title = {Time-reversal in a dipolar quantum many-body spin system},
  author = {Geier, Sebastian and Braemer, Adrian and Braun, Eduard and M\"ullenbach, Maximilian and Franz, Titus and G\"arttner, Martin and Z\"urn, Gerhard and Weidem\"uller, Matthias},
  journal = {Phys. Rev. Res.},
  volume = {6},
  issue = {3},
  pages = {033197},
  numpages = {8},
  year = {2024},
  publisher = {American Physical Society},
  doi = {10.1103/PhysRevResearch.6.033197},
  url = {https://link.aps.org/doi/10.1103/PhysRevResearch.6.033197}
}

@article{DDReview,
  title = {Dynamical decoupling for superconducting qubits: A performance survey},
  author = {Ezzell, Nic and Pokharel, Bibek and Tewala, Lina and Quiroz, Gregory and Lidar, Daniel A.},
  journal = {Phys. Rev. Appl.},
  volume = {20},
  issue = {6},
  pages = {064027},
  numpages = {42},
  year = {2023},
  publisher = {American Physical Society},
  doi = {10.1103/PhysRevApplied.20.064027},
  url = {https://link.aps.org/doi/10.1103/PhysRevApplied.20.064027}
}

@article{DDNV,
  author = {Bar-Gill, N. and Pham, L. and Jarmola, A. and Budker, D. and Walsworth, R. L.},
  title = {Solid-state electronic spin coherence time approaching one second},
  journal = {Nature Communications},
  volume = {4},
  pages = {1743},
  year = {2013},
  doi = {10.1038/ncomms2771},
  url = {https://doi.org/10.1038/ncomms2771},
  date_received = {2012-11-27},
  date_accepted = {2013-03-20},
  date_published = {2013-04-23}
}

@article{Loschmidt0,
  title = {Environment-Independent Decoherence Rate in Classically Chaotic Systems},
  author = {Jalabert, Rodolfo A. and Pastawski, Horacio M.},
  journal = {Phys. Rev. Lett.},
  volume = {86},
  issue = {12},
  pages = {2490--2493},
  numpages = {0},
  year = {2001},
  publisher = {American Physical Society},
  doi = {10.1103/PhysRevLett.86.2490},
  url = {https://link.aps.org/doi/10.1103/PhysRevLett.86.2490}
}

@article{Loschmidt1,
  title = {Decoherence and the Loschmidt Echo},
  author = {Cucchietti, F. M. and Dalvit, D. A. R. and Paz, J. P. and Zurek, W. H.},
  journal = {Phys. Rev. Lett.},
  volume = {91},
  issue = {21},
  pages = {210403},
  numpages = {4},
  year = {2003},
  publisher = {American Physical Society},
  doi = {10.1103/PhysRevLett.91.210403},
  url = {https://link.aps.org/doi/10.1103/PhysRevLett.91.210403}
}

@article{Loschmidt2,
  title = {Loschmidt echo for quantum metrology},
  author = {Macr\`{\i}, Tommaso and Smerzi, Augusto and Pezz\`e, Luca},
  journal = {Phys. Rev. A},
  volume = {94},
  issue = {1},
  pages = {010102},
  numpages = {6},
  year = {2016},
  publisher = {American Physical Society},
  doi = {10.1103/PhysRevA.94.010102},
  url = {https://link.aps.org/doi/10.1103/PhysRevA.94.010102}
}

@article{Loschmidt3,
  author       = {Ran Liu and Ze Wu and Xiaodong Yang and Yuchen Li and Hui Zhou and Zhaokai Li and Yuquan Chen and Haidong Yuan and Xinhua Peng},
  title        = {Variational Quantum Metrology with Loschmidt Echo},
  journal      = {National Science Review},
  shortjournal = {Natl Sci Rev},
  year         = {2025},
  doi          = {10.1093/nsr/nwaf091},
  eprint       = {nwaf091},
  issn         = {2095-5138},
  url          = {https://doi.org/10.1093/nsr/nwaf091},
  urldate      = {2025-04-08}
}

@article{crypto1,
  author    = {Pilaszewicz, C. and Muth, L.R. and Margraf, M.},
  title     = {A black-box attack on fixed-unitary quantum encryption schemes},
  journal   = {Discover Computing},
  volume    = {27},
  number    = {14},
  year      = {2024},
  doi       = {10.1007/s10791-024-09448-3},
  url       = {https://doi.org/10.1007/s10791-024-09448-3},
  received  = {2024-04-19},
  accepted  = {2024-06-18},
  published = {2024-06-22}
}

@article{Trev0,
  title = {SU(2) and SU(1,1) interferometers},
  author = {Yurke, Bernard and McCall, Samuel L. and Klauder, John R.},
  journal = {Phys. Rev. A},
  volume = {33},
  issue = {6},
  pages = {4033--4054},
  numpages = {0},
  year = {1986},
  publisher = {American Physical Society},
  doi = {10.1103/PhysRevA.33.4033},
  url = {https://link.aps.org/doi/10.1103/PhysRevA.33.4033}
}

@article{Trev1,
  title = {Quantum advantage of time-reversed ancilla-based metrology of absorption parameters},
  author = {Wang, Jiaxuan and Filho, Ruynet L. de Matos and Agarwal, Girish S. and Davidovich, Luiz},
  journal = {Phys. Rev. Res.},
  volume = {6},
  issue = {1},
  pages = {013034},
  numpages = {11},
  year = {2024},
  publisher = {American Physical Society},
  doi = {10.1103/PhysRevResearch.6.013034},
  url = {https://link.aps.org/doi/10.1103/PhysRevResearch.6.013034}
}

@article{Trev2,
  title = {Qubit-assisted quantum metrology under a time-reversal strategy},
  author = {Chen, Peng and Jing, Jun},
  journal = {Phys. Rev. A},
  volume = {110},
  issue = {6},
  pages = {062425},
  numpages = {12},
  year = {2024},
  publisher = {American Physical Society},
  doi = {10.1103/PhysRevA.110.062425},
  url = {https://link.aps.org/doi/10.1103/PhysRevA.110.062425}
}

@misc{Trev3,
      title={Time-reversal assisted quantum metrology with an optimal control}, 
      author={Da-Wei Luo and Ting Yu},
      year={2023},
      eprint={2312.14443},
      archivePrefix={arXiv},
      primaryClass={quant-ph},
      url={https://arxiv.org/abs/2312.14443}, 
}

@article{Yoshida2023,
  title = {Reversing Unknown Qubit-Unitary Operation, Deterministically and Exactly},
  author = {Yoshida, Satoshi and Soeda, Akihito and Murao, Mio},
  journal = {Phys. Rev. Lett.},
  volume = {131},
  issue = {12},
  pages = {120602},
  numpages = {6},
  year = {2023},
  publisher = {American Physical Society},
  doi = {10.1103/PhysRevLett.131.120602},
  url = {https://link.aps.org/doi/10.1103/PhysRevLett.131.120602}
}

@misc{chen2024,
      title={Quantum Advantage in Reversing Unknown Unitary Evolutions}, 
      author={Yu-Ao Chen and Yin Mo and Yingjian Liu and Lei Zhang and Xin Wang},
      year={2024},
      eprint={2403.04704},
      archivePrefix={arXiv},
      primaryClass={quant-ph},
      url={https://arxiv.org/abs/2403.04704}, 
}

@article{Stromberg2024,
  title = {Experimental superposition of a quantum evolution with its time reverse},
  author = {Str\"omberg, Teodor and Schiansky, Peter and Quintino, Marco T\'ulio and Antesberger, Michael and Rozema, Lee A. and Agresti, Iris and Brukner, \ifmmode \check{C}\else \v{C}\fi{}aslav and Walther, Philip},
  journal = {Phys. Rev. Res.},
  volume = {6},
  issue = {2},
  pages = {023071},
  numpages = {12},
  year = {2024},
  publisher = {American Physical Society},
  doi = {10.1103/PhysRevResearch.6.023071},
  url = {https://link.aps.org/doi/10.1103/PhysRevResearch.6.023071}
}

@article{Martyn2021,
  title = {Grand Unification of Quantum Algorithms},
  author = {Martyn, John M. and Rossi, Zane M. and Tan, Andrew K. and Chuang, Isaac L.},
  journal = {PRX Quantum},
  volume = {2},
  issue = {4},
  pages = {040203},
  numpages = {40},
  year = {2021},
  publisher = {American Physical Society},
  doi = {10.1103/PRXQuantum.2.040203},
  url = {https://link.aps.org/doi/10.1103/PRXQuantum.2.040203}
}

@inproceedings{Gily_n_2019,
   title={Quantum singular value transformation and beyond: exponential improvements for quantum matrix arithmetics},
   url={http://dx.doi.org/10.1145/3313276.3316366},
   DOI={10.1145/3313276.3316366},
   booktitle={Proceedings of the 51st Annual ACM SIGACT Symposium on Theory of Computing},
   publisher={ACM},
   author={Gilyén, András and Su, Yuan and Low, Guang Hao and Wiebe, Nathan},
   year={2019},
pages={193–204},
   collection={STOC ’19} }

@article{Bisio_2019,
   title={Theoretical framework for higher-order quantum theory},
   volume={475},
   ISSN={1471-2946},
   url={http://dx.doi.org/10.1098/rspa.2018.0706},
   DOI={10.1098/rspa.2018.0706},
   number={2225},
   journal={Proceedings of the Royal Society A: Mathematical, Physical and Engineering Sciences},
   publisher={The Royal Society},
   author={Bisio, Alessandro and Perinotti, Paolo},
   year={2019},
pages={20180706} }

@article{Apellaniz_2018,
   title={Precision bounds for gradient magnetometry with atomic ensembles},
   volume={97},
   ISSN={2469-9934},
   url={http://dx.doi.org/10.1103/PhysRevA.97.053603},
   DOI={10.1103/physreva.97.053603},
   number={5},
   journal={Physical Review A},
   publisher={American Physical Society (APS)},
   author={Apellaniz, Iagoba and Urizar-Lanz, Iñigo and Zimborás, Zoltán and Hyllus, Philipp and Tóth, Géza},
   year={2018}}

@article{Ruster_2017,
   title={Entanglement-Based dc Magnetometry with Separated Ions},
   volume={7},
   ISSN={2160-3308},
   url={http://dx.doi.org/10.1103/PhysRevX.7.031050},
   DOI={10.1103/physrevx.7.031050},
   number={3},
   journal={Physical Review X},
   publisher={American Physical Society (APS)},
   author={Ruster, T. and Kaufmann, H. and Luda, M. A. and Kaushal, V. and Schmiegelow, C. T. and Schmidt-Kaler, F. and Poschinger, U. G.},
   year={2017}}

@article{Colombo_2022,
   title={Time-reversal-based quantum metrology with many-body entangled states},
   volume={18},
   ISSN={1745-2481},
   url={http://dx.doi.org/10.1038/s41567-022-01653-5},
   DOI={10.1038/s41567-022-01653-5},
   number={8},
   journal={Nature Physics},
   publisher={Springer Science and Business Media LLC},
   author={Colombo, Simone and Pedrozo-Peñafiel, Edwin and Adiyatullin, Albert F. and Li, Zeyang and Mendez, Enrique and Shu, Chi and Vuletić, Vladan},
   year={2022},
pages={925–930} }

@article{feynman1982,
  author    = {Feynman, Richard P.},
  title     = {Simulating physics with computers},
  journal   = {International Journal of Theoretical Physics},
  volume    = {21},
  number    = {6-7},
  pages     = {467--488},
  year      = {1982},
  doi       = {10.1007/BF02650179}
}

@article{Song2024,
  title = {Agnostic Phase Estimation},
  author = {Song, Xingrui and Salvati, Flavio and Gaikwad, Chandrashekhar and {Yunger Halpern}, Nicole and Arvidsson-Shukur, David R. M. and Murch, Kater},
  journal = {Phys. Rev. Lett.},
  volume = {132},
  issue = {26},
  pages = {260801},
  numpages = {7},
  year = {2024},
  publisher = {American Physical Society},
  doi = {10.1103/PhysRevLett.132.260801},
  url = {https://link.aps.org/doi/10.1103/PhysRevLett.132.260801}
}

@book{helstrom76,
  title={Quantum Detection and Estimation Theory},
  author={Helstrom, Carl W.},
  volume={123},
  year={1976},
  publisher={Elsevier},
  note={1st Edition},
  isbn={9780080956329}
}

@article{Ranadive2022,
	Author = {Ranadive, Arpit and Esposito, Martina and Planat, Luca and Bonet, Edgar and Naud, C{\'e}cile and Buisson, Olivier and Guichard, Wiebke and Roch, Nicolas},
	Da = {2022/04/01},
	Doi = {10.1038/s41467-022-29375-5},
	Id = {Ranadive2022},
	Isbn = {2041-1723},
	Journal = {Nature Communications},
	Number = {1},
	Pages = {1737},
	Title = {Kerr reversal in Josephson meta-material and traveling wave parametric amplification},
	Ty = {JOUR},
	Url = {https://doi.org/10.1038/s41467-022-29375-5},
	Volume = {13},
	Year = {2022}}

@article{Wei21,
AUTHOR = {Wei, Songrui and Liao, Xiaoqi and Zhang, Han and Pang, Jianhua and Zhou, Yan},
TITLE = {Recent Progress of Fluxgate Magnetic Sensors: Basic Research and Application},
JOURNAL = {Sensors},
VOLUME = {21},
YEAR = {2021},
NUMBER = {4},
ARTICLE-NUMBER = {1500},
URL = {https://www.mdpi.com/1424-8220/21/4/1500},
PubMedID = {33671507},
ISSN = {1424-8220},
DOI = {10.3390/s21041500}
}

@article{Stankevic23,
AUTHOR = {Stankevi\v{c}, Voitech and Ker\v{s}ulis, Skirmantas and Dilys, Justas and Bleizgys, Vytautas and Vilinas, Mindaugas and Vertelis, Viliuns and Maneikis, Andrius and Rudokas, Vakaris and Plau\v{s}inaitien\.e, Valentina and \v{Z}urauskien\.e, Nerija},
TITLE = {Measurement System for Short-Pulsed Magnetic Fields},
JOURNAL = {Sensors},
VOLUME = {23},
YEAR = {2023},
NUMBER = {3},
ARTICLE-NUMBER = {1435},
URL = {https://www.mdpi.com/1424-8220/23/3/1435},
PubMedID = {36772475},
ISSN = {1424-8220},
DOI = {10.3390/s23031435}
}

@article{Tang18,
AUTHOR = {Tang, Wei and Lyu, Fei and Wang, Dunhui and Pan, Hongbing},
TITLE = {A New Design of a Single-Device 3D Hall Sensor: Cross-Shaped 3D Hall Sensor},
JOURNAL = {Sensors},
VOLUME = {18},
YEAR = {2018},
NUMBER = {4},
ARTICLE-NUMBER = {1065},
URL = {https://www.mdpi.com/1424-8220/18/4/1065},
PubMedID = {29614815},
ISSN = {1424-8220}
}

@misc{smith2025riskminimizingstatesquantumphaseestimationalgorithm,
      title={Risk-minimizing states for the quantum-phase-estimation algorithm}, 
      author={Joseph G. Smith and Crispin H. W. Barnes and David R. M. Arvidsson-Shukur},
      year={2025},
      eprint={2505.00764},
      archivePrefix={arXiv},
      primaryClass={quant-ph},
      url={https://arxiv.org/abs/2505.00764}, 
}

@misc{rovny2025multiqubitnanoscalesensingentanglement,
      title={Multi-qubit nanoscale sensing with entanglement as a resource}, 
      author={Jared Rovny and Shimon Kolkowitz and Nathalie P. de Leon},
      year={2025},
      eprint={2504.12533},
      archivePrefix={arXiv},
      primaryClass={quant-ph},
      url={https://arxiv.org/abs/2504.12533}, 
}

@article{Smith24,
  title = {Adaptive Bayesian quantum algorithm for phase estimation},
  author = {Smith, Joseph G. and Barnes, Crispin H. W. and Arvidsson-Shukur, David R. M.},
  journal = {Phys. Rev. A},
  volume = {109},
  issue = {4},
  pages = {042412},
  numpages = {8},
  year = {2024},
  publisher = {American Physical Society},
  doi = {10.1103/PhysRevA.109.042412},
  url = {https://link.aps.org/doi/10.1103/PhysRevA.109.042412}
}

@article{Gaikwad_2024,
  title = {Entanglement Assisted Probe of the Non-Markovian to Markovian Transition in Open Quantum System Dynamics},
  author = {Gaikwad, Chandrashekhar and Kowsari, Daria and Brame, Carson and Song, Xingrui and Zhang, Haimeng and Esposito, Martina and Ranadive, Arpit and Cappelli, Giulio and Roch, Nicolas and Levenson-Falk, Eli M. and Murch, Kater W.},
  journal = {Phys. Rev. Lett.},
  volume = {132},
  issue = {20},
  pages = {200401},
  numpages = {7},
  year = {2024},
  publisher = {American Physical Society},
  doi = {10.1103/PhysRevLett.132.200401},
  url = {https://link.aps.org/doi/10.1103/PhysRevLett.132.200401}
}

@article{Walter2017,
  title = {Rapid High-Fidelity Single-Shot Dispersive Readout of Superconducting Qubits},
  author = {Walter, T. and Kurpiers, P. and Gasparinetti, S. and Magnard, P. and Poto\ifmmode \check{c}\else \v{c}\fi{}nik, A. and Salath\'e, Y. and Pechal, M. and Mondal, M. and Oppliger, M. and Eichler, C. and Wallraff, A.},
  journal = {Phys. Rev. Appl.},
  volume = {7},
  issue = {5},
  pages = {054020},
  numpages = {11},
  year = {2017},
  publisher = {American Physical Society},
  doi = {10.1103/PhysRevApplied.7.054020},
  url = {https://link.aps.org/doi/10.1103/PhysRevApplied.7.054020}
}

@article{Boissonneault2009,
  title = {Dispersive regime of circuit QED: Photon-dependent qubit dephasing and relaxation rates},
  volume = {79},
  ISSN = {1094-1622},
  url = {http://dx.doi.org/10.1103/PhysRevA.79.013819},
  DOI = {10.1103/physreva.79.013819},
  number = {1},
  journal = {Physical Review A},
  publisher = {American Physical Society (APS)},
  author = {Boissonneault,  Maxime and Gambetta,  J. M. and Blais,  Alexandre},
  year = {2009}
}

@article{Braunstein94,
  title = {Statistical distance and the geometry of quantum states},
  author = {Braunstein, Samuel L. and Caves, Carlton M.},
  journal = {Phys. Rev. Lett.},
  volume = {72},
  issue = {22},
  pages = {3439--3443},
  numpages = {0},
  year = {1994},
  publisher = {American Physical Society},
  doi = {10.1103/PhysRevLett.72.3439},
  url = {https://link.aps.org/doi/10.1103/PhysRevLett.72.3439}
}

@article{ArvidssonShukur23,
  title = {Nonclassical Advantage in Metrology Established via Quantum Simulations of Hypothetical Closed Timelike Curves},
  author = {Arvidsson-Shukur, David R. M. and McConnell, Aidan G. and Yunger Halpern, Nicole},
  journal = {Phys. Rev. Lett.},
  volume = {131},
  issue = {15},
  pages = {150202},
  numpages = {6},
  year = {2023},
  publisher = {American Physical Society},
  doi = {10.1103/PhysRevLett.131.150202},
  url = {https://link.aps.org/doi/10.1103/PhysRevLett.131.150202}
}

@book{Nielsen11,
author = {Nielsen, Michael A. and Chuang, Isaac L.},
 title = {Quantum Computation and Quantum Information: 10th Anniversary Edition},
 year = {2011},
 isbn = {1107002176, 9781107002173},
 edition = {10th},
 publisher = {Cambridge University Press},
 address = {New York, NY, USA}
}

@article{Chuang97,
author = {Isaac L. Chuang and M. A. Nielsen},
title = {Prescription for experimental determination of the dynamics of a quantum black box},
journal = {Journal of Modern Optics},
volume = {44},
number = {11-12},
pages = {2455-2467},
year = {1997},
publisher = {Taylor & Francis},
doi = {10.1080/09500349708231894}}

@article{Altepeter03,
  title = {Ancilla-Assisted Quantum Process Tomography},
  author = {Altepeter, J. B. and Branning, D. and Jeffrey, E. and Wei, T. C. and Kwiat, P. G. and Thew, R. T. and O'Brien, J. L. and Nielsen, M. A. and White, A. G.},
  journal = {Phys. Rev. Lett.},
  volume = {90},
  issue = {19},
  pages = {193601},
  numpages = {4},
  year = {2003},
  publisher = {American Physical Society},
  doi = {10.1103/PhysRevLett.90.193601},
  url = {https://link.aps.org/doi/10.1103/PhysRevLett.90.193601}
}

@article{DAriano01,
  title = {Quantum Tomography for Measuring Experimentally the Matrix Elements of an Arbitrary Quantum Operation},
  author = {D'Ariano, G. M. and Lo Presti, P.},
  journal = {Phys. Rev. Lett.},
  volume = {86},
  issue = {19},
  pages = {4195--4198},
  numpages = {0},
  year = {2001},
  publisher = {American Physical Society},
  doi = {10.1103/PhysRevLett.86.4195},
  url = {https://link.aps.org/doi/10.1103/PhysRevLett.86.4195}
}

@article{Liu2019,
	doi = {10.1088/1751-8121/ab5d4d},
	url = {https://doi.org/10.1088/1751-8121/ab5d4d},
	year = 2019,
	publisher = {IOP Publishing},
	volume = {53},
	number = {2},
	pages = {023001},
	author = {Liu, J. and Yuan, H. and Lu, X. and Wang, X.},
	title = {Quantum {F}isher information matrix and multiparameter estimation},
	journal = {J. Phys. A Math.},
}

@book{Nielsen2012,
  title = {Quantum Computation and Quantum Information: 10th Anniversary Edition},
  ISBN = {9780511976667},
  url = {http://dx.doi.org/10.1017/CBO9780511976667},
  DOI = {10.1017/cbo9780511976667},
  publisher = {Cambridge University Press},
  author = {Nielsen,  Michael A. and Chuang,  Isaac L.},
  year = {2012}
}

@article{Nachman2020,
  title = {Unfolding quantum computer readout noise},
  volume = {6},
  ISSN = {2056-6387},
  url = {http://dx.doi.org/10.1038/s41534-020-00309-7},
  DOI = {10.1038/s41534-020-00309-7},
  number = {1},
  journal = {npj Quantum Information},
  publisher = {Springer Science and Business Media LLC},
  author = {Nachman,  Benjamin and Urbanek,  Miroslav and de Jong,  Wibe A. and Bauer,  Christian W.},
  year = {2020}
}

@article{scikit-learn,
  title={Scikit-learn: Machine Learning in {P}ython},
  author={Pedregosa, F. and Varoquaux, G. and Gramfort, A. and Michel, V.
          and Thirion, B. and Grisel, O. and Blondel, M. and Prettenhofer, P.
          and Weiss, R. and Dubourg, V. and Vanderplas, J. and Passos, A. and
          Cournapeau, D. and Brucher, M. and Perrot, M. and Duchesnay, E.},
  journal={Journal of Machine Learning Research},
  volume={12},
  pages={2825--2830},
  year={2011}
}

@misc{nevergrad,
    author = {J. Rapin and O. Teytaud},
    title = {{Nevergrad - A gradient-free optimization platform}},
    year = {2018},
    publisher = {GitHub},
    journal = {GitHub repository},
    howpublished = {\url{https://GitHub.com/FacebookResearch/Nevergrad}},
}

@article{Song2021,
  title = {Quantum process inference for a single-qubit Maxwell demon},
  volume = {104},
  ISSN = {2469-9934},
  url = {http://dx.doi.org/10.1103/PhysRevA.104.022211},
  DOI = {10.1103/physreva.104.022211},
  number = {2},
  journal = {Physical Review A},
  publisher = {American Physical Society (APS)},
  author = {Song,  Xingrui and Naghiloo,  Mahdi and Murch,  Kater},
  year = {2021}
}

@article{James2001,
  title = {Measurement of qubits},
  volume = {64},
  ISSN = {1094-1622},
  url = {http://dx.doi.org/10.1103/PhysRevA.64.052312},
  DOI = {10.1103/physreva.64.052312},
  number = {5},
  journal = {Physical Review A},
  publisher = {American Physical Society (APS)},
  author = {James,  Daniel F. V. and Kwiat,  Paul G. and Munro,  William J. and White,  Andrew G.},
  year = {2001}
}

@article{Carroll2022,
  title = {Dynamics of superconducting qubit relaxation times},
  volume = {8},
  ISSN = {2056-6387},
  url = {https://doi.org/10.1038/s41534-022-00643-y},
  DOI = {10.1038/s41534-022-00643-y},
  number = {1},
  journal = {npj Quantum Information},
  publisher = {Springer Nature},
  author = {Carroll, M. and Rosenblatt, S. and Jurcevic, P. and Lauer, I. and Kandala, A.},
  year = {2022}
}

@article{42_Stuckelberg,
   author = {St\"{u}ckelberg, E. C. C.},
   journal = "Helv. Phys. Acta",
   year = "1942",
   volume = "15",
   pages = "23-37",
   title = "La Mecanique du point materiel en theorie de relativite et en theorie des quanta"}

@article{48_Feynman_Relativistic,
  title = {A relativistic cut-Off for classical electrodynamics},
  author = {Feynman, R. P.},
  journal = {Phys. Rev.},
  volume = {74},
  issue = {8},
  pages = {939--946},
  numpages = {0},
  year = {1948},
  publisher = {American Physical Society},
  doi = {10.1103/PhysRev.74.939},
  url = {https://link.aps.org/doi/10.1103/PhysRev.74.939}
}

@misc{65_Feynman_Nobel,
   title = "The development of the space-time view of quantum electrodynamics",
   year = "1965",
   author = "Feynman, R. P.",
   note = "Nobel Lecture"
}

@article{49_Feynman_Theory,
  title = {The theory of positrons},
  author = {Feynman, R. P.},
  journal = {Phys. Rev.},
  volume = {76},
  issue = {6},
  pages = {749--759},
  numpages = {0},
  year = {1949},
  publisher = {American Physical Society},
  doi = {10.1103/PhysRev.76.749},
  url = {https://link.aps.org/doi/10.1103/PhysRev.76.749}
}

@phdthesis{xingrui_thesis,
  author       = {Song, Xingrui},
  title        = {Entanglement-Enhanced Metrology With Superconducting Circuits},
  school       = {Washington University in St. Louis},
  year         = {2025},
  type         = {PhD Dissertation},
  address      = {St. Louis, Missouri}
}

@misc{Wang25,
      title={Proposal for antimatter sensor for quantum metrology}, 
      author={Wang, Y.-X. and Salvati, Flavio and Doser, Michael and Murch, Kater and Arvidsson-Shukur, David R. M. and {Yunger Halpern}, Nicole},
      year={in prep}
}

@article{Cassidy2006,
    author = {Cassidy, D. B. and Deng, S. H. M. and Tanaka, H. K. M. and {A. P. Mills, Jr.}},
    title = "{Single shot positron annihilation lifetime spectroscopy}",
    journal = {Applied Physics Letters},
    volume = {88},
    number = {19},
    pages = {194105},
    year = {2006},
    issn = {0003-6951},
    doi = {10.1063/1.2203336},
    url = {https://doi.org/10.1063/1.2203336}
}

@article{AEgiS2024,
  title = {Positronium Laser Cooling via the ${1}^{3}S\text{\ensuremath{-}}{2}^{3}P$ Transition with a Broadband Laser Pulse},
  author = {Gl\"oggler, L. T. and Gusakova, N. and Rien\"acker, B. and Camper, A. and Caravita, R. and Huck, S. and Volponi, M. and Wolz, T. and Penasa, L. and Krumins, V. and Gustafsson, F. P. and Comparat, D. and Auzins, M. and Bergmann, B. and Burian, P. and Brusa, R. S. and Castelli, F. and Cerchiari, G. and Ciury\l{}o, R. and Consolati, G. and Doser, M. and Graczykowski, \L{}. and Grosbart, M. and Guatieri, F. and Haider, S. and Janik, M. A. and Kasprowicz, G. and Khatri, G. and K\l{}osowski, \L{}. and Kornakov, G. and Lappo, L. and Linek, A. and Malamant, J. and Mariazzi, S. and Petracek, V. and Piwi\ifmmode \acute{n}\else \'{n}\fi{}ski, M. and Posp\'{\i}\ifmmode \check{s}\else \v{s}\fi{}il, S. and Povolo, L. and Prelz, F. and Rangwala, S. A. and Rauschendorfer, T. and Rawat, B. S. and Rodin, V. and R\o{}hne, O. M. and Sandaker, H. and Smolyanskiy, P. and Sowi\ifmmode \acute{n}\else \'{n}\fi{}ski, T. and Tefelski, D. and Vafeiadis, T. and Welsch, C. P. and Zawada, M. and Zielinski, J. and Zurlo, N.},
  collaboration = {AE\ifmmode \bar{g}\else \={g}\fi{}IS Collaboration},
  journal = {Phys. Rev. Lett.},
  volume = {132},
  issue = {8},
  pages = {083402},
  numpages = {7},
  year = {2024},
  publisher = {American Physical Society},
  doi = {10.1103/PhysRevLett.132.083402},
  url = {https://link.aps.org/doi/10.1103/PhysRevLett.132.083402}
}

@article{paris2009quantum,
  title={Quantum estimation for quantum technology},
  author={Paris, Matteo G. A.},
  journal={Int. J. Quantum Inf.},
  volume={7},
  number={supp01},
  pages={125--137},
  year={2009},
  publisher={World Scientific},
  doi = {https://doi.org/10.1142/S0219749909004839},
  url = {https://doi.org/10.1142/S0219749909004839}
}

@article{smith2023adaptive,
  title = {Iterative quantum-phase-estimation protocol for shallow circuits},
  author = {Smith, Joseph G. and Barnes, Crispin H. W. and Arvidsson-Shukur, David R. M.},
  journal = {Phys. Rev. A},
  volume = {106},
  issue = {6},
  pages = {062615},
  numpages = {7},
  year = {2022},
  publisher = {American Physical Society},
  doi = {10.1103/PhysRevA.106.062615},
  url = {https://link.aps.org/doi/10.1103/PhysRevA.106.062615}
}

@article{18_NYH_Quasiprobability,
  title = {Quasiprobability behind the out-of-time-ordered correlator},
  author = {{Yunger Halpern}, Nicole and Swingle, Brian and Dressel, Justin},
  journal = {Phys. Rev. A},
  volume = {97},
  issue = {4},
  pages = {042105},
  numpages = {37},
  year = {2018},
  publisher = {American Physical Society},
  doi = {10.1103/PhysRevA.97.042105},
  url = {https://link.aps.org/doi/10.1103/PhysRevA.97.042105}
}

@article{16_Swingle_Measuring,
  title = {Measuring the scrambling of quantum information},
  author = {Swingle, Brian and Bentsen, Gregory and Schleier-Smith, Monika and Hayden, Patrick},
  journal = {Phys. Rev. A},
  volume = {94},
  issue = {4},
  pages = {040302},
  numpages = {6},
  year = {2016},
  publisher = {American Physical Society},
  doi = {10.1103/PhysRevA.94.040302},
  url = {https://link.aps.org/doi/10.1103/PhysRevA.94.040302}
}

@article{16_Zhu_Measurement,
  title = {Measurement of many-body chaos using a quantum clock},
  author = {Zhu, Guanyu and Hafezi, Mohammad and Grover, Tarun},
  journal = {Phys. Rev. A},
  volume = {94},
  issue = {6},
  pages = {062329},
  numpages = {16},
  year = {2016},
  publisher = {American Physical Society},
  doi = {10.1103/PhysRevA.94.062329},
  url = {https://link.aps.org/doi/10.1103/PhysRevA.94.062329}
}

@article{Hayashi2020, 
    title={Quantum state estimation with nuisance parameters}, 
    author={Suzuki, Jun and Yang, Yuxiang and Hayashi, Masahito}, 
    volume={53}, 
    ISSN={1751-8121}, 
    url={http://dx.doi.org/10.1088/1751-8121/ab8b78}, 
    DOI={10.1088/1751-8121/ab8b78}, 
    number={45}, 
    journal={Journal of Physics A: Mathematical and Theoretical}, 
    publisher={IOP Publishing}, 
    year={2020}, 
    pages={453001} 
}

@article{Bennett1993Teleportation,
  title = {Teleporting an Unknown Quantum State via Dual Classical and Einstein-Podolsky-Rosen Channels},
  author = {Bennett, Charles H. and Brassard, Gilles and Cr{\'e}peau, Claude and Jozsa, Richard and Peres, Asher and Wootters, William K.},
  journal = {Physical Review Letters},
  volume = {70},
  number = {13},
  pages = {1895--1899},
  year = {1993},
  doi = {10.1103/PhysRevLett.70.1895}
}

@article{Pirandola2015TeleportationReview,
  title = {Advances in Quantum Teleportation},
  author = {Pirandola, Stefano and Eisert, Jens and Weedbrook, Christian and Furusawa, Akira and Braunstein, Samuel L.},
  journal = {Nature Photonics},
  volume = {9},
  pages = {641--652},
  year = {2015},
  doi = {10.1038/nphoton.2015.154}
}

@article{10_Saffman_Quantum,
  title = {Quantum information with Rydberg atoms},
  author = {Saffman, M. and Walker, T. G. and M\o{}lmer, K.},
  journal = {Rev. Mod. Phys.},
  volume = {82},
  issue = {3},
  pages = {2313--2363},
  numpages = {0},
  year = {2010},
  publisher = {American Physical Society},
  doi = {10.1103/RevModPhys.82.2313},
  url = {https://link.aps.org/doi/10.1103/RevModPhys.82.2313}
}

@article{19_Bruzewicz_Trapped,
    author = {Bruzewicz, Colin D. and Chiaverini, John and McConnell, Robert and Sage, Jeremy M.},
    title = {Trapped-ion quantum computing: Progress and challenges},
    journal = {Applied Physics Reviews},
    volume = {6},
    number = {2},
    pages = {021314},
    year = {2019},
    issn = {1931-9401},
    doi = {10.1063/1.5088164},
    url = {https://doi.org/10.1063/1.5088164},
}

@article{25_Rovny_Multi,
  title={Multi-qubit nanoscale sensing with entanglement as a resource},
  author={Rovny, Jared and Kolkowitz, Shimon and de Leon, Nathalie P},
  journal={Nature},
  year={2025},
  volume = "647",
  pages = "876–882"
}

@article{14_Schirhagl_Nitrogen,
   author = "Schirhagl, Romana and Chang, Kevin and Loretz, Michael and Degen, Christian L.",
   title = "Nitrogen-Vacancy Centers in Diamond: Nanoscale Sensors for Physics and Biology", 
   journal= "Annual Review of Physical Chemistry",
   year = "2014",
   volume = "65",
   number = "Volume 65, 2014",
   pages = "83-105",
   doi = "https://doi.org/10.1146/annurev-physchem-040513-103659",
   url = "https://www.annualreviews.org/content/journals/10.1146/annurev-physchem-040513-103659",
   publisher = "Annual Reviews",
   issn = "1545-1593"
   }

@ARTICLE{25_Barends_Performance,
       author = {{Barends}, R. and {Wilhelm}, F.~K.},
        title = "{Performance-centric roadmap for building a superconducting quantum computer}",
      journal = {arXiv e-prints},
         year = 2025,
          eid = {arXiv:2506.23178},
        pages = {arXiv:2506.23178},
          doi = {10.48550/arXiv.2506.23178},
archivePrefix = {arXiv},
       eprint = {2506.23178},
 primaryClass = {quant-ph},
       adsurl = {https://ui.adsabs.harvard.edu/abs/2025arXiv250623178B}
}

@misc{IBM_roadmap,
   howpublished = "https://www.ibm.com/roadmaps/quantum/",
   title = "Quantum Roadmap",
   author = "IBM"
}

@ARTICLE{Microsoft_roadmap,
       author = {{Aasen}, David and {Aghaee}, Morteza and {Alam}, Zulfi and {Andrzejczuk}, Mariusz and {Antipov}, Andrey and {Astafev}, Mikhail and {Avilovas}, Lukas and {Barzegar}, Amin and {Bauer}, Bela and {Becker}, Jonathan and {Bello-Rivas}, Juan M. and {Bhaskar}, Umesh and {Bocharov}, Alex and {Boddapati}, Srini and {Bohn}, David and {Bommer}, Jouri and {Bonderson}, Parsa and {Borovsky}, Jan and {Bourdet}, Leo and {Boutin}, Samuel and {Brown}, Tom and {Campbell}, Gary and {Casparis}, Lucas and {Chakravarthi}, Srivatsa and {Chao}, Rui and {Chapman}, Benjamin J. and {Chatoor}, Sohail and {Christensen}, Anna Wulff and {Codd}, Patrick and {Cole}, William and {Cooper}, Paul and {Corsetti}, Fabiano and {Cui}, Ajuan and {van Dam}, Wim and {Dandachi}, Tareq El and {Daraeizadeh}, Sahar and {Dumitrascu}, Adrian and {Ekefj{\"a}rd}, Andreas and {Fallahi}, Saeed and {Galletti}, Luca and {Gardner}, Geoff and {Gatta}, Raghu and {Gavranovic}, Haris and {Goulding}, Michael and {Govender}, Deshan and {Griggio}, Flavio and {Grigoryan}, Ruben and {Grijalva}, Sebastian and {Gronin}, Sergei and {Gukelberger}, Jan and {Haah}, Jeongwan and {Hamdast}, Marzie and {Hansen}, Esben Bork and {Hastings}, Matthew and {Heedt}, Sebastian and {Ho}, Samantha and {Hogaboam}, Justin and {Holgaard}, Laurens and {Van Hoogdalem}, Kevin and {Indrapiromkul}, Jinnapat and {Ingerslev}, Henrik and {Ivancevic}, Lovro and {Jablonski}, Sarah and {Jensen}, Thomas and {Jhoja}, Jaspreet and {Jones}, Jeffrey and {Kalashnikov}, Kostya and {Kallaher}, Ray and {Kalra}, Rachpon and {Karimi}, Farhad and {Karzig}, Torsten and {Kimes}, Seth and {Kliuchnikov}, Vadym and {Kloster}, Maren Elisabeth and {Knapp}, Christina and {Knee}, Derek and {Koski}, Jonne and {Kostamo}, Pasi and {Kuesel}, Jamie and {Lackey}, Brad and {Laeven}, Tom and {Lai}, Jeffrey and {de Lange}, Gijs and {Larsen}, Thorvald and {Lee}, Jason and {Lee}, Kyunghoon and {Leum}, Grant and {Li}, Kongyi and {Lindemann}, Tyler and {Lucas}, Marijn and {Lutchyn}, Roman and {Madsen}, Morten Hannibal and {Madulid}, Nash and {Manfra}, Michael and {Markussen}, Signe Brynold and {Martinez}, Esteban and {Mattila}, Marco and {Mattinson}, Jake and {McNeil}, Robert and {Mei}, Antonio Rodolph and {Mishmash}, Ryan V. and {Mohandas}, Gopakumar and {Mollgaard}, Christian and {de Moor}, Michiel and {Morgan}, Trevor and {Moussa}, George and {Narla}, Anirudh and {Nayak}, Chetan and {Nielsen}, Jens Hedegaard and {Nielsen}, William Hvidtfelt Padk{\ae}r and {Nolet}, Fr{\'e}d{\'e}ric and {Nystrom}, Mike and {O'Farrell}, Eoin and {Otani}, Keita and {Paetznick}, Adam and {Papon}, Camille and {Paz}, Andres and {Petersson}, Karl and {Petit}, Luca and {Pikulin}, Dima and {Pons}, Diego Olivier Fernandez and {Quinn}, Sam and {Rajpalke}, Mohana and {Ramirez}, Alejandro Alcaraz and {Rasmussen}, Katrine and {Razmadze}, David and {Reichardt}, Ben and {Ren}, Yuan and {Reneris}, Ken and {Riccomini}, Roy and {Sadovskyy}, Ivan and {Sainiemi}, Lauri and {Salda{\~n}a}, Juan Carlos Estrada and {Sanlorenzo}, Irene and {Schaal}, Simon and {Schmidgall}, Emma and {Sfiligoj}, Cristina and {da Silva}, Marcus P. and {Singh}, Shilpi and {Sinha}, Sarat and {Soeken}, Mathias and {Sohr}, Patrick and {Stankevic}, Tomas and {Stek}, Lieuwe and {Str{\o}m-Hansen}, Patrick and {Stuppard}, Eric and {Sundaram}, Aarthi and {Suominen}, Henri and {Suter}, Judith and {Suzuki}, Satoshi and {Svore}, Krysta and {Teicher}, Sam and {Thiyagarajah}, Nivetha and {Tholapi}, Raj and {Thomas}, Mason and {Tom}, Dennis and {Toomey}, Emily and {Tracy}, Josh and {Troyer}, Matthias and {Turley}, Michelle and {Turner}, Matthew D. and {Upadhyay}, Shivendra and {Urban}, Ivan and {Vaschillo}, Alexander and {Viazmitinov}, Dmitrii and {Vogel}, Dominik and {Wang}, Zhenghan and {Watson}, John and {Webster}, Alex and {Weston}, Joseph and {Williamson}, Timothy and {Winkler}, Georg W. and {van Woerkom}, David J. and {W{\"u}tz}, Brian Paquelet and {Yang}, Chung Kai and {Yu}, Richard and {Yucelen}, Emrah and {Zamorano}, Jes{\'u}s Herranz and {Zeisel}, Roland and {Zheng}, Guoji and {Zilke}, Justin and {Zimmerman}, Andrew},
        title = "{Blueprint for fault-tolerant quantum computation with topological qubit arrays}",
      journal = {Physical Review Research},
         year = 2025,
       volume = {7},
       number = {4},
          eid = {041002},
        pages = {041002},
          doi = {10.1103/qx36-4rv1},
archivePrefix = {arXiv},
       eprint = {2502.12252},
 primaryClass = {quant-ph},
       adsurl = {https://ui.adsabs.harvard.edu/abs/2025PhRvR...7d1002A}
}

@misc{Google_roadmap,
   title = "Our quantum computing roadmap",
   howpublished = "https://quantumai.google/roadmap",
   author = "Google"
}

@article{25_Gardner_Stochastic,
  title = {Stochastic Waveform Estimation at the Fundamental Quantum Limit},
  author = {Gardner, James W. and Gefen, Tuvia and Haine, Simon A. and Hope, Joseph J. and Preskill, John and Chen, Yanbei and McCuller, Lee},
  journal = {PRX Quantum},
  volume = {6},
  issue = {3},
  pages = {030311},
  numpages = {42},
  year = {2025},
  publisher = {American Physical Society},
  doi = {10.1103/h91r-4ws9},
  url = {https://link.aps.org/doi/10.1103/h91r-4ws9}
}

@article{Cerf1997,
  author    = {Cerf, Nicolas J. and Adami, Christoph},
  title     = {Negative Entropy and Information in Quantum Mechanics},
  journal   = {Physical Review Letters},
  volume    = {79},
  number    = {26},
  pages     = {5194--5197},
  year      = {1997},
  month     = dec,
  publisher = {American Physical Society},
  doi       = {10.1103/PhysRevLett.79.5194},
  url       = {https://doi.org/10.1103/PhysRevLett.79.5194}
}

\end{document}